\documentclass[aps, superscriptaddress,  twocolumn]{revtex4}
\usepackage{CJK}
\usepackage{mathtools}
\usepackage[normalem]{ulem}
\usepackage{graphicx}% Include figure files
\usepackage{dcolumn}% Align table columns on decimal point
\usepackage[mathlines]{lineno}
\usepackage[hidelinks]{hyperref}
\usepackage{bm}% bold math
\usepackage{amssymb}
\usepackage{amsmath}
\usepackage{braket}
\usepackage{color}
\usepackage{xcolor}
\usepackage{soul} 
\usepackage{bbm}
\usepackage{dsfont}
\usepackage{epstopdf} %converting to PDF
\usepackage{titlecaps}
\usepackage{titlesec}
\newcommand{\IN}[1]{\textcolor{black}{ #1}}
\newcommand{\BS}[1]{\textcolor{black}{ #1}}

\newcommand{\av}[1]{\textcolor{black}{ #1}}

\newcommand{\revise}[1]{\textcolor{black}{ #1}}

\newcommand{\rvs}[1]{\textcolor{black}{ #1}}

\newcommand{\beginsupplement}{%
        \setcounter{table}{0}
        \renewcommand{\tablename}{Supplementary Table}%
        \setcounter{figure}{0}
        \renewcommand{\figurename}{Supplementary Fig.}%
         \setcounter{equation}{0}
        \renewcommand{\theequation}{S\arabic{equation}}%
         \setcounter{section}{0}
        \renewcommand{\thesection}{Supplementary Note \arabic{section} } 
     }

\makeatletter
\renewcommand\@seccntformat[1]{\csname the#1\endcsname - }
\makeatother

\begin{document}
\title{Quantum transport of high-dimensional spatial information with a nonlinear detector}

\author{Bereneice Sephton}
\affiliation{School of Physics, University of the Witwatersrand, Private Bag 3, Wits 2050, South Africa}

\author{Adam Vall\'es}
\email{adam.valles@icfo.eu}
\affiliation{School of Physics, University of the Witwatersrand, Private Bag 3, Wits 2050, South Africa}
\affiliation{Molecular Chirality Research Center, Chiba University, 1-33 Yayoi-cho, Inage-ku, Chiba 263-8522, Japan}
\affiliation{ICFO - Institut de Ciencies Fotoniques, The Barcelona Institute of Science and Technology, Castelldefels (Barcelona) 08860, Spain}

\author{Isaac Nape}
\affiliation{School of Physics, University of the Witwatersrand, Private Bag 3, Wits 2050, South Africa}

\author{Mitchell A. Cox}
\affiliation{School of Electrical and Information Engineering, University of the Witwatersrand, Johannesburg, South Africa}

\author{Fabian Steinlechner}
\affiliation{Fraunhofer Institute for Applied Optics and Precision Engineering, Albert-Einstein-Str. 7, 07745 Jena, Germany}
\affiliation{Friedrich Schiller University Jena, Abbe Center of Photonics, Albert-Einstein-Str. 6, 07745 Jena, Germany}

\author{Thomas~Konrad}
\affiliation{School of Chemistry and Physics, University of KwaZulu-Natal, Durban, South Africa}
\affiliation{National Institute of Theoretical and Computational Sciences (NITheCS), KwaZulu-Natal, South Africa}

\author{Juan~P.~Torres}
\affiliation{ICFO - Institut de Ciencies Fotoniques, The Barcelona Institute of Science and Technology, Castelldefels (Barcelona) 08860, Spain}
\affiliation{Department of Signal Theory and Communications, Universitat Politecnica de Catalunya, Campus Nord D3, 08034 Barcelona, Spain}

\author{Filippus S. Roux}
\affiliation{National Metrology Institute of South Africa, Meiring Naud\'e Road, Brummeria, Pretoria 0040, South Africa}

\author{Andrew Forbes}
\email{andrew.forbes@wits.ac.za}
\affiliation{School of Physics, University of the Witwatersrand, Private Bag 3, Wits 2050, South Africa}

\begin{abstract}
\end{abstract}
\maketitle

%%%%%%%%%%%%%%%%%%%%%%%%%%%%%%%%%%%%%%%%%%%%%%%%%%
% \section{Introduction}
%%%%%%%%%%%%%%%%%%%%%%%%%%%%%%%%%%%%%%%%%%%%%%%%%%
\noindent 

\noindent \textbf{\revise{Information exchange between two distant parties, where information is shared without physically transporting it, is a crucial resource in future quantum networks.  Doing so with high-dimensional states offers the promise of higher information capacity and improved resilience to noise, but progress to date has been limited.  Here we demonstrate how a nonlinear parametric process allows for arbitrary high-dimensional state projections in the spatial degree of freedom, where a strong coherent field enhances the probability of the process.  This allows us to experimentally realise quantum transport of high-dimensional spatial information facilitated by a quantum channel with a single entangled pair and a nonlinear spatial mode detector. Using sum frequency generation we upconvert one of the photons from an entangled pair resulting in high-dimensional spatial information transported to the other.  We realise a $d=15$ quantum channel for arbitrary photonic spatial modes which we demonstrate by faithfully transferring information encoded into orbital angular momentum, Hermite-Gaussian and arbitrary spatial mode superpositions, without requiring knowledge of the state to be sent. Our demonstration merges the nascent fields of nonlinear control of structured light with quantum processes, offering a new approach to harnessing high-dimensional quantum states, and may be extended to other degrees of freedom too.}} 
%While our quantum transport process is stimulated with a coherent source, it could in future be performed with a single photon input by advances in nonlinear optics without changing its single entangled pair configuration, offering the first versatile and scalable approach for high-dimensional teleportation of information.} %Our new approach to quantum transport is both versatile and scaleable, converging to high-dimensional quantum teleportation in the limit of a single photon input.

%\AFout{Quantum teleporation can facilitate this through the sharing of entangled photons and a classical communication channel, but is limited by the commonly used linear optical detection schemes that require the number of ancillary photons to grow with dimension.}  
%Teleportation \AV{is an established tool} for information exchange between two distant parties, facilitated by the sharing of entangled photons and a classical communication channel. With increased dimensionality, teleportation offers the promise of higher information capacity and improved resilience to noise, but is limited by the commonly used linear optical detection schemes that require the number of ancillary photons to grow with dimension. 
%Additionally, quantum teleportation offers a simple way to explore phenomenon such as closed time-like curves \cite{lloyd2011closed}. 

Information exchange is the backbone of modern society, with our world connected by global networks of fibre and terrestrial links.  Quantum technologies allow this exchange to be fundamentally secure, fueling the nascent quantum global network \cite{kimble2008quantum}.  For example, quantum key distribution exchanges a key from peer to peer (usually Alice and Bob) to decode the information transmitted between communicating parties \cite{pirandola2020advances}, quantum secret sharing splits such a key amongst many nodes \cite{hillery1999quantum} and quantum secure direct communication sends it without a key, but rather encoded in a transmitted quantum state \cite{pan2023free}. In all these schemes, like its classical counterpart, the information is sent across a physical link between the sender and receiver. Remote state preparation \cite{bennett2001remote,bennett2005remote} allows information exchange between parties without transmitting the information physically across the link, but the sender (Alice) must know the information to be sent.  Teleportation \cite{bennett1993teleporting,nielsen2002quantum,wilde2013quantum,weedbrook2012gaussian} allows protected information exchange between distant parties without the need for a physical link \cite{RevModPhys.74.145}, facilitated by the sharing of entangled photons and a classical communication channel, where the information sent must not be known by Alice.  

\revise{All the aforementioned schemes would benefit from using high dimensional quantum states, offering higher channel capacity \cite{barreiro2008beating}, security \cite{bouchard2017high}, or resilience to noise \cite{ecker2019overcoming}.  In the context of spatial modes of light as a basis, orbital angular momentum (OAM) has proven particularly useful and topical \cite{mair2001entanglement,molina2007twisted, erhard2018twisted}, as has path \cite{kysela2020path} and pixels \cite{valencia2020high}, as potential routes towards high-dimensions. Yet experimental progress has been slow, with sharing keys shown up to $d = 6$ in optical fibre \cite{cozzolino2019high} and $d = 7$ in free-space \cite{mirhosseini2015high}, and sharing secrets up to $d=11$ \cite{pinnell2020experimental}.  Our interest is in schemes where the information is remotely shared and not physically sent, such as teleportation, which has been limited to $d=2$ using OAM \cite{Leach2017,liu2020orbital,Wang2015} and $d=3$ using the path degree of freedom \cite{luo2019quantum, PhysRevLett.125.230501}. So far all of these approaches have used linear optics for their state control and detection, which has known limitations in the context of high-dimensional states \cite{calsamiglia2002generalized}.  More recently, nonlinear optics has emerged as an exciting creation, control and detection tool for spatially structured classical light \cite{buono2022nonlinear}, but has not found its way to controlling spatially structured quantum states beyond polarisation qubit measurement \cite{PhysRevLett.86.1370}. Although theoretical schemes have been proposed to use nonlinear approaches for high-dimensional quantum information processing and communication \cite{molotkov1998quantum, walborn2010spatial, humble2010spectral}, none have yet been demonstrated experimentally.} 

\begin{figure}[h!]
\centering
\includegraphics[width=0.95\linewidth]{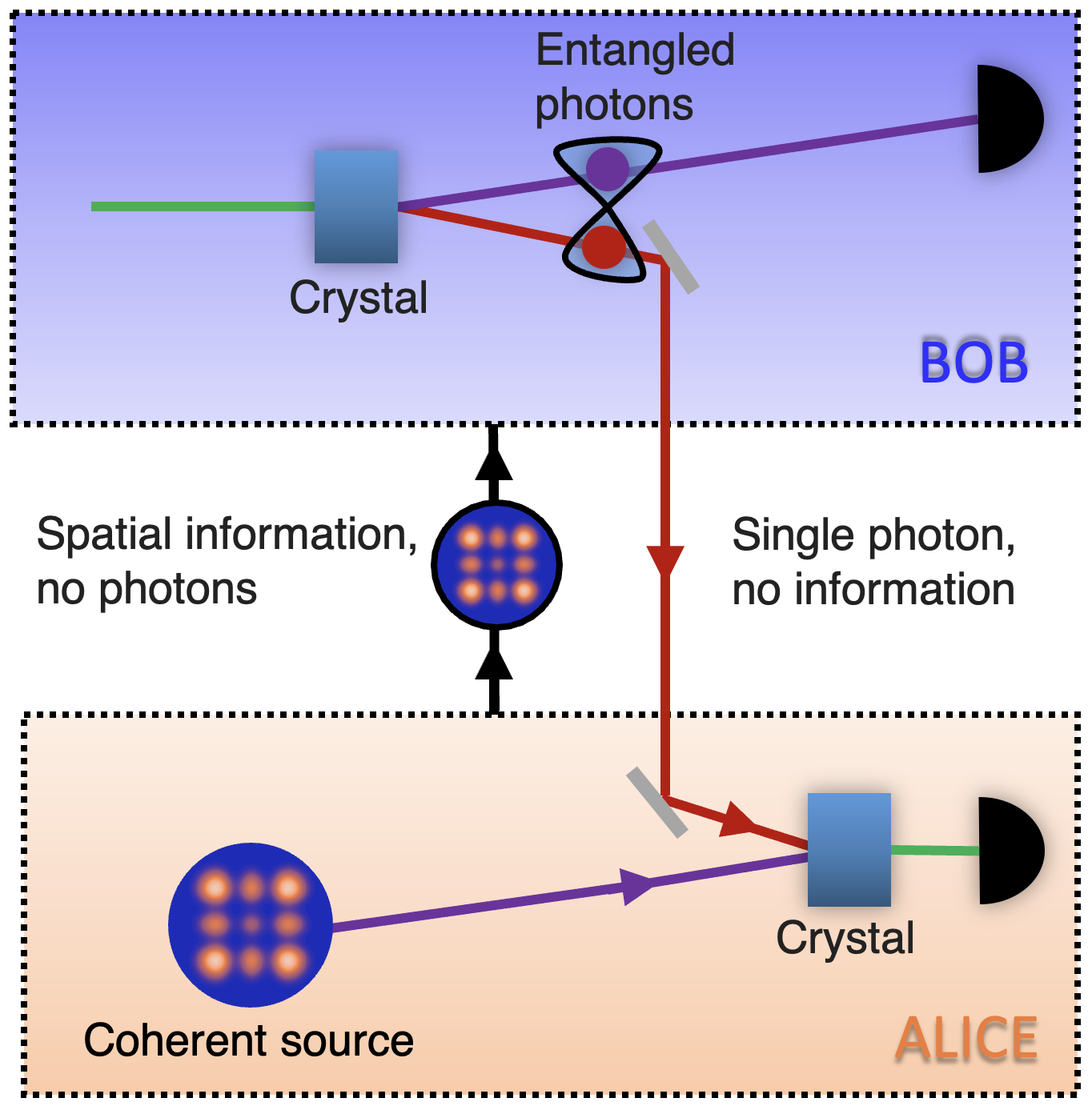} %was Fig1_v6.png
\caption{\revise{\textbf{High-dimensional quantum transport enabled by nonlinear detection.} In our concept, information is encoded on a coherent source and overlapped with a single photon from an entangled pair in a nonlinear crystal for up-conversion by sum frequency generation, the latter acting as a nonlinear spatial mode detector. The bright source is necessary to achieve the efficiency required for the nonlinear detection.  Information and photons flow in opposite directions: one of Bob's entangled photons is sent to Alice and has no information, while a measurement on the other in coincidence with the upconverted photon establishes the transport of information across the quantum link.  Alice need not know this information for the process to work, while the nonlinearity allows the state to be arbitrary and unknown in dimension and basis.}}
\label{fig:concept1}
\end{figure}

Here, we experimentally demonstrate \revise{a nonlinear spatial quantum transport scheme} for arbitrary dimensions using two entangled photons to form the quantum channel and a bright coherent source for information encoding. One of the photons from the entangled pair is upconverted in a nonlinear crystal using the coherent beam both as the information carrier and efficiency enhancer, with a successful single photon detection resulting in information transported to the other photon enabled by a bi-photon coincidence measurement. \revise{Our system works for spatial information in a manner that is dimension and basis independent, with the modal capacity of our quantum channel easily controlled by parameters such as beam size and crystal properties, which we outline theoretically and confirm experimentally up to $d=15$ dimensions. Using the spatial modes of light as our encoding basis, we use this channel to transfer information expressed across many spatial bases, including OAM, Hermite-Gaussian and their superpositions. Our experiment is supported by a full theoretical treatment and offers a new approach to harnessing high-dimensional structured quantum states by nonlinear optical control and detection.}
\begin{figure*}[]
\centering
\includegraphics[width=0.95\linewidth]{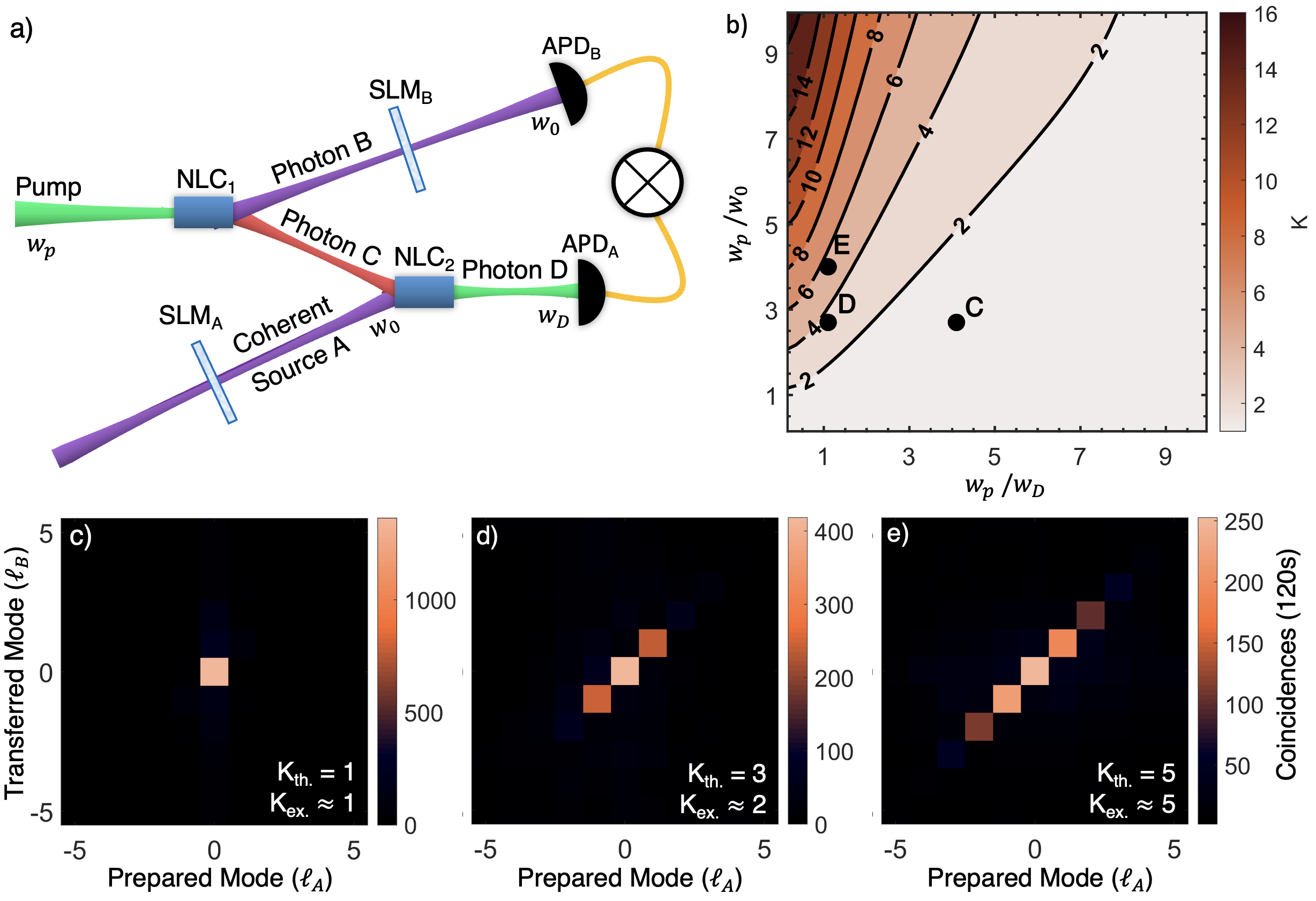} %was TeleFig_setup2.png
\caption{\textbf{Realising a quantum transport channel.} (a) A pump photon ($\lambda_p = 532$ nm) undergoes spontaneous parametric downconversion (SPDC) in a nonlinear crystal (NLC$_1$), producing a pair of entangled photons (signal B and idler C), at wavelengths of $\lambda_B = 1565$ nm and $\lambda_C = 808$ nm, respectively. Photon B is directed to a spatial mode detector comprising a spatial light modulator (SLM$_B$) and a single mode fibre coupled avalanche photo-diode detector (APD).  The state to be transferred is prepared as a \revise{coherent source A} using SLM$_A$ ($\lambda_A = 1565$ nm), and is overlapped in a second nonlinear crystal (NLC$_2$) with photon C, resulting in an upconverted photon D which is sent to a single mode fibre coupled APD. Photons B and D are measured in coincidence to find the joint probability of the prepared and measured states using the two SLMs. (b) The quantum transport channel's theoretical modal bandwidth ($K$) as a function of the pump ($w_p$) and detected photons' ($w_0$ and $w_D$) radii, with experimental confirmation shown in (c) through (e) corresponding to parameter positions $C$, $D$ and $E$ in (b). $K_\text{th}$ and $K_\text{ex}$ are the theoretical and experimental quantum transport channel capacities, respectively. The cross-talk plots are shown as orbital angular momentum (OAM) modes prepared and transferred. The raw data is reported with no noise suppression or background subtraction, and considering the same pump power conditions in all three configurations}.
\label{fig:ConceptFig}
\end{figure*}
%%%%%%%%%%
%Entanglement is therefore transferred to D and the spatial state of the pump ($\ket{\varphi}$) is teleported to B to within a unitary rotation (depending on the state of the photon upconverted) which can be preformed by a spatial light modulator, SLM$_B$. 
%%%%%%%%%%%%%%%%%%%%%%%%%%%%%%%%%%%%%%%%%%%%%%%%%%
\section*{Results}
%%%%%%%%%%%%%%%%%%%%%%%%%%%%%%%%%%%%%%%%%%%%%%%%%%
\noindent \textbf{Concept.}  \revise{A schematic of our concept is shown in Figure \ref{fig:concept1}} together with the experimental realisation in Figure \ref{fig:ConceptFig} (a), with full details provided in Supplementary Note 1. Two entangled photons, B and C, are produced from a nonlinear crystal (NLC$_1$) configured for collinear non-degenerate spontaneous parametric downconversion (SPDC).  Photon C is sent to interact with the state to be transferred \revise{(coherent source A)}, as prepared using a spatial light modulator (SLM$_A$), while photon B \revise{is measured by spatial projection with a spatial light modulator (SLM$_B$) and SMF.} %In prior qubit demonstrations of quantum transport, e.g., teleportation, the interaction of photons A and C takes the form of a Bell state analyser with linear optical elements (often combination of beam splitters and polarising beam splitters), that renders the success of the teleportation a probabilistic process. However, beyond qubits, linear optical solutions result in mixed states \cite{Leach2017} unless additional photons (ancillary photons) are added \cite{luo2019quantum, PhysRevLett.125.230501}.  

In our scheme, we \revise{overlap photons from the coherent source A with single photon C} in a second nonlinear crystal (NLC$_2$), and detect the upconverted photon D, generated by means of sum frequency generation (SFG). The success of the process is conditioned on the measurement of the \textit{single} photon D \revise{(due to the \textit{single} photon C from the entangled pair)} in coincidence with the \textit{single} photon B from the entangled pair.  We use a coherent state as input to enhance the probability for up-conversion, where all the photons carry the same modal information which we want to transport. %One such photon then takes part in the up-conversion process to produce the photon that is detected to signal the transport of the modal information to photon B. One can consider the coherent input state and the output single photon state as \emph{carriers} of the quantum information to be transported, i.e., coherent superpositions of photonic spatial modes.

To understand the process better, it is instructive to use OAM modes as an example; a full basis-independent theoretical treatment is given in Supplementary Notes 2 through 4. We pump the SPDC crystal with a Laguerre-Gaussian mode of azimuthal and radial indices $\ell_p = 0$ and $p_p=0$, respectively.  OAM is conserved in the SPDC process \cite{mair2001entanglement} so that $\ell_p = 0 = \ell_B + \ell_C$.  The up-conversion process also conserves OAM \cite{zhou2016orbital}, so if the detection is by a single mode fibre (SMF) that supports only spatial modes with $\ell_D = 0$, then $\ell_D = 0 = \ell_A + \ell_C$.  One can immediately see that a coincidence is only detected when both A and B are conjugate to C, $ \ell_A = \ell_B = -\ell_C$, and thus the prepared state (A) matches the transported state (B).  One can show more generally (see Supplementary Note~2) that if the detection of photon D is configured to be into the same mode as the initial SPDC pump (we may call photon D the anti-pump), then the up-conversion process acts as the conjugate of the SPDC process, \revise{and the state of each photon in the coherent source A that is involved in the up-conversion is transported to that of photon B. To keep the language clear, we will refer to those photons in coherent source A that take part in the up-conversion as \textit{photon-state A}}\revise{, as in the SPDC process where only one pump photon is considered to take part in the down-conversion process, ignoring the vacuum term in both cases since they do not give rise to coincidences in our process.}  However, up-conversion aided quantum transport only takes place under pertinent experimental conditions, namely, perfect anti-correlations between the signal and idler photons from the SPDC process in the chosen basis, and an up-conversion crystal with length and phase-matching to ensure for anti-correlations between \revise{photon-state A and photon C} (see Supplementary Note 3 for full details).

%\AF{While the initial $\ell_p = 0$ pump results in the SPDC state,
%$\ket{\ell=0}\rightarrow\ket{\Psi}_\text{spdc}=\sum c_\ell\ket{\ell}_B \ket{-\ell}_C\,$,  the upconversion can be seen as the reverse, where the input photons in state $\ket{\Psi}_{CA}$ are upconverted to a single photon state given by $\ket{\ell=0}\leftarrow\ket{\Psi}_{CA}=\sum c_l\ket{\ell}_A\ket{-\ell}_C$. This allows us to realise a Bell filter by conditioning detection of the up-converted photon on $\ell_D =0$. A projection on $\ket{\Psi}_{CA}$ is thus made, since $\bra{\ell=0} U_{\mathrm{\tiny SFG}}\ket{\Gamma}_{CA} =  \bra{\ell=0} U_{\mathrm{\tiny SPDC}}^\dagger\ket{\Gamma}_{CA} = \braket{\Psi| \Gamma}_{CA}$  where the unitary evolution operators correspond to the inverse processes, and $\ket{\Gamma}_{ca}$ is an arbitrary input state carried by C and A.}

\begin{figure}[]
\centering
\includegraphics[width=1.0\linewidth]{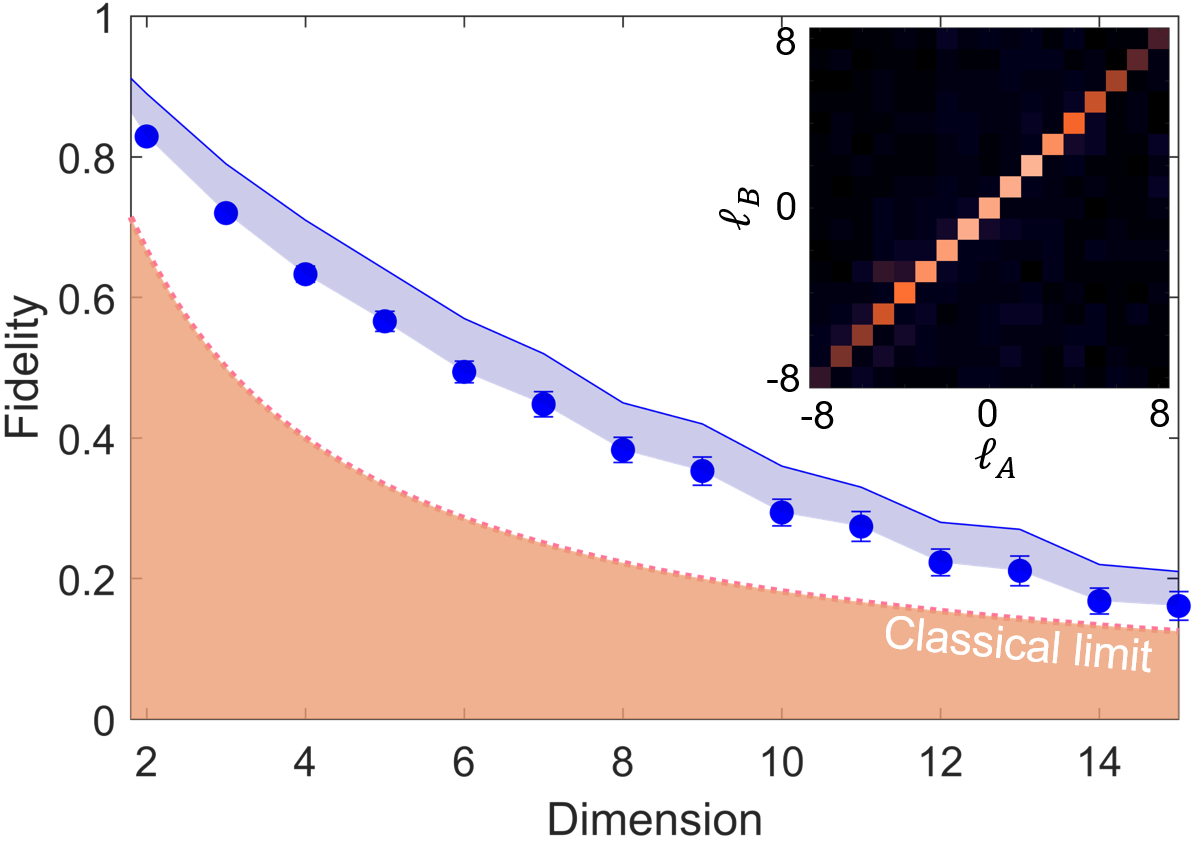} %was Fig2v2.png
\caption{\textbf{Quality of the quantum transport process.} Experimental fidelities (points) for our channel dimensions up to the maximum achievable channel capacity of \BS{$K = 15 \pm 1$}, all well above the classical limit (dashed line). \BS{The solid line forms a maximum fidelity for the measured transferred state.} The inset shows the measured OAM modal spectrum of the optimised quantum transport channel with maximum coincidences of \BS{320 per second for a 5 minute integration time}. The raw data is reported with no noise suppression or background subtraction. }
\label{fig:DimFid}
\end{figure}

To find a bound on the modal capacity of the channel, one can treat the process as a communication channel with an associated channel operator.  This, in turn, can be treated as an entangled state, courtesy of the Choi-Jamoilkowski state-channel duality \cite{jiang2013channel}, from which a Schmidt number ($K$) can be calculated. We interpret this as the effective number of modes the channel can transfer (its modal capacity), given by

\begin{equation}
  K = \frac{\left[\int T^2(\mathbf{q}_A,\mathbf{q}_B) \text{d}^2\mathbf{q}_A \text{d}^2\mathbf{q}_B\right]^2}{\int \left[ \int T(\mathbf{q}_A,\mathbf{q}_C)T(\mathbf{q}_C,\mathbf{q}_B) \text{d}^2\mathbf{q}_C\right]^2\text{d}^2\mathbf{q}_A \text{d}^2\mathbf{q}_B},
  \label{eq:band}
\end{equation} 
\noindent where 
\begin{equation}
  T(\mathbf{q}_A,\mathbf{q}_B)= \int \psi_{\text{\tiny SFG}}^*(\mathbf{q}_A,\mathbf{q}_C)  \psi_{\text{\tiny SPDC}}(\mathbf{q}_C,\mathbf{q}_B) \text{d}^2\mathbf{q}_C,
  \nonumber
\end{equation} 
\noindent with the SFG and SPDC wave functions expressed in the momentum ($\mathbf{q}$) basis.
%\Note{Under some simplifying assumptions this reduces to
%\begin{equation}
%    \tilde{K} \approx \frac{n_A n_B w_D^2 w_p^2}{(w_D^2 + w_p^2)(n_A\lambda_B + n_B \lambda_A) L}
%    \nonumber
%\end{equation}
%}
%\Note{where $n$ is the crystal refractive index at wavelength $\lambda$ for photons A and B (subscripts), and $L$ is the length of the crystals (assumed to be approximately the same for both NLC$_{1\&2}$).} 
Full details are given in Supplementary Note 4.  The controllable parameters are the beam radii of the pump ($w_p$), and the spatially filtered photons D ($w_D$) and B ($w_0$).  Using Equation (\ref{eq:band}), we calculated the channel capacity for OAM modes, with the results shown in Figure \ref{fig:ConceptFig} (b), revealing that a large pump mode relative to the detected transferred modes is optimal for capacity. A large pump mode with respect to the crystal length also increases the channel capacity, consistent with the well-known thin-crystal approximation.  However, this comes at the expense of coincidence events, the probability of detecting the desired OAM mode, which must be balanced with the noise threshold in the system.  We show three experimental examples of this trade-off in Figures \ref{fig:ConceptFig} (c), (d) and (e), where the parameters for each can be deduced from the corresponding labelled positions in Figure \ref{fig:ConceptFig} (b).  Good agreement between theoretical ($K_\text{th}$) and experimentally measured ($K_\text{ex}$) capacities validates the theory.  Using the theory, we adjust the experimental parameters to optimise the quantum transport channel, reaching a maximum of $K \approx \BS{15}$ for OAM modes, as shown in the inset of Figure \ref{fig:DimFid}. This limit is not fundamental and is set only by our experimental resources.  We are able to establish a quantum transport setup where the channel supports at least \BS{15} OAM modes.  The balance of channel capacity with noise is shown in Figure \ref{fig:DimFid}. Using a probe of purity and dimension \cite{nape2021measuring} we use a traditional measure and estimate a channel fidelity which decreases with channel dimension, but is always well above the upper bound of the achievable fidelity for the classical  case, i.e., having no entanglement between photons B and C, given by $F_\text{classical} \leq \frac{2}{d + 1}$ for $d$ dimensions and shown as the classical limit (dashed line) in Figure~\ref{fig:DimFid}. Blue points show the quantum transport fidelity, measured from Eq.~(\ref{eq:telefidelity}), using the channel fidelity $F_{Ch}$. Here, the channel fidelity measures the quality of the correlations that can be established between \revise{photon-state A} and photon B over the two particle $d^2$ subspace while the quantum transport fidelity, $\mathcal{F}$ measures how well SLM$_B$ and APD$_B$ can measure states transmitted over the channel, requiring measurements over a single particle $d$ dimensional space. Since $F_{Ch} \leq \mathcal{F}=\frac{F_{Ch} d +1}{d+1}$ \cite{horodecki1999general}, it follows that $\mathcal{F}$, shown as the solid line above the shaded region in Fig.~\ref{fig:DimFid}, sets the upper-bound for the quantum transport fidelity and is therefore the highest achievable fidelity for our system (See Methods for further details). \revise{Note that we use a measurement of a two particle system because we condition on coincidence events between \emph{single} photons B and D.}
% Using a probe of purity and dimension \cite{nape2021measuring} we measure a decreasing fidelity with channel dimension, but always well above the upper bound of the fidelity achievable for classical teleportation, given by $F_\text{classical} \leq \frac{2}{d + 1}$ for $d$ dimensions, shown as the classical limit (dashed line) in Figure \ref{fig:DimFid} (full details in the Methods section).
%%% figure

\begin{figure*}[hbtp]
\centering
\includegraphics[width=\linewidth]{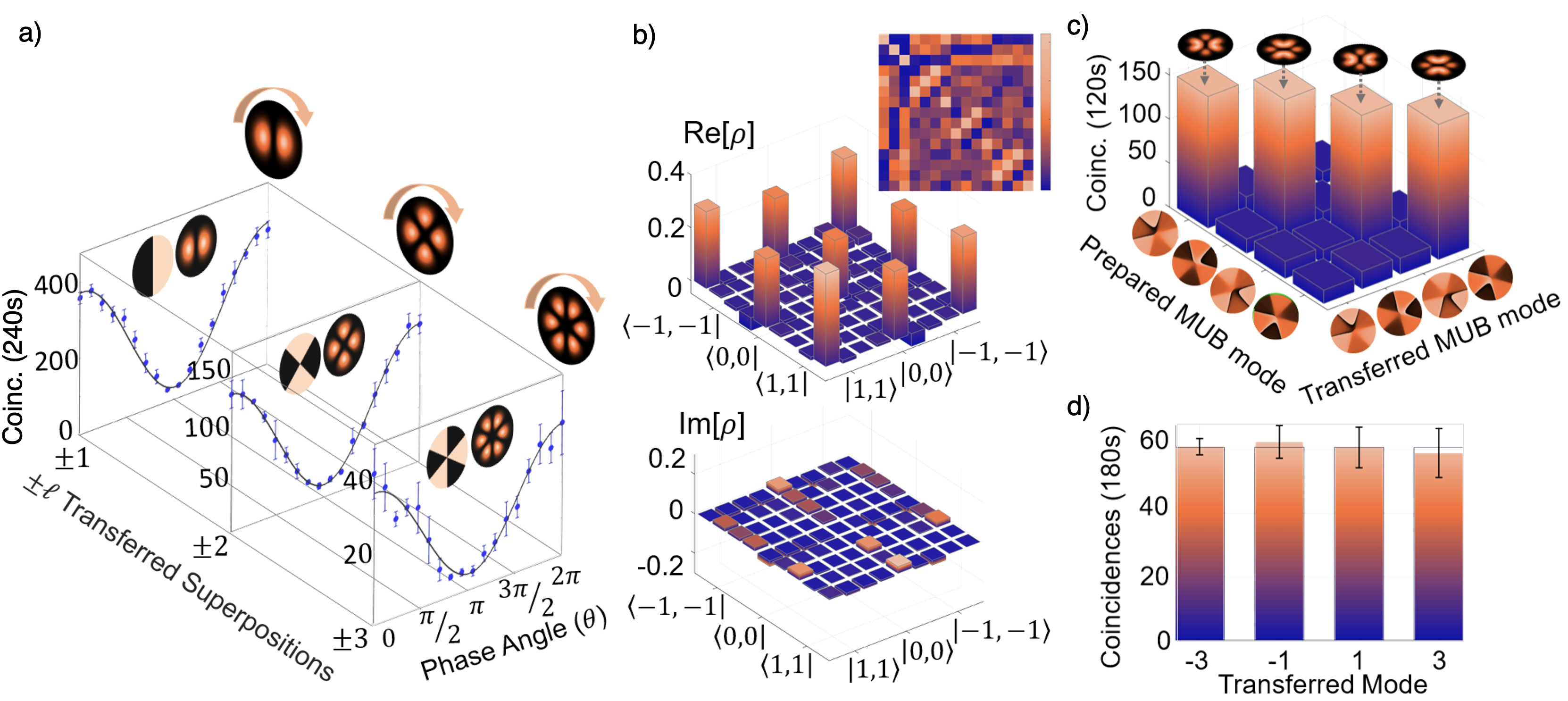} %was newFig3_v2.png
\caption{\textbf{Visibilities and quantum state tomography.} (a) Measured coincidences (points) and fitted curve (solid) as a function of the phase angle ($\theta$) of the corresponding detection analyser for the state $\ket{\phi} = \ket{\ell} + \exp{(i\theta)}\ket{-\ell}$, for three OAM subspaces of $\ell = \pm 1$, $\pm 2$, and $\pm 3$ (further details in the Supplementary Note 5). (b) The real (Re[$\rho$]) and imaginary (Im[$\rho$]) parts of the density matrix ($\rho$) for the qutrit state $\ket{\Psi} = \ket{-1} + \ket{0} + \ket{1}$ as reconstructed by quantum state tomography. The inset shows the raw coincidences with maximum coincidences of \BS{220} detected per second from the tomographic projections (full details in the Supplementary Notes 6 \av{and 13}). (c) Measurements for the quantum transport of a 4-dimensional state, constructed from the states $\ell = \{\pm1,\pm3\}$. (d) Measurements showing the detection (solid bars) of all the prepared (transparent bars) OAM states comprising one of the MUB states. The raw data is reported with no noise suppression or background subtraction.}
\label{fig:VisTomoDim}
\end{figure*}

\vspace{0.5cm}
\noindent \textbf{Quantum transport results.} In Figure \ref{fig:VisTomoDim} (a) we show results for the quantum transport channel in two, three and four dimensions.  We confirm quantum transport beyond just the computational basis by introducing a modal phase angle, $\theta$, on photon B relative to \revise{\revise{photon-state A}} ($\theta = 0$) for the two-dimensional state $\ket{\Psi} = \ket{\ell} + \exp (i \theta) \ket{-\ell}$ (we omit the normalization throughout the text for simplicity).  We vary the phase angle while measuring the resulting coincidences for three example OAM subspaces, $\ell = \pm 1$, $\pm 2$ and $\pm 3$.  The raw coincidences, without any noise subtraction, are plotted as a function of the phase angle in Figure \ref{fig:VisTomoDim} (a), confirming the quantum transport across all bases. The resulting \revise{visibilities (V) allow us to determine the fidelities \cite{gisin2002quantum} from $F = \tfrac{1}{2}(1+V)$, with raw values varying from 90\% to 93\%, and background subtracted all above 98\% (see Supplementary Notes 7 through 9).} Example results for the qutrit state $\ket{\Psi} = \ket{-1} + \ket{0} + \ket{1}$ are shown in Figure \ref{fig:VisTomoDim} (b) as the real and imaginary parts of the density matrix, reconstructed by quantum state tomography, obtaining a transferred qutrit with an average channel fidelity of 0.82 $\pm$ 0.016 (see Supplementary Note 13 for all detailed measurements with the raw coincidences from the projections in all orthogonal and mutually unbiased basis). Further analysis of judiciously chosen transferred states themselves lead to even higher values (see Supplementary Notes 11 and 14).

% \av{[Bereneice, do you mean that we performed this fidelity measurement with the conditions from Fig. 1(e) ($K = 5$ channel) and inset of Fig. 2 ($K = 10$ channel)?]} Without any noise subtraction the fidelity is $F = 0.66$ ($K = 5$ channel) and $F = 0.78$ ($K = 10$ channel), both well above the classical limit for a three-dimensional state.  The values are consistent with the notion of noise increasing as the teleported state dimension approaches the channel capacity.  With background subtraction the $K=5$ channel fidelity increases to $F = 0.75$, comparable to the lower noise $K=10$ channel.

\revise{Next,} we proceed to illustrate the potential of the quantum transport channel by sending four-dimensional states of the form $\ket{\Psi} = \ket{-3} + \exp (i \theta_1) \ket{-1} + \exp (i \theta_2) \ket{1} + \exp (i \theta_3) \ket{3}$, with inter-modal phases of $\{\theta_1, \theta_2, \theta_3\} = \{-\pi/2, -\pi, -\pi/2\}, \{-\pi/2, 0, \pi/2\}, \{\pi/2, \pi, \pi/2\}$ and $\{\pi/2, 0, -\pi/2\}$. All possible outcomes from these mutually unbiased basis (MUBs) are shown in Figure \ref{fig:VisTomoDim} (c). We encoded each superposition (one at a time) in SLM$_A$ and projected photon B in each of the four states. \revise{The strong diagonal with little cross-talk in the off-diagonal terms confirms quantum transport across all states.} Figure \ref{fig:VisTomoDim} (d) shows an \revise{exemplary detection of one such MUB state in the OAM basis}: the transferred state (solid bars) with the prepared state (transparent bars), for a similarity of \BS{$S = 0.98 \pm0.047$} (see description used in the Methods section). \rvs{Note that the prepared states (transparent bars) in the figures throughout the letter are obtained by the averaged sum of all measured values involved, facilitating comparison with the raw coincidences.} \av{Furthermore, we have also transferred various unbalanced superpositions of OAM states (see Supplementary Note 12 and Suppl. Fig. \ref{UnevenStates} for full details), being able to assign different weightings. The encoded states are the following: $\ket{\Psi} = 2\ket{-1} + 3\ket{0} + \ket{1}$, $\ket{\Psi} = 2\ket{-2} + 3\ket{0} + \ket{2}$, $\ket{\Psi} = \ket{-2} + 2\ket{0} + \ket{2}$, and $\ket{\Psi} = 2\ket{-3} + \ket{-1} + \ket{1} + 2\ket{4}$.}
% $\ket{\Psi} = \frac{2}{\sqrt{6}}\ket{-1} + \frac{3}{\sqrt{6}}\ket{0} + \frac{1}{\sqrt{6}}\ket{1}$, $\ket{\Psi} = \frac{2}{\sqrt{6}}\ket{-2} + \frac{3}{\sqrt{6}}\ket{0} + \frac{1}{\sqrt{6}}\ket{2}$, $\ket{\Psi} = \frac{1}{\sqrt{4}}\ket{-2} + \frac{2}{\sqrt{4}}\ket{0} + \frac{1}{\sqrt{4}}\ket{2}$, and $\ket{\Psi} = \frac{2}{\sqrt{4}}\ket{-3} + \frac{1}{\sqrt{4}}\ket{-1} + \frac{1}{\sqrt{4}}\ket{1} + \frac{2}{\sqrt{4}}\ket{4}$.} %and $S = 0.982$ with background subtraction

\begin{figure}
\centering
\includegraphics[width=\linewidth]{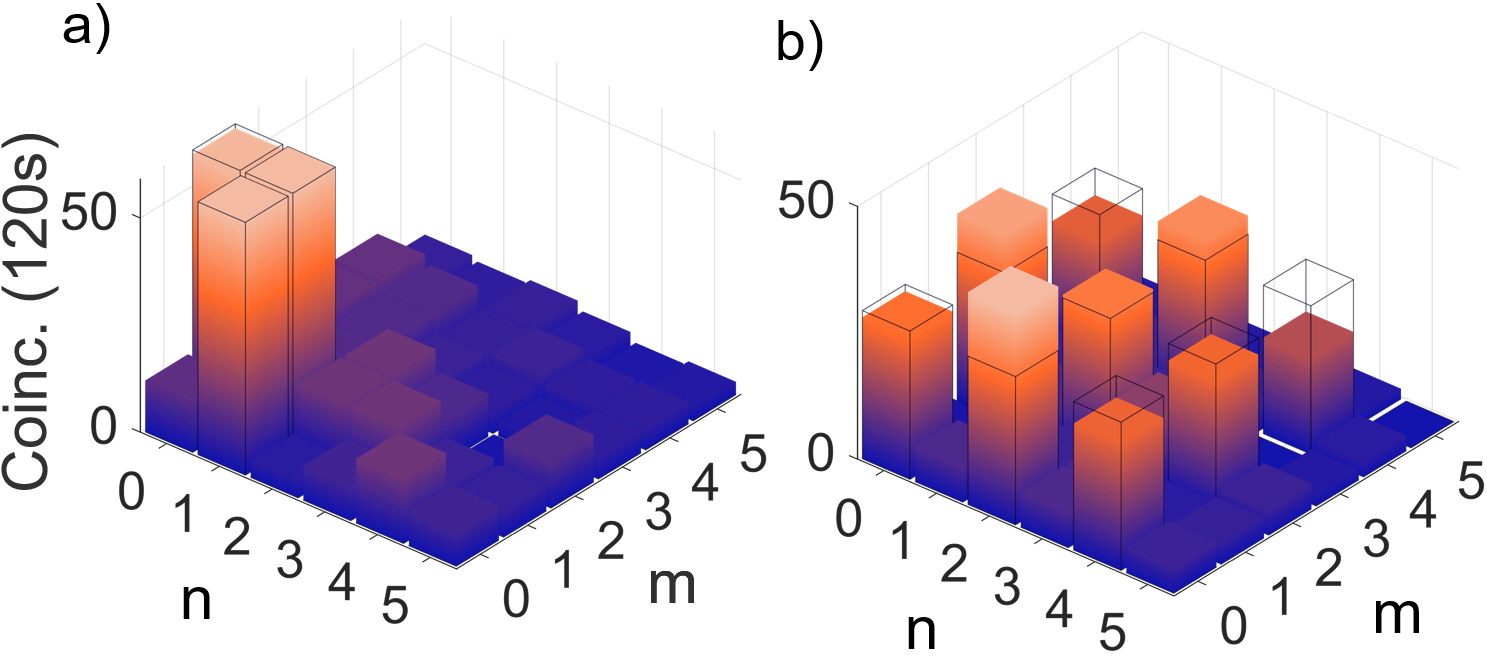} %wasFig4.png
\caption{\textbf{Quantum transport in the Hermite-Gaussian basis}. Coincidence measurements for quantum transport of (a) a 3-dimensional and (b) a 9-dimensional $\text{HG}_{n,m}$ state, constructed from the states $(n,m) = \{(0,1),(1,0),(1,1)\}$ and $(n,m) = \{(0,0),(2,0),(0,2),(2,2),(2,4),(4,2),(4,4)\}$, respectively. \revise{The weights of the diagonal elements of the density operator of the transported state (solid bars) are in good agreement with the weights of the prepared state (transparent bars). The raw data is reported with no noise suppression or background subtraction.}}
\label{fig:HGtele}
\end{figure}

\begin{figure*}
\centering
\includegraphics[width=\linewidth]{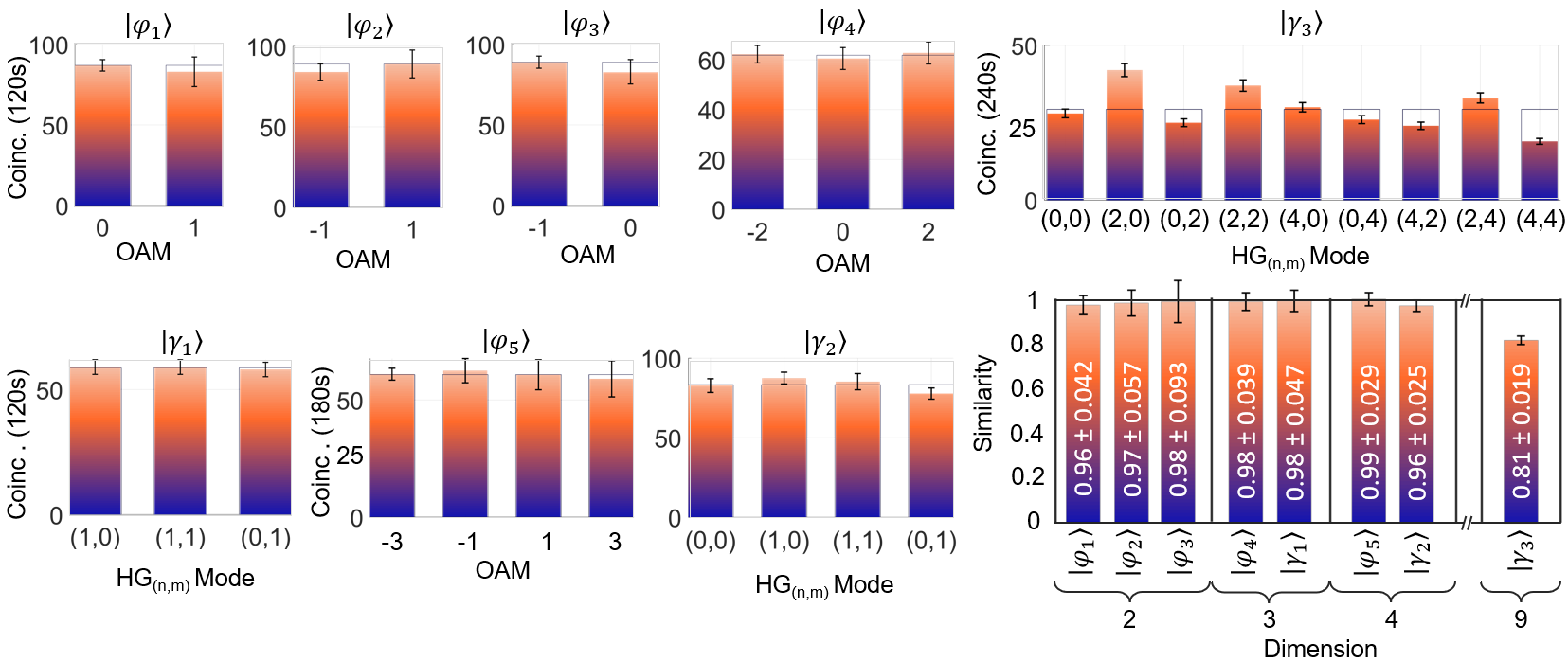} %was Fig5.png
\caption{\textbf{Summary of transferred states}. Similarities for transport of a 2,3,4 and 9-dimensional superposition states in the OAM (represented as $\ket{\varphi}$) and HG (represented as $\ket{\gamma}$) bases shown and labelled to the left. Transferred states are 
$\ket{\varphi_1}=\ket{0}+\ket{-1}$, $\ket{\varphi_2}=\ket{-1}+\ket{1}$, $\ket{\varphi_3}=\ket{0}-\ket{1}$, $\ket{\varphi_4}=\ket{-2}+\ket{0}+\ket{2}$, $\ket{\gamma_1}=\ket{HG_{1,0}}+\ket{HG_{1,1}}+\ket{HG_{0,1}}$, $\ket{\varphi_5}=\ket{-3}-i\ket{-1}+\ket{1}+i\ket{3}$, $\ket{\gamma_2}=\ket{HG_{0,0}}+\ket{HG_{1,0}}+\ket{HG_{1,1}}+\ket{HG_{0,1}}$ and $\ket{\gamma_3}=\ket{HG_{0,0}}+\ket{HG_{2,0}}+\ket{HG_{0,2}}+\ket{HG_{2,2}}+\ket{HG_{4,0}}+\ket{HG_{0,4}}+\ket{HG_{4,2}}+\ket{HG_{2,4}}+\ket{HG_{4,4}}$. \revise{The similarity of diagonal elements of the density matrix together with prior phase information confirms coherent transport up to $d=4$ but not for $d=9$, where only the diagonal elements are assessed.} Raw data are reported without noise suppression or background subtraction.}
\label{fig:TeleSims}
\end{figure*}

The result in Figure \ref{fig:VisTomoDim} (c) also confirms that the channel is not basis dependent, since this superposition of OAM states is not itself an OAM eigenmode.  To reinforce this message, we proceed to transfer $d = 3$ and $d = 9$ states in the Hermite-Gaussian (HG$_{n,m}$) basis with indices $n$ and $m$, with the results shown in Figure \ref{fig:HGtele}. 
% In this basis the quantum transport channel has a maximum capacity of $K=25$. 
In both cases the \revise{measured state (solid bars) is in very good agreement with the prepared state (transparent bars). Note that the results only confirm that the diagonal terms of the density matrices of the input states were transported successfully and so cannot confirm the transportation of coherences (off-diagonal elements of the density matrices) before and after quantum transport.}  
\revise{The good agreement between the diagonal elements of the initial and final states is evidence that the quantum transport works for these elements, corroborated by the full phase information already confirmed up to $d=4$ and a channel capacity (that includes phases) up to $d=15$.  To quantify the final state's diagonal terms for $d=9$ we make use of similarity as a measure (see Methods) because of the prohibitive time (due to low counts) to determine a Fidelity from a quantum state tomography, but note that this measure does not account for modal phases in the prepared and measured state.} A final summary of example transferred states is shown in Figure \ref{fig:TeleSims}, covering dimensions two through nine, and across many bases. The prepared (transparent bars) and transferred (solid bars) states are in good agreement, \revise{as determined from the similarity,} confirming the quality of the channel. \revise{Note that the coincidence counts are given for the detected OAM states (solid bars). The weightings of the prepared ones (transparent bars) are intended to show the normalized probabilities for visualization purposes.}

%%% figure
%%%%%

%%% figure
%%%%%
% $\ket{\varphi_1}=\frac{1}{\sqrt{2}}[\ket{0}+\ket{-1}]$, $\ket{\varphi_2}=\frac{1}{\sqrt{2}}[\ket{-1}+\ket{1}]$, $\ket{\varphi_3}=\frac{1}{\sqrt{2}}[\ket{0}-\ket{1}]$, $\ket{\varphi_4}=\frac{1}{\sqrt{3}}[\ket{-2}+\ket{0}+\ket{2}]$, $\ket{\gamma_1}=\frac{1}{\sqrt{3}}[\ket{HG_{1,0}}+\ket{HG_{1,1}}+\ket{HG_{0,1}}]$, $\ket{\varphi_5}=\frac{1}{\sqrt{4}}[\ket{-3}-i\ket{-1}+\ket{1}+i\ket{3}]$, $\ket{\gamma_2}=\frac{1}{\sqrt{4}}[\ket{HG_{0,0}}+\ket{HG_{1,0}}+\ket{HG_{1,1}}+\ket{HG_{0,1}}]$ and $\ket{\gamma_3}=\frac{1}{\sqrt{9}}[\ket{HG_{0,0}}+\ket{HG_{2,0}}+\ket{HG_{0,2}}+\ket{HG_{2,2}}+\ket{HG_{4,0}}+\ket{HG_{0,4}}+\ket{HG_{4,2}}+\ket{HG_{2,4}}+\ket{HG_{4,4}}]$. Raw data are reported without noise suppression or background subtraction.}
% \label{fig:TeleSims}
% \end{figure*}
% %%%%%

%%%%%%%%%%%%%%%%%%%%%%%%%%%%%%%%%%%%%%%%%%%%%%%%%%
%\section*{From transport to teleport}
%%%%%%%%%%%%%%%%%%%%%%%%%%%%%%%%%%%%%%%%%%%%%%%%%%

\vspace{0.5cm}

%%%%%%%%%%%%%%%%%%%%%%%%%%%%%%%%%%%%%%%%%%%%%%%%%%
\section*{Discussion}
%%%%%%%%%%%%%%%%%%%%%%%%%%%%%%%%%%%%%%%%%%%%%%%%%%

\revise{Structured quantum light has gained traction of late \cite{forbes2021structured,forbes2019quantum,nape2023quantum}, promising a larger Hilbert space for information processing and communication. The use of nonlinear optics in the \textit{creation} of high-dimensional quantum states is exhaustive (SPDC, photonic crystals, resonant metasurfaces and so on), while the preservation of entanglement and coherence in nonlinear processes \cite{huang1992observation} has seen it used for efficient photon detection \cite{vandevender2004high}, particularly for measurement of telecom wavelength photons \cite{zaske2012visible}.  Full harnessing and controlling high-dimensional quantum states by nonlinear processes has however remained elusive.  Notable exceptions include advances made in the time-frequency domain \cite{ansari2018tailoring}, another degree of freedom to harness high-dimensional states, such as the demonstration of quantum pulse gates \cite{eckstein2011quantum} for efficient demultiplexing of temporal modes as well as for tomographic measurements \cite{ansari2018tomography,donohue2013coherent}, the inverse process of multiplexing by difference frequency generation \cite{allgaier2020pulse}, quantum interference of spectrally distinguishable sources \cite{ates2012two}, high-dimensional information encoding \cite{lukens2014orthogonal} and simultaneous temporal shaping and detection of quantum wavefunctions \cite{pe2005temporal}.  To the best of our knowledge, our work is the first in the spatial domain, offering an exciting resource for controlling and processing spatial quantum information by nonlinear processes. Combining advances in high-dimensional spectral-temporal state control \cite{kues2017chip} and on-chip nonlinear solutions \cite{baboux2023nonlinear} with the spatial degree of freedom could herald new prospects in quantum information processing beyond qubits.}

In conclusion, we have demonstrated an elegant way to perform a projection of an unknown state using a nonlinear detector, facilitating quantum information in high dimensions, and across many spatial bases, to be transferred with just one entangled pair as the quantum resource. Our results validate the non-classical nature of the channel without any noise suppression or background subtraction. While our quantum transport scheme cannot teleport entanglement due to the need of encoding the state to be transferred in many copies, it nevertheless securely transfers the state of the laser photons to the distant and previously entangled photon, and it does this without using knowledge of the state of the laser photons \revise{(see Supplementary Notes 10 and 15 discussing the challenges to move from transport to teleport).} Importantly, our comprehensive theoretical treatment outlines the tuneable parameters that determine the modal capacity of the quantum transport channel, such as modal sizes at the SPDC crystal and detectors, requiring only minor experimental adjustments (for example, the focal length of the lenses). The modal capacity of our channel was limited only by experimental resources, while future research could target an increase of the number of transferred modes by optimising the choice of the relevant parameters and improved nonlinear processes. \revise{Our work highlights the exciting prospect this approach holds for the quantum transport of unknown high-dimensional spatial states, and could in the future be extended to mixed degrees of freedom, for instance, hybrid entangled (polarization and space) and hyper entangled (space and time) states, for multi-degree-of-freedom and high-dimensional quantum control.}
 
 %\revise{Our proof-of-principle experiment sets a new state-of-the-art and offers a new roadmap towards versatile teleportation solutions based on nonlinear optics.} %for \av{future quantum networks \cite{kimble2008quantum}}.

%Ultimately, the resolution and image quality might be improved by using squeezed states as an entanglement resource to teleport bright states of light. \
%%It is important to note, however, that advances in non-linear technology could facilitate single photon input states, mitigating security issues pertaining to the sender keeping a better copy of the state than that transferred. That would enable a natural transition from quantum {\it transport} to quantum {\it teleportation} with no actual change to the experimental setup currently used, as described in Supplementary Note 10.}

% \BS{\sout{with background subtraction we find excellent fidelities as reported in Supplementary Note 7}}. 
\vspace{0.2 cm}

%%%%%%%%%%%%%%%%%%%%%%%%%%%%%%%%%%%%%%%%%%%%%%%%%%
\section*{Methods}\label{sec:holo_gen}
%%%%%%%%%%%%%%%%%%%%%%%%%%%%%%%%%%%%%%%%%%%%%%%%%%
%%%%%%%%%%%%%%%%%%%%%%%%%%%%%%%%%%%%%%%%%%%%

\noindent \textbf{Fidelity}. To quantify the quality of the quantum transport process, we use {\em fidelity}. It is defined for pure states as the squared magnitude of the overlap between the initial state that was to be transferred $\ket{\psi_{A}}$ and the final transferred state that was received by SLM$_B$ and APD$_B$ $\ket{\psi_{B}}$:
\begin{equation}
F = |\braket{\psi_{A}|\psi_{B}}|^2 .
\end{equation}
In the ideal case, where the transferred state is $\ket{\psi_{B}} = \int \alpha(\mathbf{q}) \hat{a}_{B}^{\dag}(\mathbf{q}) \ket{\text{vac}}d\mathbf{q}$ (with detailed description in Supplementary Note 2), the fidelity is $F=1$. However, in a practical experiment, the conditions for the ideal case cannot be met exactly. Therefore, the fidelity is given more generally by
\begin{align}
F = & \int \alpha^{*}(\mathbf{q}_{B}) \beta(\mathbf{q}_{B})\ d^2 q_{B} \nonumber \\
 = & \int \alpha^{*}(\mathbf{q}_{B}) U^{*}(\mathbf{q}_{D}) g(\mathbf{q}_{A},\mathbf{q}_{D}-\mathbf{q}_{A},\mathbf{q}_{D}) \nonumber \\
& \times f(\mathbf{q}_{B},\mathbf{q}_{D}-\mathbf{q}_{A}) \alpha(\mathbf{q}_{A})\ d^2 q_{B}\ d^2 q_{A}\ d^2 q_{D} .
\label{eqn:Fid1}
\end{align}
\IN{Here,  $f(\mathbf{q}_{B},\mathbf{q}_{C})$ is the two photon wave-function of the SPDC state, while $g(\mathbf{q}_{A}, \mathbf{q}_{C}, \mathbf{q}_{D})$ and $U^{*}( \mathbf{q}_{D})$  are the SFG kernel and projection mode for photon $D$ (the up-converted photon), respectively.}  It is possible to envisage a classical implementation of the state-transfer process. One would make a complete measurement of the initial state, send the information and then prepare photon B with the same state. To ensure that the quantum transport process can outperform this classical state-transfer process, the fidelity of the process must be better than the maximum fidelity that the classical quantum transport process can obtain.

In order to determine the classical bound on the fidelity by which we measure the transferred state $\ket{\psi}$, we define the probability of measuring a value $a$ by
\begin{equation}
P_\psi(a) = \bra{\psi}\hat{E}_a\ket{\psi} ,
\end{equation}
where $\hat{E}_a$ is an element of the positive operator valued measure (POVM) for the measurement of the initial state. These elements obey the condition.
\begin{equation}
\sum_a \hat{E}_a = \mathbb{I} ,
\end{equation}
where $\mathbb{I}$ is the identity operator. The estimated state associated with such a measurement result is represented by $\ket{\psi_a}$.

For the classical bound, we consider the average fidelity that would be obtained for all possible initial states. This average fidelity is given by
\begin{align}
\mathcal{F} = & \int \sum_a P_\psi(a) |\braket{\psi|\psi_a}|^2\ d\psi, \nonumber\\
 = & \int \sum_a \bra{\psi}\hat{E}_a\ket{\psi} |\braket{\psi|\psi_a}|^2\ d\psi ,
\end{align}
where $d\psi$ represents an integration measure on the Hilbert space of all possible input state. We assume that this space is finite-dimensional but larger than just two-dimensional. Since all the states in this Hilbert space are normalized, the space is represented by a hypersphere. A convenient way to represent such an integral is with the aid of the Haar measure. For this purpose, we represent an arbitrary state in the Hilbert space as a unitary transformation from some fixed state in the Hilbert space $\ket{\psi}\rightarrow \hat{U}\ket{\psi_0}$, so that $d\psi\rightarrow dU$. The average fidelity then becomes the following
\begin{align}
\mathcal{F} = & \int \sum_a \bra{\psi_a}\hat{U}\ket{\psi_0} \bra{\psi_0}\hat{U}^{\dag}\hat{E}_a\hat{U}\ket{\psi_0} \bra{\psi_0}\hat{U}^{\dag}\ket{\psi_a}\ dU \nonumber\\
= & \int \sum_a \text{tr}\{\hat{\rho}_a\hat{U}\hat{\rho}_0\hat{U}^{\dag}\hat{E}_a\hat{U}\hat{\rho}_0\hat{U}^{\dag}\}\ dU.
\label{AveF}
\end{align}
The general expression for the integral of the tensor product of four such unitary transformations, represented as matrices, is given by
\begin{align}
& \int U_{ij} (U^{\dag})_{kl} U_{mn} (U^{\dag})_{pq}\ dU \nonumber \\
= & \frac{1}{d^2-1} \left(\delta_{il} \delta_{jk} \delta_{mq} \delta_{np}
+ \delta_{iq} \delta_{jp} \delta_{ml} \delta_{nk} \right) \nonumber \\
& - \frac{1}{(d^2-1)d} \left(\delta_{il} \delta_{jp} \delta_{mq} \delta_{nk}
+ \delta_{iq} \delta_{jk} \delta_{ml} \delta_{np} \right) .
\end{align}
Using this result in Eq.~(\ref{AveF}), we obtain
\begin{equation}
% \mathcal{F} = &
% \sum_a \frac{1}{d^2-1} \left( \text{tr}\{\hat{\rho}_a\hat{E}_a\} \text{tr}\{\hat{\rho}_0\} \text{tr}\{\hat{\rho}_0\}
% + \text{tr}\{\hat{\rho}_a\} \text{tr}\{\hat{\rho}_0 \hat{\rho}_0\} \text{tr}\{\hat{E}_a\} \right) \nonumber \\
% & - \frac{1}{(d^2-1)d} \left( \text{tr}\{\hat{\rho}_a \hat{E}_a \} \text{tr}\{\hat{\rho}_0\hat{\rho}_0 \}
% +\text{tr}\{\hat{\rho}_a\} \text{tr}\{\hat{\rho}_0\} \text{tr}\{\hat{E}_a\} \text{tr}\{\hat{\rho}_0\} \right) \nonumber \\
% = & \sum_a
% \frac{1}{d^2-1} \left( \text{tr}\{\hat{\rho}_a\hat{E}_a\}
% + \text{tr}\{\hat{\rho}_a\} \text{tr}\{\hat{E}_a\} \right) \nonumber \\
% & - \frac{1}{(d^2-1)d} \left( \text{tr}\{\hat{\rho}_a \hat{E}_a \}
% +\text{tr}\{\hat{\rho}_a\} \text{tr}\{\hat{E}_a\} \right) \nonumber \\
% = &
% \frac{d-1}{(d^2-1)d} \left(\sum_a \text{tr}\{\hat{\rho}_a\hat{E}_a\}
% + \sum_a \text{tr}\{\hat{E}_a\} \right) \nonumber \\
% = &
% \frac{1}{(d+1)d} \left(\sum_a \text{tr}\{\hat{\rho}_a\hat{E}_a\}
% + \text{tr}\{\mathbb{I}\} \right) \nonumber \\
\mathcal{F} = \frac{1}{(d+1)d} \left(d + \sum_a\bra{\psi_a}\hat{E}_a\ket{\psi_a}\right) ,
\end{equation}
where $d$ is the dimension of the Hilbert space and where we imposed $\text{tr}\{\hat{\rho}_0\}=\text{tr}\{\hat{\rho}_0^2\}=\text{tr}\{\hat{\rho}_a\}=1$. We see that $\mathcal{F}$ is maximal if $E_a$ represents rank 1 projectors and $E_a\ket{\psi_a} = \ket{\psi_a}$, that is, $E_a = \ket{\psi_a}\bra{\psi_a}$. Then $\sum_a\bra{\psi_a}\hat{E}_a\ket{\psi_a}=d$. \av{It follows that the upper bound of the fidelity achievable for the classical state-transfer process is given by \cite{horodecki1999general}}
\begin{equation}
\mathcal{F} \leq \frac{2}{d+1} .
\label{Eq:FidelityBound}
\end{equation}
The fidelity obtained in quantum transport needs to be better than this bound to outperform the classical scheme.

\IN{Furthermore, we can consider the particular quantum transport of a subspace smaller than the supported by the quantum transport channel capacity (see more details in Supplementary Note 4). The quantum transport fidelity of the channel for each subspace $d'$ within the $d$ dimensional state, $\rho$, can be computed by truncating the density matrix and overlapping it with a channel state that has perfect correlations. The theoretical fidelity is given by the expression \cite{horodecki1999general}}
\begin{equation}
    F_{Ch} =\frac{d'( p' - 1 ) + d'^2 }{d'^2},
\end{equation}
\IN{where $p'$ and $d'$ are the purity and dimensionality of truncated states. While this assumes that the channel has a random noisy component given by $\mathbb{I}_{d'^2}/d'^2$, the photon C only has a noise component given by $\mathbb{I}_{d'}/d'$ therefore the quantum transport fidelity for each photon is given by,}
\begin{equation}
     \mathcal{F} = \frac{F_{Ch} d' + 1}{d'+1}.
     \label{eq:telefidelity}
\end{equation}
\IN{Here the separability criterion admits the classical bounds  $\frac{1}{d'^2} \leq F_{Ch} \leq \frac{1}{d'}$ and $\frac{1}{d'} \leq \mathcal{F} \leq \frac{2}{d'+1}$ for the full channel and a single state received, respectively.}

\noindent \textbf{Similarity}. We use a normalised distance measure to quantify the quality of the state being transferred, denoted the \emph{Similarity} (S), 
\begin{equation}
    S = 1 - \frac{\sum_j{|(|C_j^{Ex.}|^2-|C_j^{Th.}|^2)|}}{\sum_j{|C_j^{Ex.}|^2}+\sum_j{|C_j^{Th.}|^2}} .
\label{Eq:Similarity}
\end{equation}
Here we take the normalised intensity coefficients, $|C^{Th.}_j|^2$, encoded onto SLM$_A$ for the $jth$ basis mode comprising the state being transferred (i.e. $\ket{\Phi} = \sum_j{C_j\ket{j}}$) and compare it with the corresponding $jth$ coefficient  $|C^{Ex.}_j|^2$ detected after traversing the quantum transport channel (made with $jth$-mode projections on SLM$_B$) as described in the Supplementary Note 4. A small difference in values between encoded and detected state would result in a small 'distance' between the prepared and received value. As such, the second term in Eq. (\ref{Eq:Similarity}) diminishes with increasing likeness of the states, causing the Similarity measure to tend to 1 for unperturbed quantum transport of the state.

\noindent \textbf{Dimensionality measurements}. \revise{We employ a fast and quantitative dimensionality measure to determine the capacity of our quantum channel.  The reader is referred to Ref.~\cite{nape2021measuring} for full details, but here we provide a concise summary for convenience.} The approach coherently probes the channel with multiple superposition states $\ket{M,\theta}_n$. 
%  %%% equation
% \begin{equation}
% \ket{M,\theta}_n = \mathcal{N} \sum_{\ell} 
% e^{-i \pi \ell(n-1)/n}   A^{n}_{\ell} c_{\ell,M}(\theta) \ket{\ell},
% \label{eq:projectionState}
% \end{equation}
% %%%
% \noindent where $\mathcal{N}$ is a normalisation factor, and $
% c_{\ell,M}(\theta) = -\frac{i e^{-i \ell \theta } \sin{\left( M  \pi \right)} }{ \pi  (M -\ell)}$ and $M = n/2$, with each state indexed by $n \in \{ 1, 3, 5, ...\}$. Furthermore, for each state 
% \begin{equation}
% A^{n}_{\ell} = \left \{\begin{array}{cc}
% 1 & \text{mod}\left\{ \ell  ,n \right\} = 0 \\[2mm]
% 0 & \text{otherwise} \\ \end{array} \right. .
% \end{equation} }

We construct the projection holograms from the states
\begin{equation}\label{eq:NFracState2}
U_{n}(\phi, \theta) = \mathcal{M}\sum^{n-1}_{k=0} \exp \left( i \Phi_M\left( \phi;\beta{_k}\oplus\theta \right) \right),
\end{equation}
which are  superpositions of fractional OAM modes, 
% \cite{gotte2007quantum, oemrawsingh2005experimental},
\begin{equation}\label{eq:phaseSPP}
\exp \left(i \Phi_{M}(\phi; \theta) \right) =
\begin{cases}
e^{i M \left( 2\pi +\phi - \theta \right)} & \quad 0 \leq \phi < \theta \\
e^{i M \left( \phi - \theta \right)} & \quad \theta \leq \phi < 2\pi
\end{cases},
\end{equation}s
rotated by an angle $\beta_k\oplus\ \theta=\text{mod}\left\{ \beta_k + \theta, 2\pi \right \}$ for $\beta_k=\frac{2\pi}{n}k$. \revise{Here, $\phi$ is the azimuthal coordinate.}

\revise{While $\theta$  determines the relative phase for the projections, physically it corresponds} to the relative rotation of the holograms. After transmitting the photon imprinted with the state $\ket{M, \theta}_n$, through the quantum transport channel, $\hat{T} = \sum_{\ell} \lambda_{\ell} \ket{\ell}_{A} \bra{\ell}_{B} $, the photon is projected onto the state $\ket{M, 0}_n$. The detection probability is then given by
\begin{equation}
P_n\left( \theta \right) = |\bra{0,M}\hat{T}\ket{M, \theta}|^{2},
\end{equation}
having a peak value at $P(\theta = 0)$ and a minimum at $P(\theta = \pi/ n)$. In the experiment, there are noise contributions which can be attributed to noise from the environment, dark counts and from the down-conversion and up-conversion processes. Since the channel is isomorphic to an entangled state, i.e
\begin{equation}
\hat{T} = \sum_{\ell} \lambda_{\ell} \ket{\ell}_{A} \bra{\ell}_{B} \rightarrow \rho_{\hat{T}}:= \sum_{\ell} \lambda_{\ell} \ket{\ell}_{A} \ket{\ell}_{B},
\end{equation}
we represent the system by an isotropic state,
\begin{equation}
\rho = p\rho_{\hat{T}} + (1-p)\mathbb{I}_d^2/d^{2},
\label{eq:isotropic_Channel}
\end{equation}
where $p$ is the probability of transferring a state through the channel or equivalently the purity and $\mathbb{I}_d^2$ is a $d^2$ dimensional identity matrix. In this case, the detection probability is given by
\begin{equation}
P_n\left( \theta \right) = |\bra{0,M}\hat{T}\ket{M, \theta}|^{2} + (1-p)/d^2 \ I_n(\theta),
\end{equation}
where $I_{n}(\theta) = |\bra{0,M}\mathbb{I}_{d^2}\ket{M, \theta}|^{2}$.

We compute the visibilities
\begin{equation}
  V_n = \frac{|P_n(0) - P_n(\pi/n)|}{|P_n(0) + P_n(\pi/n)|}.
\end{equation}
Using the fact that the visibility, $V_n :=V_n(p,d)$, obtained for each analyser indexed by, $n=1,3,...,2N-1$, scales monotonically with $d$ and $p$ \cite{nape2021measuring}, we determine the optimal $(p,d)$ pair that best fit the function $V_n(p,K)$ to all $N$ measured visibilities by employing the method of least squares (LSF). \IN{The fidelity for the channel, $F_{Ch}$, can therefore be computed by overlapping the truncated subspaces of dimensions $d'$ in the $d$ dimensional state from Eq.~(\ref{eq:isotropic_Channel}), with a channel state having perfect correlations. From this we compute the quantum transport fidelity, $\mathcal{F}$ from Eq.~(\ref{eq:telefidelity}).}
%%%%%%%%%%%%%%%%%%%%%%%%%%%%%%%%%%%%%%%%%%%%%%%%%%
%%% Additional info
%%%%%%%%%%%%%%%%%%%%%%%%%%%%%%%%%%%%%%%%%%%%%%%%%%
%%%%%%
\section*{Acknowledgements}
B.S. would like to acknowledge the Department of Science and Innovation and Council for Industrial and Scientific Research (South Africa) for funding. A.V. acknowledges the MCIN with funding from European Union (QSNP, 101114043), Next Generation EU (PRTR-C17.I1), and from Generalitat de Catalunya, also the Japan Society for the Promotion of Science for funding (JSPS-KAKENHI - G21K14549). F.S. acknowledges financial support by the Fraunhofer Internal Programs under Grant No. Attract 066-604178. M.A.C., F.S.R. and A.F. thanks the National Research Foundation for funding (NRF Grant No. 121908, 118532, TTK2204011621). A.V. and J.P.T. acknowledge financial support from the “Severo Ochoa” program for Centres of Excellence CEX2019-000910-S [MICINN/ AEI/10.13039/501100011033], Fundació Cellex, Fundació Mir-Puig, and Generalitat de Catalunya through CERCA, from project 20FUN02 “POLight” funded by the EMPIR programme, and from project QUISPAMOL (PID2020-112670GB-I00).
\vspace{0.2 cm}

%%%%%
\section*{Author contributions}
The experiment was performed by B.S., A.V. and I.N., with technical support by M.A.C., and the theory developed by F.S., T.K., J.P.T., and F.S.R. Data analysis was performed by B.S., A.V., I.N. and A.F. and the experiment was conceived by A.V., F.S., T.K., J.P.T., F.S.R. and A.F. All authors contributed to the writing of the manuscript. A.F. supervised the project.

%%%%%
\section*{Competing Interests}
The authors declare no competing interests.

\section*{Data availability}
The data that supports the plots within this paper and other findings of this study are available from the corresponding authors upon request.
%%%%%%%%%%%%%%%%%%%%%%%%%%%%%%%%%%%%%%%%%%%%%%%%%%
%%% References
%%%%%%%%%%%%%%%%%%%%%%%%%%%%%%%%%%%%%%%%%%%%%%%%%%

% \begin{thebibliography}{10}
% \end{thebibliography}
% \newpage
% \bibliography{mybibfile}
% \bibliographystyle{ieeetr}

\newpage \clearpage

\setcounter{page}{1}

\onecolumngrid
\begin{center}
    \textbf{\Large Supplementary information for: Quantum transport of high-dimensional spatial information with a nonlinear detector}

    \vspace{0.5 cm}

    Bereneice Sephton,$^1$ Adam Vall\'es,$^{1,2,3}$ Isaac Nape,$^1$ Mitchell A. Cox,$^4$ Fabian Steinlechner,$^{5,6}$
Thomas Konrad,$^{7,8}$ Juan P. Torres,$^{3,9}$ Filippus S. Roux,$^{10}$ and Andrew Forbes$^1$

\vspace{0.5 cm}

$^1$School of Physics, University of the Witwatersrand, Private Bag 3, Wits 2050, South Africa

$^2$Molecular Chirality Research Center, Chiba University, 1-33 Yayoi-cho, Inage-ku, Chiba 263-8522, Japan

$^3$ICFO - Institut de Ciencies Fotoniques, Castelldefels (Barcelona) 08860, Spain

$^4$School of Electrical and Information Engineering, University of the Witwatersrand, Johannesburg, South Africa

$^5$Fraunhofer Institute for Applied Optics and Precision Engineering, Albert-Einstein-Str. 7, 07745 Jena, Germany

$^6$Friedrich Schiller University Jena, Abbe Center of Photonics, Albert-Einstein-Str. 6, 07745 Jena, Germany

$^7$School of Chemistry and Physics, University of KwaZulu-Natal, Durban, South Africa

$^8$National Institute of Theoretical and Computational Sciences (NITheCS), KwaZulu-Natal, South Africa

$^9$Department of Signal Theory and Communications, UPC - Campus Nord D3, 08034 Barcelona, Spain

$^{10}$National Metrology Institute of South Africa, Meiring Naud\'e Road, Brummeria, Pretoria 0040, South Africa

\end{center}

\vspace{0.5 cm}

\twocolumngrid

\beginsupplement{\begin{center}
    \textbf{Supplementary Note 1 - Experimental setup}
\end{center}

\noindent We refer the reader to the detailed schematic of our experiment found in Suppl. Fig. \ref{detailedSetup}. Here a 1.5 W linearly polarised continuous wave (CW) Coherent Verdi laser centred at a wavelength of $\lambda_p = 532$ nm was focused down using a $f_1$ = 750 mm lens to produce a pump spot size of $2w_p \approx 600 \mu$m in a periodically-poled potassium titanyl phosphate (PPKTP) crystal (NLC$_1$), yielding signal and idler photons at wavelengths $\lambda = 1565$ nm and 806 nm. A HWP placed before the crystal facilitated polarisation matching. A 750 nm long-pass filter (LPF) placed directly after the crystal blocked the unconverted pump beam, while a long-pass dichroic mirror (DM$_1$) centred at $\lambda = 950$ nm transmitted the $\lambda_B = 1565$ nm down-converted photon through and reflected the $\lambda_C = 806$ nm down-converted photons to the sender party. The reflected photon was relayed onto the second PPKTP crystal (NLC$_2$), with a 1:1 imaging 4$f$-system (focal lengths of $f_2$ = $f_3$ = 175 mm), for sum-frequency generation (SFG).

Both crystals used for up- and down-conversion were 1 x 2 x 5 mm PPKTP crystal with poling period 9.675 $\mu$m for type-0 phase matching. They were spatially orientated so that frequency conversion occurred for vertically polarised pump (and seed) light, producing vertically polarised photons. Phase matching for collinear generation of 1565 nm and 806 nm SPDC as well as up-conversion of 806 nm photons with the 1565 nm structured pump was achieved through control of the crystal temperatures.

\begin{figure*}[t]
    \centering
    \includegraphics[width=\linewidth]{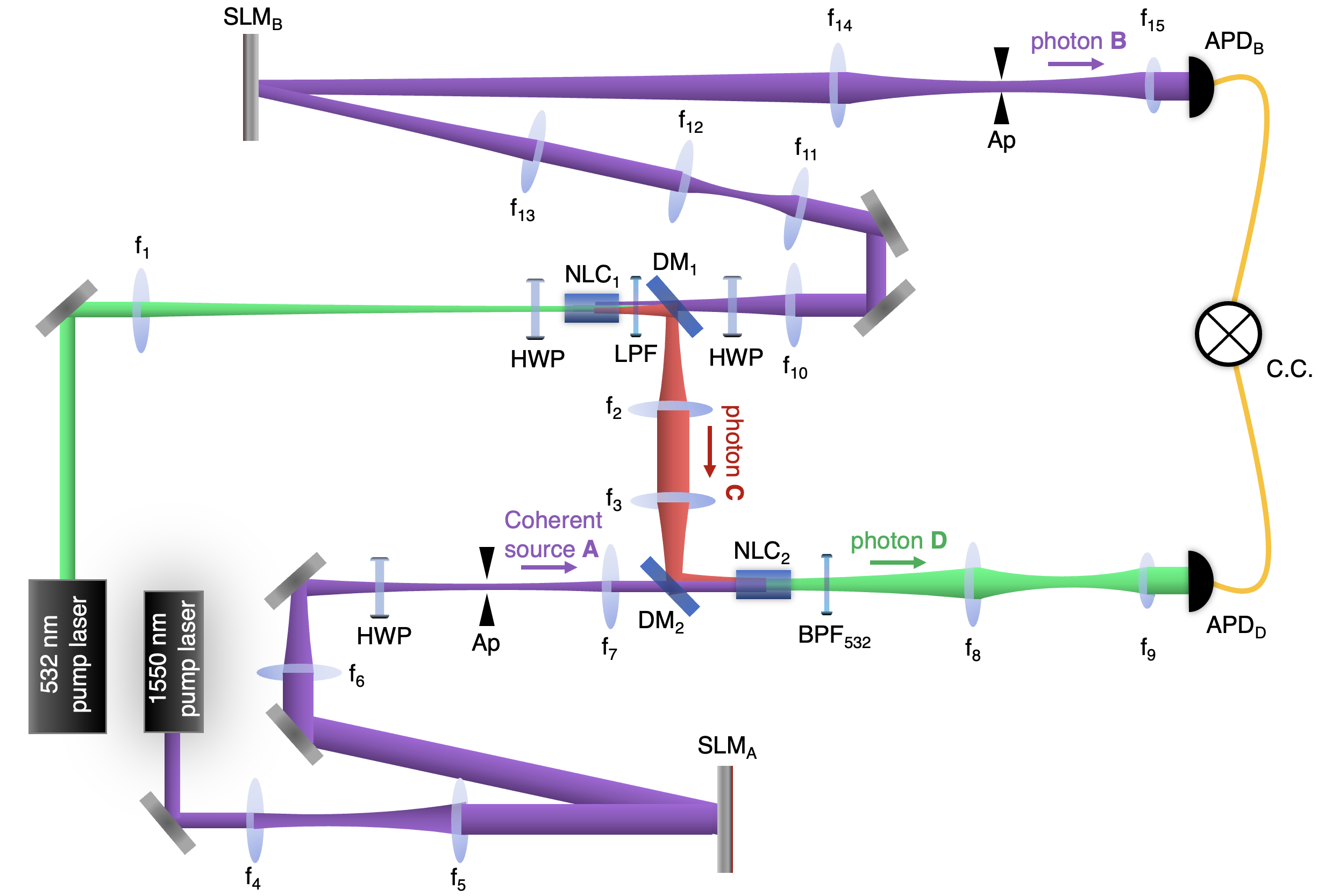}
    \caption{Detailed experimental setup description for high-dimensional spatial quantum transport without ancillary photons. \BS{Ap: Aperture; BPF: Bandpass filter; DM: Dichroic mirror; f: Lens focal length; LPF: Lowpass filter; HWP: Half-waveplate; NLC: $\chi^{(2)}$ Non-linear crystal; SLM: Spatial light modulator (phase-only).} }
    \label{detailedSetup}
\end{figure*} 
The coherent source A carrying the spatial information to be transferred was created a 3.5 W horizontally polarised EDFA amplified 1565 nm laser beam that was expanded onto SLM$_A$ with a 1:3 imaging 4$f$-system of $f_4$ = 50 mm and $f_5$ = 150 mm. The polarisation of the modulated light was rotated to vertical using a second HWP to meet the phase-matching condition for SFG. A second 10:1 imaging 4$f$-system (focal lengths $f_6$ = 750 mm and $f_7$ = 75 mm) with an aperture (Ap) in the Fourier plane resized and isolated the 1st diffraction order of the modulated beam from the SLM$_A$. The prepared state then formed a ~200 $\mu$m spot size in the second PPKTP crystal and was overlapped with the 806 nm photons by means of another long-pass dichroic mirror centered at 950 nm (DM$_2$) to generate up-converted photons of 532 nm. A 532$\pm3$ nm band-pass filter (BPF$_{532}$) after the crystal blocked the residual down-converted photons and two-photon absorption noise from the 1565 nm pump laser, allowing the up-converted photons to be coupled into a single-mode fiber (SMF) with a $f_8$ = 750 mm and $f_9$ = 4.51 mm imaging 4$f$-system. The photons were detected with a Perkin-Elmer VIS avalanche photodiode (APD) and in coincidence with the photon B.

The transmitted $\lambda = 1565$ nm down-converted photons were expanded and imaged onto a second SLM with two 4$f$-systems (focal lengths of $f_{10}$ = 100 mm, $f_{11}$ = 200 mm, $f_{12}$ = 150 mm and $f_{13}$ = 750 mm) for spatial tomographic projections of the transferred state. Here the spatially modulated photons were then filtered with an aperture and resized (4$f$-system with focal lengths $f_{14}$ = 750 mm and $f_{15}$ = 2.0 mm) for coupling into an SMF, which was detected by an IDQuantique ID220 InGaAs free-running APD. A PicoQuant Hydraharp 400 event timer allowed the projected SFG and SPDC photons to be measured in coincidences (C.C.).

\begin{center}
    \textbf{Supplementary Note 2 - Quantum transport with SFG} % --- (Juan's notes)
\end{center}

\revise{We consider only those photons in coherent state A that are involved in the SFG process, and following the main text consider these as photon-state A.  Considering that the state to be transferred is a high-dimensional single-photon state after its post-selection in coincidences, i.e., the spatial mode is selected from a high-dimensional set, the superposition state to be transferred can be represented by}

% The general scheme for teleportation is used here, developed as in Refs. \cite{molotkov1998quantum, molotkov1998experimental}. However, the state that is transferred is a high-dimensional single-photon state after its post-selection in coincidences --- i.e., the spatial mode is selected from a high-dimensional set. The superposition state to be transferred can be represented by
\begin{equation}
\ket{\psi_{A}} = \int \alpha(\mathbf{q}_{A})\hat{a}_{A}^{\dag}(\mathbf{q}_{A})\ket{\text{vac}}\ d^2 q_{A},
\label{eqn:channel0}
\end{equation}
where $\alpha(\mathbf{q}_{A})$ is the angular spectrum associated with the chosen spatial mode, $\hat{a}_{A}^{\dag}(\mathbf{q}_{A})$ is the creation operator of photons with two-dimensional transverse wave vector $\mathbf{q}_{A}$ and $\ket{\text{vac}}$ is the vacuum state. It is assumed that the frequency $\omega_{A}$ is fixed.

Using SPDC, we prepare an entangled state and consider a single pair of photons with transverse wave vectors $\mathbf{q}_{B}$ and $\mathbf{q}_{C}$, respectively. The state of this photon pairs can be expressed by
\begin{align}
\ket{\psi_{BC}} = & \int f(\mathbf{q}_{B},\mathbf{q}_{C}) \hat{a}_{B}^{\dag}(\mathbf{q}_{B}) \nonumber \\
& \times \hat{a}_{C}^{\dag}(\mathbf{q}_{C}) \ket{\text{vac}}\ d^2 q_{B}\ d^2 q_{C} ,
\label{eqn:channel}
\end{align}
where $f(\mathbf{q}_{B},\mathbf{q}_{C})$ is the two-photon wave function. The state of the combined system is then given by
%\begin{widetext}
\begin{align}
\ket{\psi_{ABC}} = & \ket{\psi_{A}} \otimes \ket{\psi_{BC}} \nonumber \\
= & \int \alpha(\mathbf{q}_{A}) f(\mathbf{q}_{B},\mathbf{q}_{C}) \hat{a}_{A}^{\dag}(\mathbf{q}_{A}) \hat{a}_{B}^{\dag}(\mathbf{q}_{B}) \nonumber \\
& \times \hat{a}_{C}^{\dag}(\mathbf{q}_{C})\ket{\text{vac}}\ d^2 q_{A}\ d^2 q_{B}\ d^2 q_{C} ,
\label{eqn:channel1}
\end{align}
%\end{widetext}

The process of sum-frequency generation (SFG) is now applied to the state in Eq.~(\ref{eqn:channel1}) to produce an up-converted photon D from a pair of photons: \revise{photon-state A and photon C}. The resulting quantum state of the system becomes
\begin{align}
\ket{\psi_{BD}} = & \int g(\mathbf{q}_{A},\mathbf{q}_{C},\mathbf{q}_{D}) f(\mathbf{q}_{B},\mathbf{q}_{C}) \alpha(\mathbf{q}_{A}) \hat{a}_{B}^{\dag}(\mathbf{q}_{B}) \nonumber \\
& \times \hat{a}_{D}^{\dag}(\mathbf{q}_{D})\ket{\text{vac}}\ d^2 q_{A}\ d^2 q_{B}\ d^2 q_{C}\ d^2 q_{D} ,
\end{align}
where $g(\mathbf{q}_{A},\mathbf{q}_{C},\mathbf{q}_{D}) $ is the kernel for the SFG process.

If we assume the critical phase-matching condition $\mathbf{q}_{A} + \mathbf{q}_{C} = \mathbf{q}_{D}$, then the expression becomes
\begin{align}
\ket{\psi_{BD}} = & \int g(\mathbf{q}_{A},\mathbf{q}_{D}-\mathbf{q}_{A},\mathbf{q}_{D}) f(\mathbf{q}_{B},\mathbf{q}_{D}-\mathbf{q}_{A})
\alpha(\mathbf{q}_{A}) \nonumber \\
& \times \hat{a}_{B}^{\dag}(\mathbf{q}_{B}) \hat{a}_{D}^{\dag}(\mathbf{q}_{D}) \ket{\text{vac}}\ d^2 q_{A}\ d^2 q_{B}\ d^2 q_{D} ,
\end{align}
%\end{widetext}
where we eliminate $\mathbf{q}_{C}$ in terms of $\mathbf{q}_{A}$ and $\mathbf{q}_{D}$. From the arguments of $f$, we see that the wave vector of photon-state A is now related to that of the measured photon B.

With the aid of a projective measurement of the SFG photon D in terms of a mode $U(\mathbf{q}_{D})$, analogous to projecting into one of the Bell states, we can herald the quantum transport of the state. The state of photon B is then given by
\begin{equation}
\ket{\psi_{B}} = \int \beta(\mathbf{q}_{B}) \hat{a}_{B}^{\dag}(\mathbf{q}_{B}) \ket{\text{vac}}\ d^2 q_{B} ,
\label{eqn:pC1}
\end{equation}
where
\begin{align}
\beta(\mathbf{q}_{B}) = & \int U^{*}(\mathbf{q}_{D}) g(\mathbf{q}_{A},\mathbf{q}_{C},\mathbf{q}_{D}) \nonumber \\
& \times f(\mathbf{q}_{B},\mathbf{q}_{C}) \alpha(\mathbf{q}_{A})\ d^2 q_{A}\ d^2 q_{C}\ d^2 q_{D} .
\label{eqn:pC2}
\end{align}

A successful quantum transport process would imply that $\beta(\mathbf{q})=\alpha(\mathbf{q})$. It requires that
\begin{align}
\int U^{*}(\mathbf{q}_{D}) g(\mathbf{q}_{A},\mathbf{q}_{C},\mathbf{q}_{D}) & \nonumber \\
\times f(\mathbf{q}_{B},\mathbf{q}_{C})\ d^2 q_{C}\ d^2 q_{D} & \approx \delta(\mathbf{q}_{B}-\mathbf{q}_{A}) .
\label{eqn:tp0}
\end{align}

Under what circumstance would this condition be satisfied? First, we will assume that the mode $U(\mathbf{q})$ for the measurement of the SFG photon D (the so-called {\em anti-pump}) is the same as the mode of the pump beam. The SFG process can then be regarded as the conjugate of the SPDC process, used to produce the entangled photons. Hence,
\begin{equation}
\int U^{*}(\mathbf{q}_{D}) g(\mathbf{q}_{A},\mathbf{q}_{C},\mathbf{q}_{D})\ d^2 q_{D} \sim f^*(\mathbf{q}_{A},\mathbf{q}_{C}) .
\end{equation}
The two-photon wave function is a product of the pump mode and the phase-matching function, which is in the form of a sinc-function:
\begin{equation}
f(\mathbf{q}_{B},\mathbf{q}_{C}) \sim U(\mathbf{q}_{B}+\mathbf{q}_{C}) \text{sinc}\left(\eta|\mathbf{q}_{B}-\mathbf{q}_{C}|^2\right) ,
\label{eqn:tpwf}
\end{equation}
where $\eta$ represents a dimension parameter that determines the width of the function (see below). Under suitable experimental conditions (discussed below) the sinc-function only contributes when its argument is close to zero so that the sinc-function can be replaced by 1. Moreover, if the modes for the pump and the anti-pump are wide enough, they can be regarded as plane waves, which are represented as Dirac $\delta$ functions in the Fourier domain. Then
\begin{equation}
f(\mathbf{q}_{B},\mathbf{q}_{C}) \approx \delta(\mathbf{q}_{B}+\mathbf{q}_{C}) ,
\end{equation}
and
\begin{equation}
\int U^{*}(\mathbf{q}_{D}) g(\mathbf{q}_{A},\mathbf{q}_{C},\mathbf{q}_{D})\ d^2 q_{D}
\approx \delta(\mathbf{q}_{A}+\mathbf{q}_{C}) .
\label{eqn:sfgdel}
\end{equation}
Together, they produce the required result in Eq.~(\ref{eqn:tp0}) after the integration over $\mathbf{q}_{C}$ has been evaluated.

It follows that, by detecting the up-converted photon D, the state of photon B is heralded to be
\begin{equation}
\ket{\psi_{B}} = \int \alpha(\mathbf{q}) \hat{a}_{B}^{\dag}(\mathbf{q}) \ket{\text{vac}}\ d\mathbf{q} .
\label{eqn:TeleState}
\end{equation}
It means that the quantum transport process can be performed successfully with SFG, provided that the applied approximation are valid under the pertinent experimental conditions, which are considered next.

\begin{center}
    \textbf{Supplementary Note 3 - Experimental conditions}
\end{center}

It is well-known that SPDC produces pairs of photons (signal and idler) that are entangled in several degrees of freedom, including energy-time, position-momentum and spatial modes. A good review covering these scenarios is found in Ref. \cite{erhard2020advances}. With SPDC being a suitable source of entanglement for our protocol, we consider in more detail what the experimental conditions need to be to achieve successful quantum transport with the aid of SFG. For this purpose, we consider a collinear SPDC system with some simplifying assumptions. Even though details may be different from a more exact solution, the physics is expected to be the same.

As shown in the previous section, the success of the process requires that $\mathbf{q}_{B}=-\mathbf{q}_{C}$, provided that the pump beam is a Gaussian mode, which implies perfect anti-correlation of the wave vectors between the signal (photon B) and idler (photon C). It is achieved when (a) the argument of the sinc-function can be set to zero, which is valid under the {\em thin-crystal approximation}, and (b) the beam waist of the pump beam $w_p$ is relatively large, leading to the {\em plane-wave approximation}.

The scale of the sinc-function is inversely proportional to $\sqrt{\lambda_p L}$ where $\lambda_p$ is the wavelength of pump (or anti-pump) and $L$ is the length of the nonlinear crystal ($L$ = 5 mm in our case). To enforce the requirement that its argument is evaluated close to zero, we require that the integral only contains significant contributions in this region. Therefore, the angular spectrum of the pump mode with which it is multiplied, must be much narrower than the sinc-function. The width of the angular spectrum of the pump mode is inversely proportional to the beam waist $w_p$. Therefore, the condition requires that
\begin{equation}
\frac{1}{\lambda_p L} \gg \frac{1}{w_p^2} ~~~ \Rightarrow ~~~ 1 \gg \frac{\lambda_p L}{w_p^2} \propto \frac{L}{z_R} ,
\label{eqn:DCphotons}
\end{equation}
where $z_R$ is the Rayleigh range of the pump beam. The relationship shows that the sinc-function can be replaced by 1 if the Rayleigh range of the pump beam is much larger than the length of the nonlinear crystal, leading to the thin-crystal approximation. We see that this condition is consistent with the requirement that $w_p$ is relatively large, which is required for the plane-wave approximation.

%\section{Sum-frequency generation}

Similar conditions are required for the second nonlinear crystal that performs SFG. In that case, two input photons with angular frequencies $\omega_A$ and $\omega_C$, respectively, are annihilated to generate a photon with an angular frequency $\omega_D = \omega_A + \omega_C$, imposed by energy conservation. The size of the mode that is detected, takes on the role of $w_p$ and the length of the second nonlinear crystal replaces the length $L$ of the first crystal. The wavelength after the sum-frequency generation process is the same as that of the pump for the SPDC $\lambda_p$. The equivalent conditions for the momentum conservation impose an anti-correlation $\mathbf{q}_{A}=-\mathbf{q}_{C}$, considering we only project the upconverted photon D onto the Gaussian mode (fundamental spatial mode), as implied in Eq.~(\ref{eqn:sfgdel}).

\begin{center}
    \textbf{Supplementary Note 4 - Quantum transport channel}
\end{center}

In order to simulate the quantum transport process, one may view it as a communication channel with imperfections such as loss and a limited bandwidth. The operation that represents the quantum transport channel may be obtained by overlapping a photon from the SPDC state with one of the inputs for the SFG process, where the SPDC state is $\ket{\psi_{\text{SPDC}}}=\ket{\psi_{B,C}^{(\text{SPDC})}}$, as defined in Eq.~(\ref{eqn:channel}). The two-photon wave function, which is symbolically provided in Eq.~(\ref{eqn:tpwf}) can be represented more accurately as
\begin{align}
f_{\text{SPDC}}(\mathbf{q}_{B},\mathbf{q}_{C}) = & \mathcal{N}\exp(-\tfrac{1}{4} w_p^2 |\mathbf{q}_{B}+ \mathbf{q}_{C}|^2) \nonumber \\
& \times \text{sinc}(\tfrac{1}{2} L_p \Delta k_z) ,
\label{Eq:SPDCpsi}
\end{align}
where $\mathcal{N}$ is a normalisation constant, $w_p$ is the pump beam radius, and $L_p$ is the nonlinear crystal length. The mismatch in the z-components of the wave vectors for non-degenerate collinear quasi-phase matching is
\begin{align}
\Delta k_z = & - \frac{\lambda_p}{4\pi n_p}|\mathbf{q}_{B}+\mathbf{q}_{C}|^2 \nonumber \\
& + \frac{\lambda_B}{4\pi n_B}|{\mathbf{q}_{B}}|^2 + \frac{\lambda_C}{4\pi n_C}|{\mathbf{q}_{C}}|^2 ,
\label{Eq:dkz}
\end{align}
where, $\lambda_{B,C}$ are the down-converted wavelengths in vacuum for the signal and idler, respectively, with their associated crystal refractive indices $n_B$ and $n_C$, and $\lambda_p$ is the pump wavelength in vacuum, with its associated crystal refractive index denoted by $n_p$. The quasi-phase matching condition is implemented by periodic poling of the nonlinear medium. It implies a slight reduction in efficiency by a factor $2/\pi$, which is absorbed into the normalisation constant.

The SFG process may be thought of as the SPDC case in reverse where photon C and \revise{photon-state A} (with wave vectors $\mathbf{q}_{C}$ and ${\mathbf{q}_A}$, respectively) are up-converted to an 'anti-pump' photon D. It can thus be represented, in analogy to Eq.~(\ref{eqn:tpwf}), by the bra-vector
\begin{align}
\bra{\psi_{C,A}^{(\text{SFG})}} = & \int \bra{\text{vac}} \hat{a}_{C}(\mathbf{q}_{C}) \hat{a}_{A}(\mathbf{q}_{A}) \nonumber \\
& \times f^*(\mathbf{q}_{C},\mathbf{q}_{A})\ d^2 q_{C}\ d^2 q_{A} ,
\label{Eq:SFG}
\end{align}
where the associated two-photon wave function is given by
\begin{align}
f_{\text{SFG}}^*(\mathbf{q}_{C},\mathbf{q}_{A}) = & \mathcal{N} \exp(-\tfrac{1}{4} w_D^2|\mathbf{q}_{C}+ \mathbf{q}_{A}|^2) \nonumber \\
& \times \text{sinc}(\tfrac{1}{2} L_D \Delta k_z) ,
\label{Eq:SFGpsi}
\end{align}
with $w_D$ being the anti-pump beam waist (replacing $w_p$), and $L_D$ being the nonlinear crystal length (replacing $L_p$). The wave vector mismatch $\Delta k_z$ differs from the expression in Eq.~(\ref{Eq:dkz}) only in the replacement of $\mathbf{q}_{B}$ by $\mathbf{q}_{A}$ and corresponding different values for $\lambda$.

We can now define a {\em quantum transport channel operator} as the partial overlap between $\ket{\psi_{B,C}^{(\text{SPDC})}}$ and $\bra{\psi_{C,A}^{(\text{SFG})}}$, where only the photons associated with $C$ are contracted. The resulting operator is given by
\begin{align}
\hat{T} = & \braket{\psi_{C,A}^{(\text{SFG})}|\psi_{B,C}^{(\text{SPDC})}} \nonumber \\
= & \int \ket{\mathbf{q}_{B}} T(\mathbf{q}_{B},\mathbf{q}_{A}) \bra{\mathbf{q}_{A}}\ d^2 q_{A}\ d^2 q_{B} ,
%\label{Eq:SFGpsi}
\end{align}
where $\ket{\mathbf{q}_{B}}=\hat{a}_{B}^{\dag}(\mathbf{q}_{B})\ket{\text{vac}}$, and $\bra{\mathbf{q}_{A}}=\bra{\text{vac}} \hat{a}_{A}(\mathbf{q}_{A})$. The kernel for the channel is given by
\begin{equation}
T(\mathbf{q}_{B},\mathbf{q}_{A})
= \int f_{\text{SFG}}^*(\mathbf{q}_{C},\mathbf{q}_{A}) f_{\text{SPDC}}(\mathbf{q}_{B},\mathbf{q}_{C})\ d^2 q_{C} .
\label{Eq:channelt}
\end{equation}
It describes how spatial information is transferred by the quantum transport process, implemented with SFG.

The quantum transport process can be simplified by using the thin-crystal approximation, discussed above. The Rayleigh ranges of the pump beam and anti-pump beam are made much larger than their respective crystal lengths. Therefore, $L/z_R\rightarrow 0$, for both the pump and the anti-pump. It allows us to approximate the phase-matching sinc-functions in Eqs.~(\ref{Eq:SPDCpsi}) and (\ref{Eq:SFGpsi}) as Gaussian functions \cite{law2004analysis}
\begin{equation}
\text{sinc}(\tfrac{1}{2}L \Delta k_z) \rightarrow \exp(- L \Delta k_z) .
\label{Eq:thinLimit}
\end{equation}
The wave functions then become
\begin{align}
f_{\text{SPDC}}(\mathbf{q}_{B},\mathbf{q}_{C}) = & \mathcal{N} \exp(-\tfrac{1}{4} w_p^2 |\mathbf{q}_{B}+ \mathbf{q}_{C}|^2) \nonumber \\
& \times \exp[-L_p \Delta k_z(\mathbf{q}_{B},\mathbf{q}_{C})] ,
\label{Eq:SPDCpsi_thin}
\end{align}
and
\begin{align}
f_{\text{SFG}}^*(\mathbf{q}_{C},\mathbf{q}_{A}) = & \mathcal{N} \exp(-\tfrac{1}{4} w_D^2|\mathbf{q}_{C}+ \mathbf{q}_{A}|^2) \nonumber \\
& \times \exp[-L_D \Delta k_z(\mathbf{q}_{C},\mathbf{q}_{A})] ,
\label{Eq:SFGpsi_thin}
\end{align}
where $\Delta k_z$ is given by Eq.~(\ref{Eq:dkz}).

Substituting Eq.~(\ref{Eq:SPDCpsi_thin}) and (\ref{Eq:SFGpsi_thin}) into Eq.~(\ref{Eq:channelt}), we obtain
\begin{align}
T(\mathbf{q}_B,\mathbf{q}_A)
 = & \mathcal{N}^2 \int \exp\left[-\tfrac{1}{4} w_p^2 |\mathbf{q}_{B}+\mathbf{q}_{C}|^2 \right. \nonumber \\
 & -\tfrac{1}{4} w_D^2 |\mathbf{q}_{C}+\mathbf{q}_{A}|^2-L_p \Delta k_z(\mathbf{q}_B,\mathbf{q}_C) \nonumber \\
 & \left. -L_D \Delta k_z(\mathbf{q}_C,\mathbf{q}_A)\right]\ d^2 q_{C} .
\label{Eq:ChanMod0}
\end{align}
If we set $L_p=L_D=0$ and evaluate the integral, we obtain the thin-crystal limit expression
\begin{align}
T(\mathbf{q}_B,\mathbf{q}_A)
 = & \frac{\mathcal{N}^2}{\pi (w_D^2+w_p^2)} \nonumber \\
 & \times \exp\left[-\frac{w_D^2 w_p^2}{4 (w_D^2+w_p^2)} |\mathbf{q}_B-\mathbf{q}_A|^2 \right] \nonumber \\
 = & T'(\mathbf{q}_B-\mathbf{q}_A) .
\label{Eq:ChannelMod}
\end{align}

According to the Choi-Jamoilkowski (state-channel) duality, we can treat the channel operation in Eq.~(\ref{Eq:ChanMod0}) as an entangled state. It can thus be used to calculate a Schmidt number for the state, which can be interpreted as the effective number of modes that the channel can transfer. For this purpose, we set $L_p=L_D=L$. The result is
\begin{equation}
K = \frac{n_{A} n_{B} w_D^2 w_p^2}{(w_D^2+w_p^2)(n_{A}\lambda_{B}+n_{B}\lambda_{A})L}.
\end{equation}
Although the Schmidt number provides an indication of the number of modes that can be transferred by the quantum transport process, it does not tell us what the modes are that can be transferred. For this purpose, we investigate the system numerically.

\begin{center}
    \textbf{Supplementary Note 5 - Numerical simulation}
\end{center}
\label{NumSec}

It follows that Eq.~(\ref{Eq:ChannelMod}) can be used to simulate the conditional probabilities for encoding and detecting spatial modes using \revise{photon-state A} and photon C, respectively. A summary of the experiment with the relevant parameters is given in Suppl. Fig. \ref{TeleAnaSetup}. Here the up-conversion beam waist ($w_0$) is set by the mode field diameter (MDF) of the SMF.

\begin{figure}
  \centering
  \includegraphics[width=\linewidth]{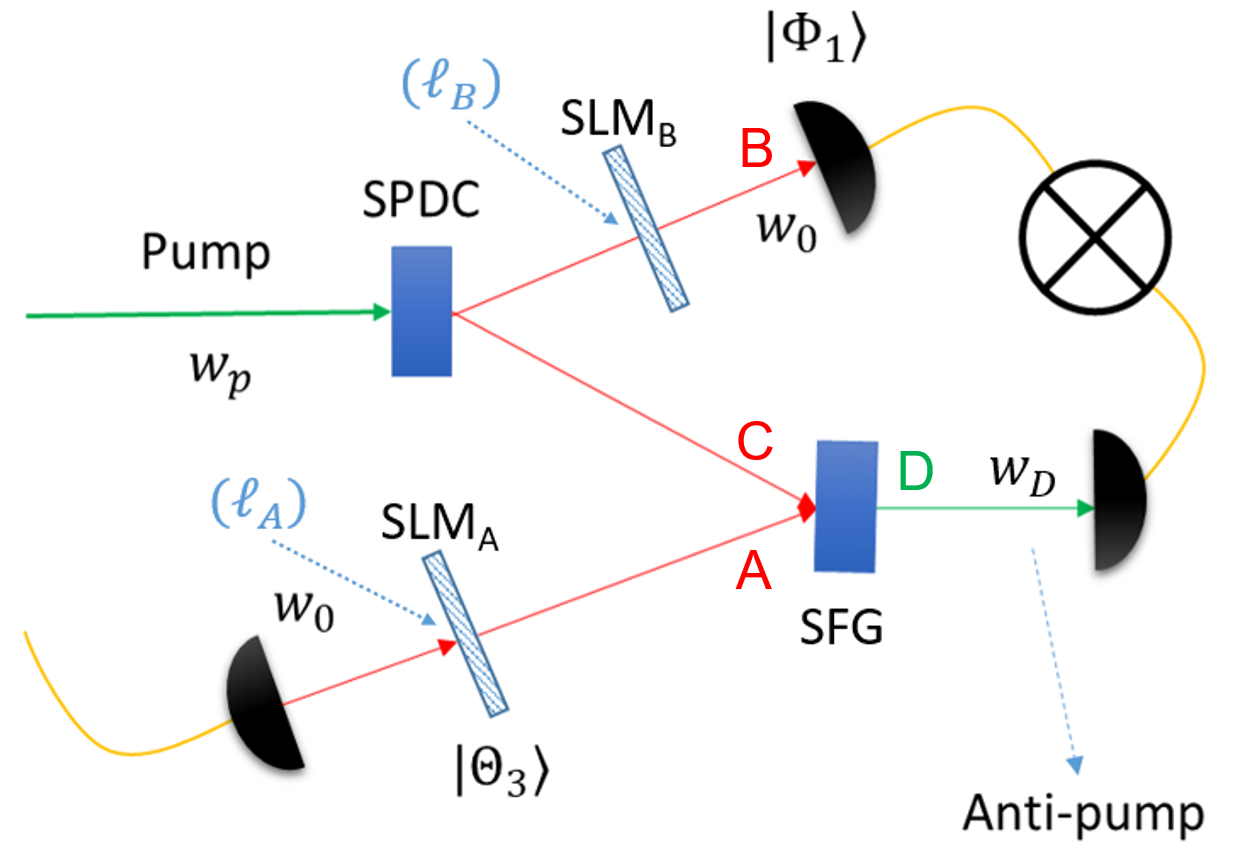}
  \caption{\textbf{Quantum transport channel scheme with optimisation parameters.} A pump photon with a waist size of $w_p$ impinges on a nonlinear crystal, generating two photons, photon B and photon C. Photon C is sent to a second crystal for SFG where it is absorbed with another independent photon encoded with the mode $\ket{\Theta_A}$, corresponding to a mode field with a waist size of $w_0$. We will call this independent photon, \revise{photon-state A}. The resulting photon D from the up-conversion process, with a waist size of $w_D$, is coupled into a SMF and measured in coincidences with photon B. To recover the spatial information of photon-state A, we scan the spatial mode of photon B with spatial projections mapping onto the state $\ket{\Phi_B}$ with a corresponding mode field that also has a waist size of $w_0$. $\ell_A$ and $\ell_B$ refer to the encoded and projected vortex states displayed on the SLMs.}
  \label{TeleAnaSetup}
\end{figure}

\begin{figure}
  \centering
\includegraphics[width=\linewidth]{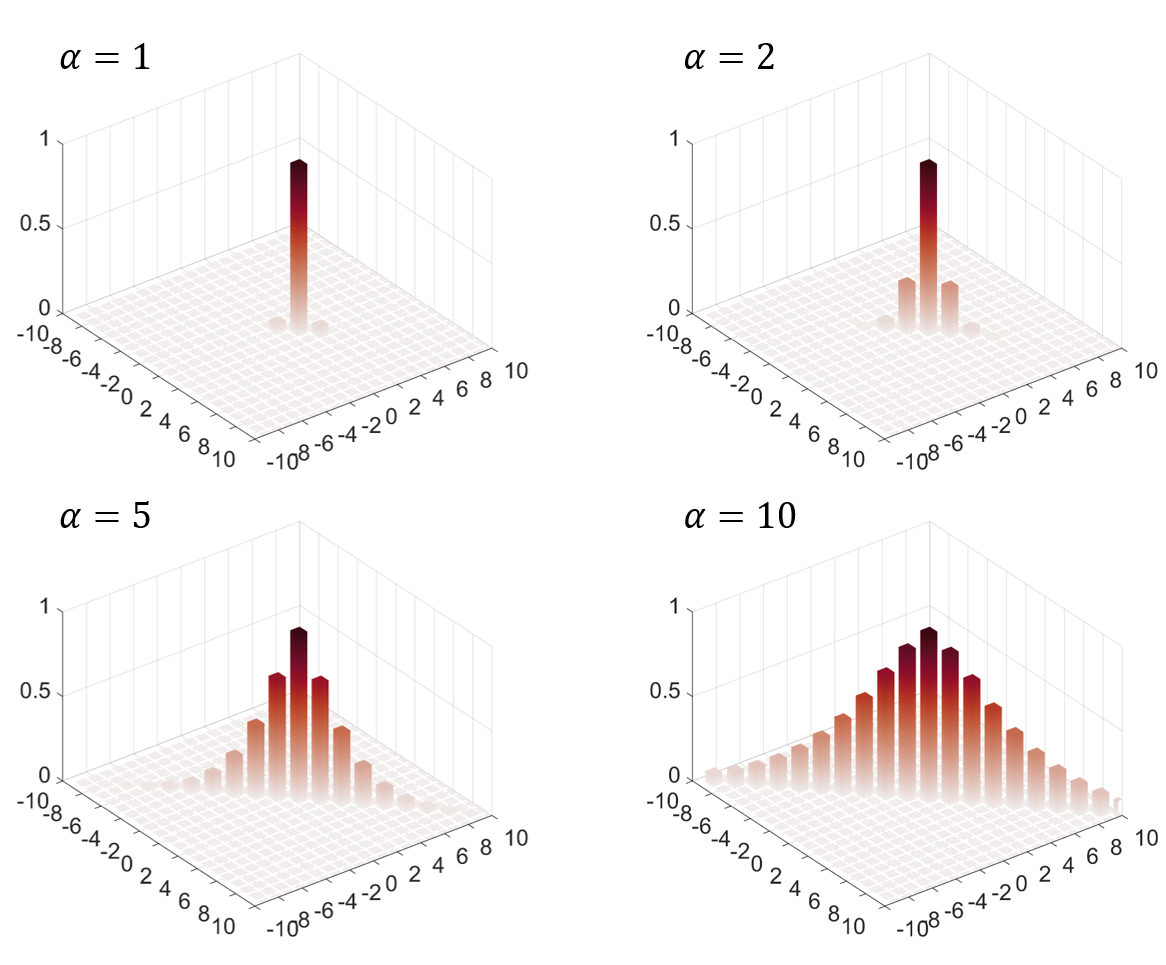}
 \caption{\textbf{Simulated modal spectrum from measurements of photon B for the encoded states of \revise{photon-state A}.} Vortex modes for various $\alpha = w_p/w_0$ with a fixed $\beta = w_p/w_c = 1$ were used as the transferred states. The modal spectrum shows non-zeros probabilities for $\ell_A=\ell_B$. Moreover, the spectrum becomes wider with increasing $\alpha$. This means that the transferred and detected mode sizes must be significantly smaller than the SPDC mode to see a wider spectrum.}
  \label{fig:spec}
\end{figure}

\begin{figure*}
  \centering
  \includegraphics[width=\linewidth]{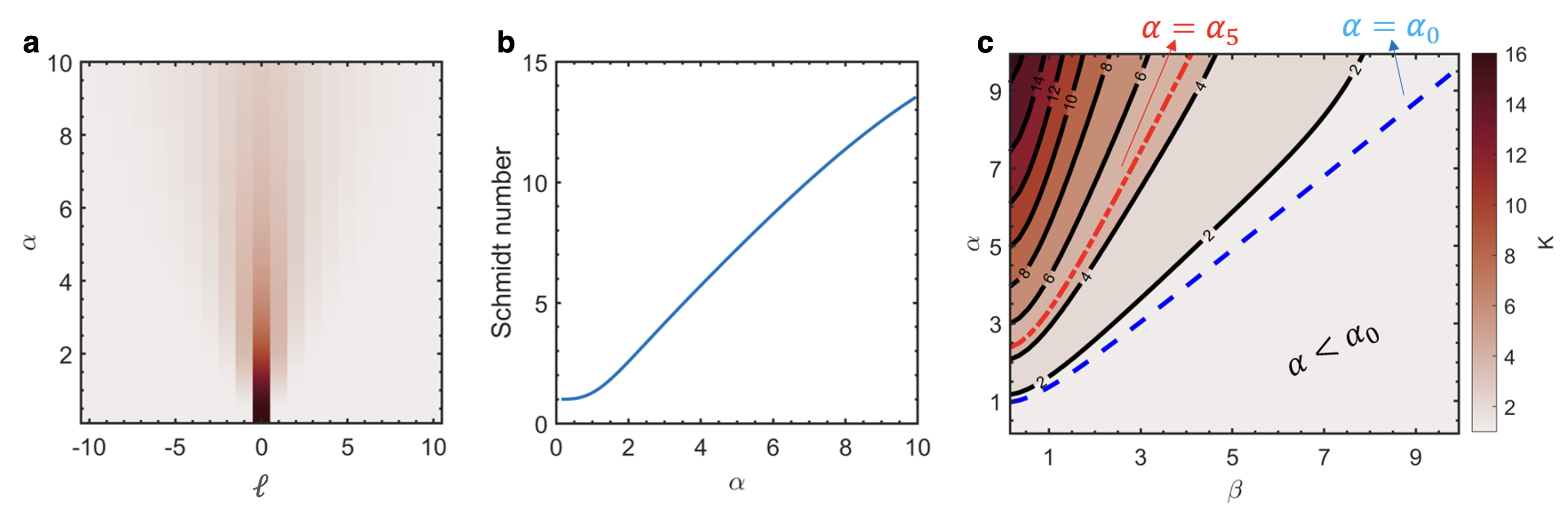}
 \caption{\textbf{Quantum transport channel capacity analysis.} (a) Density plot of the spiral spectrum as a function of $\alpha$ and $\ell$. Only the diagonal is shown. (b) The dimensionality (K), measured from the Schmidt number vs $\alpha = w_p/w_0$ with a fixed $\beta=w_p/w_c = 1$. (c) Contour plot of the dimensionality (K) as a function of $\beta$ and $\alpha$. For higher dimensionality we need a small $\beta<1$ and large $\alpha$. The blue single dash line corresponds to the minimum $\alpha=\alpha_0$ for transferring a spatial mode through the setup. The red double dashed line corresponds to the minimum $\alpha=\alpha_5$ for transferring OAM modes with $\ell = [-2,2]$ giving access to no more than $K=5$ dimensions.}
  \label{fig:Dimen}
\end{figure*}

Now supposing we want to transfer the spatial information of \revise{photon-state A} to photon B, let the modes corresponding to each photon be expressed as
\begin{equation}
\ket{\Phi_B} = \int \phi(\mathbf{q}_{B}) \ket{\mathbf{q}_{B}}\ d^2 q_{B} ,
\end{equation}
and
\begin{equation}
\ket{\Theta_A} = \int \theta(\mathbf{q}_{A}) \ket{\mathbf{q}_{A}}\ d^2 q_{A} ,
\end{equation}
where $\phi(\cdot)$ and $\theta(\cdot)$ are the field amplitudes. The overlap probability amplitude, given the quantum transport matrix presented earlier, is therefore
\begin{align}
\bra{\Phi_B} \hat{T} \ket{\Theta_A} = & \int \phi^{\dag}(\mathbf{q}_B) \theta(\mathbf{q}_A) \nonumber \\
 & \times T'(\mathbf{q}_B-\mathbf{q}_A)\ d^2 q_{B}\ d^2 q_{A} .
\end{align}

Since the weighting function of the channel matrix depends only on the relative momenta, we can simplify the integral
\begin{equation}
\bra{\Phi_B} \hat{T} \ket{\Theta_A} = \int \phi^{\dag} (\mathbf{q}_B) \theta'(\mathbf{q}_B)\ d^2 q_{B},
\end{equation}
where $\theta'(\mathbf{q}) = \theta*T $ is a simple convolution.

For the numerical calculation, vortex modes will be considered, which are basis modes with orbital angular momentum (OAM or $\ell$), i.e $\ket{\Phi_B}, \ket{\Theta_A} \in \{ \ket{\ell}, \ell \in \mathcal{Z} \}$.

Photon-state A is then encoded with the vortex modes, using phase-only modulation:
\begin{equation}
\ket{\ell} = \int G(\mathbf{q};w_0) \exp(i\ell\phi_q)\ \ket{\mathbf{q}} d^2 q,
\end{equation}
where $\ell$ is the topological charge of the mode, $\phi$ is the azimuth coordinate and $G(\mathbf{q};w_0)$ is a Gaussian mode with a transverse waist of $w_0$ at the crystal plane. The photon $B$ is projected onto these vortex modes. To ascertain the best experimental settings for measuring a large spectrum of OAM modes through the channel, the parameters $\alpha = w_p / w_0$ and $\beta = w_p / w_D$ are considered.

In Suppl. Fig. \ref{fig:spec}, the conditional probabilities
\begin{equation}
P_{\ell_B, \ell_A}(\alpha) = |\bra{\ell_B} \hat{T} \ket{\ell_A}|^2,
\end{equation}
are presented for various $\alpha$ values with a fixed $\beta=1$ ($w_D=w_p$), i.e the anti-pump and SPDC pump modes are the same size. Here, larger values of $\alpha$ widen the modal spectrum, which can be seen in Suppl. Fig. \ref{fig:Dimen}(a) where only the diagonals are extracted. Therefore, larger values of $\alpha$ increases the dimensionality of the system. The dimensions can be quantitatively measured using the Schmidt number
\begin{equation}
  K(\alpha) = \frac{1}{\sum_\ell P_{\ell}^2(\alpha)},
\end{equation}
where $P_{\ell}(\alpha) = |\bra{\ell} \hat{T} \ket{\ell}|^2$.

The subsequent dimensionality $K$ is given in Suppl. Fig. \ref{fig:Dimen}(b) as a function of $\alpha$. It can be seen that an increase in the dimensionality of the modes requires a large $\alpha$ or $w_p > w_0$. Supplementary Figure \ref{fig:Dimen}(c) further shows the dimensionality as a function of $\alpha$ and $\beta$ in a contour plot. Here, a smaller value for $\beta$ yields a larger accessible dimensionality. Consequently, for detection of a dimensionality larger than two,
\begin{equation}
\alpha > \alpha_0 = \frac{n_A}{n_B} \sqrt{\beta + 1}.
\end{equation}
The blue dashed line in Suppl. Fig. \ref{fig:Dimen}(c) corresponds to $\alpha_0$ for various $\beta$ values. Indeed the dimensionality below this region is less than $K=2$. This is due to $w_0$ corresponding to the Gaussian argument of the vortex modes and not the optimal mode size of the generated or detected vortex mode.

To detect higher dimensional states, the scaling of higher order modes must be taken into account. Therefore, by noting that OAM basis modes increase in size by a factor of $M_\ell = \sqrt{|\ell| + 1}$ the relation $\alpha >\alpha_{\ell}$ where $\alpha_\ell = \sqrt{\beta + 1}M_\ell$ should be satisfied. This observation is illustrated for $\alpha_5$ as the red dashed line in Suppl. Fig. \ref{fig:Dimen}(c). Below this line, only states with less than $K=5$ dimensions are accessible. Accordingly, $\alpha_\ell$ sets a restriction on the upper limit of the dimensions accessible with the quantum transport system.

Varying these parameters in the experimental setup, we obtained the spiral bandwidths shown in Fig. \ref{fig:ConceptFig} (c-e) of the main text for the experimental parameters given in Suppl. Table \ref{tab:DimPara} and marked on the contour plot in Fig. \ref{fig:ConceptFig} (b) of the main text. Note that the same pump power conditions were considered for the three tested configurations.
% Varying these parameters in the experimental setup, we obtained the spiral bandwidths shown in Fig. \ref{NumPara}(b-e) for the experimental parameters given in Table \ref{tab:DimPara}. These values are marked on the contour plot in Fig. \ref{NumPara}(a).

\begin{table}[h!]
\begin{tabular}{ |c|c|c| }
%\caption{Variations in beam size and detection parameters}
\hline
Fig. \ref{fig:ConceptFig} (main text) & $\beta$ & $\alpha$ \\
\hline
c&4.1 & 2.7 \\
d&1.1 & 2.7 \\
e&1.1 & 4.1 \\
\hline
\end{tabular}
\caption{\textbf{Experimentally tested parameters.} Parameters values used experimentally to test the numerically simulated dimensionality trends.}
\label{tab:DimPara}
\end{table}

It follows that a large $\beta$ generates a very small bandwidth with only one OAM mode discernibly present in (c). Changing $\beta$ to be near 1 showed more modes present (see Fig. \ref{fig:ConceptFig} (d) in main text).  In the experiment this means that we must ensure that the SPDC pump mode size is smaller than the anti-pump's while significantly larger than the detection modes. Further optimising the parameters with an increase in $\alpha$ then allowed an additional increase in the spiral bandwidth, shown in the inset of Fig.~2 in the main text.

\begin{center}
    \textbf{Supplementary Note 6 - Procrustean filtering}
\end{center}

Experimental factors required compensation when evaluating the transferred results in the OAM basis and required the application of correction to the detected coincidences. These were the result of a convolution of corrections resulting from a non-flat spiral bandwidth from the SPDC photons \cite{torres2003quantum,law2004analysis}, variation in the overlap of the down-converted 806 nm photons and the 1565 nm photons in the SFG process (as shown by the quantum transport operator) and the fixed-size Gaussian filter resulting from detection with a SMF \cite{roux2014projective}. Supplementary Figure \ref{Flat}(a) shows the spiral bandwidth (at 2 minute integration time per point) resulting from these factors with a (i) density plot and the associated (ii) correlated modes diagonal as well as a (iii) 3D-representation, highlighting the non-flat spectrum.

As a flat spiral bandwidth is preferable for unbiased quantum transport of states, the modal weights were equalised by a mode-specific decrease of the grating depth for the holograms, allowing one to implement Procrustean filtering \cite{vaziri2003concentration, dada2011experimental,bennett1996concentrating} and thus sacrificing signal for the smaller $|\ell|$ values.

\begin{figure}
    \centering
    \includegraphics[width=\linewidth]{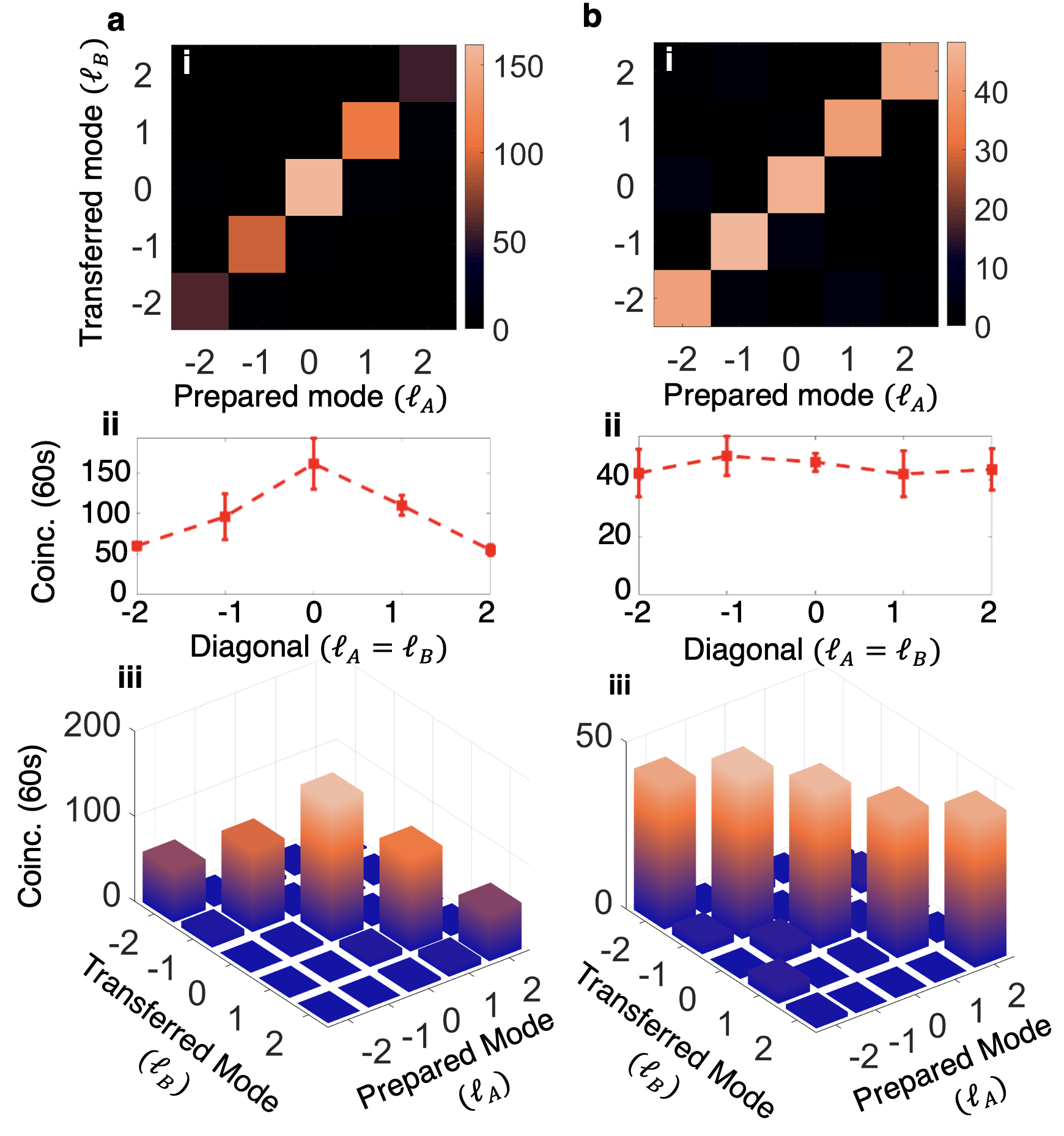}
    \caption{\textbf{Procrustean filtering of the OAM modes.} (a) Unflattened and (b) flattened spiral bandwidths by decreasing the grating depth for lower $\ell$-values. Here (i) gives the density plot, (ii) shows the diagonal of (i) and (iii) renders the data in 3D where the diagonal values are highlighted.}
    \label{Flat}
\end{figure} 

Supplementary Figure \ref{Flat}(b) shows the result of implementing an $\ell$-dependent grating depth compensation. Here it can be seen that the detected weights across the 5 OAM modes were flattened to within the experimental uncertainties, with a small increase in the $\ell = -1$ mode due to laser fluctuation. This, however, does come at the cost of a smaller signal-to-noise ratio as is demonstrated in the density and 3D-plots given in (b)(i) and (b)(iii), respectively (maximum coincidences are less by about a third). Supplementary Figures \ref{Flat} (ii) show the diagonals for clearer comparison of the modal weights and present noise. \revise{Such spiral flattening was used to improve the results given in Figs. \ref{fig:VisTomoDim}(c), \ref{fig:TeleSims} and Suppl. Fig. \ref{MUBFid}.}

\begin{center}
    \textbf{Supplementary Note 7 - Background subtraction}
\end{center}

\revise{Due to the low efficiencies in the up-conversion process, a low signal-to-noise ratio was an experimental factor. The noise in our system is generated by various effects, e.g. the dark counts, originated in the avalanche photodiodes (APDs), also contributing to false (accidental) coincidence events. An additional mode-dependent noise was also observed as a result of two photon absorption occurring for lower $\ell$-values as the 1565 nm pump power density is higher. See Supplementary Note 9 for a more detailed description of the different sources of error in our system. As a result, the visibilities and fidelities of the states are decreased. Here, reducing the temporal window for which the coincidences were detected aided to reduce the noise at the cost of some signal. Another method which was employed was to measure the detected 'coincidences' far away from the actual arrival window of the entangled photons. In other words, the easiest way to statistically quantify this noise is to count the coincidence events when the difference of the time of arrival between photons B and D is much larger than the coincidence window. That measurement was then taken as the background noise of the system and subtracted from the actual measured coincidences. This is illustrated in the histogram shown in Suppl. Fig. \ref{histo} of the measured coincidences vs. time delay for the signals received from both detectors. Here the blue rectangle highlights the coincidences being detected while the red highlights the values taken to be the background or noise signal.}

\begin{figure}[h]
    \centering
    \includegraphics[width=0.7\linewidth]{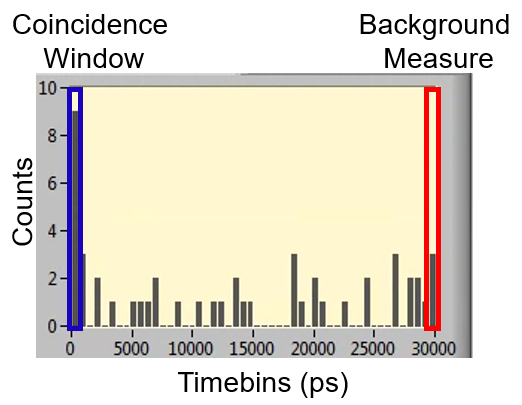}
    \caption{\textbf{Illustration of background measurement.} Histogram showing the arm delays with the coincidence windows \BS{for a 3s integration time}, demonstrating the measured background values for noise correction.}
    \label{histo}
\end{figure} 

\begin{figure*}[t]
    \centering
    \includegraphics[width=1\linewidth]{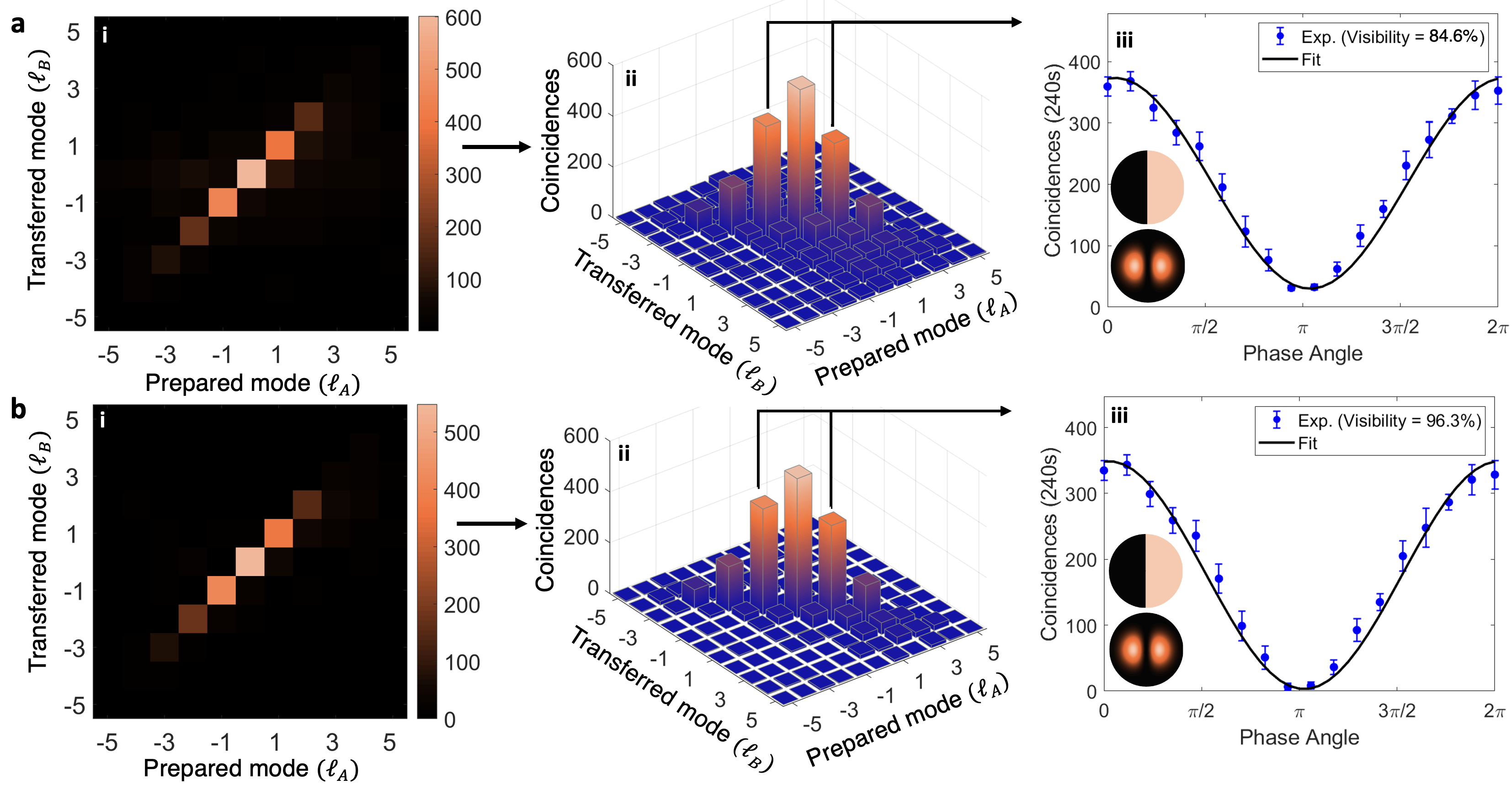}
    \caption{\textbf{Effects of applying noise correction to the results.} Plots showing the (a) raw measured coincidences and (b) coincidences corrected by subtracting the background measured in an uncorrelated time-bin for the (i) spiral bandwidth with a (ii) 3D rendering and (iii) the visibility measurable for rotating the projected state for the $\ell = \pm$ 1 transferred state.}
    \label{background}
\end{figure*} 

\revise{These measured coincidence values were consequently in the same length time bin (coincidence window = 0.5 ns) with the time delay being 30 ns outside of the actual coincidence window (20 times away from the actual coincidence window). By subtracting the noise signal, the actual coincidences from the quantum transport process could be determined. The results of this subtraction is then showcased in Suppl. Fig. \ref{background} for the spiral bandwidth and visibility from a superposition OAM state of $\ell = \pm 1$. Here, Suppl. Fig. \ref{background}(a) shows the raw measured results, while (b) is the only plot that illustrates the effects of subtracting the measured noise from the coincidences as described in Suppl. Fig. \ref{histo}. The spiral bandwidth is shown in Suppl. Fig. \ref{background} (i), ranging from $\ell=-5$ to $\ell = 5$ and with a 5 minute integration time per projection measurement. The 3D rendering of the measurements is shown in (ii), so that the noise can be easily identified. And the $\ell = \pm 1$ superposition state is shown in (iii), where the projection state was rotated by adjusting the inter-modal phase from $\theta = [0,2\pi]$. In all cases a clear improvement in the measured states can be seen with particular attention to the increase in visibility of Suppl. Fig. \ref{background} (b, iii) from (a,iii), reaching almost a perfect fidelity of the transferred state.}

A summary of the difference in visibilities for rotations of the different projected modes shown in Fig. \ref{fig:VisTomoDim}(a) of the main text is further provided in Suppl. Table \ref{tab:Visibilities}, along with the other results given throughout the paper.

\begin{table}[h!]
	\begin{tabular}{|c|c|c|} 
		\hline
		 OAM Superposition	& Raw \revise{Fidelity} & B. Sub. \revise{Fidelity} \\
		\hline
		$\ket{1}+\ket{-1}$ & revise{0.925 $\pm$0.015} & \revise{0.98 $\pm$0.022} \\
		\hline
		$\ket{2}+\ket{-2}$ & \revise{0.915 $\pm$0.05} & \revise{0.98 $\pm$0.06} \\
	    \hline
	 	$\ket{3}+\ket{-3}$ & \revise{0.895 $\pm$0.10} & \revise{0.985 $\pm$0.12} \\
	 	\hline
	 	$\ket{4}+\ket{-4}$ & \revise{0.825 $\pm$0.12} & \revise{0.97 $\pm$0.18} \\
	    \hline
		 3D Tomography	& Raw Fidelity & B. Sub. Fidelity \\
		\hline
		$\ket{-1}+\ket{0}+\ket{1}$ & \BS{0.82 $\pm$0.016} & \BS{0.92 $\pm$0.017} \\
		\hline
		2D OAM Superposition& Raw Similarity & B. Sub. Similarity \\
		\hline
		$\ket{\varphi_1}$ & \BS{0.96 $\pm$0.042} & \BS{0.96 $\pm$0.051} \\
		\hline
		$\ket{\varphi_2}$ & \BS{0.97 $\pm$0.057} & \BS{0.97 $\pm$0.069} \\
		\hline
		$\ket{\varphi_3}$ & \BS{0.98 $\pm$0.093} & \BS{0.98 $\pm$0.10} \\
		\hline
		3D OAM Superposition& Raw Similarity & B. Sub. Similarity \\
		\hline
		$\ket{\varphi_4}$ & \BS{0.98 $\pm$0.039} & \BS{0.96 $\pm$0.06} \\
		\hline
		 4D OAM Superposition& Raw Similarity & B. Sub. Similarity \\
		\hline
		$\ket{\varphi_5}$ & \BS{0.98 $\pm$0.047} & \BS{0.97 $\pm$0.065} \\
		\hline
		 3D HG Superposition& Raw Similarity & B. Sub. Similarity \\
		\hline
		$\ket{\gamma_1}$ & \BS{0.99 $\pm$0.029} & \BS{0.99 $\pm$0.042} \\
		\hline
		 4D HG Superposition& Raw Similarity & B. Sub. Similarity \\
		\hline
		$\ket{\gamma_2}$ & \BS{0.96 $\pm$0.025} &\BS{0.95 $\pm$0.037} \\
		\hline
		 9D HG Superposition& Raw Similarity & B. Sub. Similarity \\
		\hline
		$\ket{\gamma_3}$ & \BS{0.81$\pm$0.019} & \BS{0.80 $\pm$0.025} \\
		\hline
	\end{tabular}
\caption{\textbf{Results summary of background subtracted and raw data.} Experimental visibilities, fidelities and similarities calculated for the quantum transport channel and transferred states comparing raw and background subtracted (B. Sub.) outcomes. Abbreviated states are: $\ket{\varphi_1}=\ket{0}+\ket{-1}$, $\ket{\varphi_2}=\ket{-1}+\ket{1}$, $\ket{\varphi_3}=\ket{0}-\ket{1}$, $\ket{\varphi_4}=\ket{-2}+\ket{0}+\ket{2}$, $\ket{\gamma_1}=\ket{HG_{1,0}}+\ket{HG_{1,1}}+\ket{HG_{0,1}}$, $\ket{\varphi_5}=\ket{-3}-i\ket{-1}+\ket{1}+i\ket{3}$, $\ket{\gamma_2}=\ket{HG_{0,0}}+\ket{HG_{1,0}}+\ket{HG_{1,1}}+\ket{HG_{0,1}}$ and $\ket{\gamma_3}=\ket{HG_{0,0}}+\ket{HG_{2,0}}+\ket{HG_{0,2}}+\ket{HG_{2,2}}+\ket{HG_{4,0}}+\ket{HG_{0,4}}+\ket{HG_{4,2}}+\ket{HG_{2,4}}+\ket{HG_{4,4}}$, as given in the main text. \BS{Slightly better similarities may be noted in some cases for the raw values of the superposition on states as the Procrustean filtering applied was optimised for the raw data.}}
% $\ket{\varphi_1}=\frac{1}{\sqrt{2}}[\ket{0}+\ket{-1}]$, $\ket{\varphi_2}=\frac{1}{\sqrt{2}}[\ket{-1}+\ket{1}]$, $\ket{\varphi_3}=\frac{1}{\sqrt{2}}[\ket{0}-\ket{1}]$, $\ket{\varphi_4}=\frac{1}{\sqrt{3}}[\ket{-2}+\ket{0}+\ket{2}]$, $\ket{\gamma_1}=\frac{1}{\sqrt{3}}[\ket{HG_{1,0}}+\ket{HG_{1,1}}+\ket{HG_{0,1}}]$, $\ket{\varphi_5}=\frac{1}{\sqrt{4}}[\ket{-3}-i\ket{-1}+\ket{1}+i\ket{3}]$, $\ket{\gamma_2}=\frac{1}{\sqrt{4}}[\ket{HG_{0,0}}+\ket{HG_{1,0}}+\ket{HG_{1,1}}+\ket{HG_{0,1}}]$ and $\ket{\gamma_3}=\frac{1}{\sqrt{9}}[\ket{HG_{0,0}}+\ket{HG_{2,0}}+\ket{HG_{0,2}}+\ket{HG_{2,2}}+\ket{HG_{4,0}}+\ket{HG_{0,4}}+\ket{HG_{4,2}}+\ket{HG_{2,4}}+\ket{HG_{4,4}}]$, as given in the main text. \BS{Slightly better similarities may be noted in some cases for the raw values of the superposition on states as the Procrustean filtering applied was optimised for the raw data.}}
	\label{tab:Visibilities}
\end{table}

% \begin{table}
% \begin{tabular}{|c|c|c|c|}\hline
% \multicolumn{2}{|c|}{\multirow{$\ell = \pm 1$ Superposition }}
%      &Raw Visibility&0.85\\ \cline{3-4}
% \multicolumn{2}{|c|}{}
%      &Corrected Visibility&0.96\\ \hline
% \multicolumn{2}{|c|}{\multirow{$\ell = \pm 2$ Superposition }}
%      &Raw Visibility&0.83\\ \cline{3-4}
% \multicolumn{2}{|c|}{}
%      &Corrected Visibility&0.96\\ \hline
% \multicolumn{2}{|c|}{\multirow{$\ell = \pm 3$ Superposition }}
%      &Raw Visibility&0.79\\ \cline{3-4}
% \multicolumn{2}{|c|}{}
%      &Corrected Visibility&0.97\\ \hline
% \multicolumn{2}{|c|}{\multirow{3D Tomography }}
%      &Raw Fidelity&0.66\\ \cline{3-4}
% \multicolumn{2}{|c|}{}
%      &Corrected Fidelity&0.75\\ \hline
% \multicolumn{2}{|c|}{\multirow{4D OAM Superposition }}
%      &Raw Similarity&0.73\\ \cline{3-4}
% \multicolumn{2}{|c|}{}
%      &Corrected Similarity&0.94\\ \hline
% \multicolumn{2}{|c|}{\multirow{3D HG Superposition }}
%      &Raw Similarity&0.4\\ \cline{3-4}
% \multicolumn{2}{|c|}{}
%      &Corrected Similarity&0.66\\ \hline
% \multicolumn{2}{|c|}{\multirow{3D HG Superposition }}
%      &Raw Similarity&0.71\\ \cline{3-4}
% \multicolumn{2}{|c|}{}
%      &Corrected Similarity&0.83\\ \hline
% \end{tabular}
% \caption{\textbf{Results summary of backround subtracted and raw data.} Experimental visibilities, fidelities and similarities calculated for the teleportation channel and teleported states comparing raw and background subtracted (B. Sub.) outcomes.}
% \label{tab:DimPara}
% \end{table}

\begin{center}
    \textbf{Supplementary Note 8 - Process efficiencies}
\end{center}

Since the quantum transport protocol presented here, is based on single photon pairs, the efficiency of the nonlinear processes is required to be well characterised and controlled. Under such conditions, the complete state produced by SPDC can be represented by
\begin{equation}
\ket{\psi_{\text{SPDC}}} \approx \ket{\text{vac}} + \ket{\psi_{BC}} \sigma + O\{\sigma^2\} ,
\end{equation}
where $\sigma\ll 1$ is the nonlinear coefficient which is determined by
\begin{equation}
\sigma = \chi^{(2)} \sqrt{\frac{\hbar \omega_p \omega_B \omega_C}{8 \epsilon_0 c^3 n_p n_B n_C} \frac{F_0}{A_p}} .
\label{eqn:nonlinCoeff}
\end{equation}
Here, $\chi^{(2)}$ is the second-order nonlinear susceptibility coefficient of the nonlinear material for a given phase-matching condition, the subscript $p$ refers to the pump, $F_0/A_p$ is the number of pump photons per second per area or flux rate (photons/s/$m^2$), $n$ refers to the respective refractive indices, $\omega$ refers to the respective central angular frequencies, $c$ is the speed of light, and $\epsilon_0$ is the vacuum permittivity.

\av{In order to predict the behaviour of the nonlinear crystals in our experiment, accurate knowledge of the properties of the material is required. The knowledge of the specific wavelengths generated by quasi-phase matched crystals relies on the ability to determine the respective refractive indices for the desired input and output wavelengths involved in the parametric processes. As the refractive index varies with the wavelength of the light incident on the material, the values can be calculated from Sellmeier equations when the coefficients have been experimentally determined. For a KTP crystal, it has been reported \cite{fan1987second,fradkin1999tunable} that we can accurately determine this by using the two-pole Sellmeier equation}
%%%%%%%%%%%%%%%%%%%%%%%%%%%%%%%%%%%%%%%%%%%%%%%%%%%%%%%%%%%%%
\begin{equation}
n(\lambda)^2 = A + \frac{B}{1-\frac{C}{\lambda^2}} + \frac{D}{1-\frac{E}{\lambda^2}} - F\lambda^2.
\label{SellmeierEq}
\end{equation}
%%%%%%%%%%%%%%%%%%%%%%%%%%%%%%%%%%%%%%%%%%%%%%%%%%%%%%%%%%%%%
\av{Here, $\lambda$ is the wavelength, $n(\lambda)$ is the refractive index and $A-F$ are the experimentally determined coefficients which depend on the centred wavelength, e.g. $\lambda \le 1~\mu$m \cite{fan1987second} or above it \cite{fradkin1999tunable}.}

\av{We can thus use the calculated refractive indices to determine the efficiency ($\eta$) of the SFG for a single photon input ($\lambda_C = 806$ nm) to an output ($\lambda_D = 532$ nm) for a high pump power ($\lambda_A = 1565$ nm) from the relation \cite{albota2004efficient}}

\begin{equation}
\eta_{SFG} = \sin^2{\left(\frac{\pi}{2} \sqrt{\frac{P_A}{P_{max}}}\right)},
\label{Efficiency}
\end{equation}
\av{where $P_A$ is the input pump power centred at $\lambda_A$ and}
\begin{equation}
P_{max} = \frac{c \epsilon_0 n_C n_A \lambda_C \lambda_A \lambda_D}{128 (d_{eff})^2 L h_m},
\end{equation}
\av{is the pump power (of $\lambda_A$) required to achieve 100\% up-conversion of the input single photons $\lambda_C$. Here, $n_i$ are the respective refractive indices in the nonlinear crystal, $d_{eff}$ is the effective nonlinear coefficient, L is the length of the crystal and $h_m$ is a reduction factor for focused Gaussian beams \cite{boyd1968parametric}, which depends mainly on a focusing parameter $L/b$, determined by the confocal parameter $b = 2z_R$ (two times the Raleigh range). In our case, we expect this variable to be small ($h_m \approx 0.06$), due to the poor ratio between the crystal length and confocal parameter, considering also the mode mismatch between the SFG pump beam waist ($w_A \sim 100$ $\mu$m for the $\ell = 0$ case) and the input photon C (with a similar beam waist as the Gaussian beam pumping the SPDC process: $w_p \sim 300$ $\mu$m). For a type-0 periodically poled KTP crystal, $d_{eff} = \frac{2}{\pi}d_{33} \approx 10$ pm/V \cite{raicol} (a factor 2/$\pi$ is required when considering quasi phase-matching). Hence, we can up-convert the photon C into photon D with an efficiency of $\eta_{SFG} = 0.3\%$, considering that we pump with $P_A$ = 3.5 W of optical power, the two input modes are Gaussian modes and we do not consider the losses in the system. It follows that this relation should allow us to ascertain how the efficiency of the system (and thus detected counts) should scale with a change in the length of the crystal, nonlinear efficiencies or higher modal mismatch (higher OAM modes).}

\begin{center}
    \textbf{Supplementary Note 9 - Constraints and sources of error}
\end{center}

\BS{A feature of this demonstration involved optimisation of the experimental parameters to allow the access to higher dimensions, while maintaining enough signal for detection and minimising noise contributions. Accordingly, the experimental constraints in the system can be categorised into sources of noise, sources of experimental error and limitations imposed by the experimental parameters.}

\BS{\textit{Limitations imposed by experimental parameters.} As eluded to with the numerical simulation of parameters in Supplementary Note 5, an interplay between the detection and pump waists changes the dimensionality accessible in our system. A byproduct of altering these sizes for higher dimensionality is the reduction in the efficiency at which the lower-order modes are detected. This factor is illustrated in Suppl. Fig. \ref{modalEfficiency}. Here, the relative efficiency of detection for the lowest order mode ($\ell = 0$) is shown as a factor of the parameter space used to optimise the dimensionality. It follows that the increase in dimensionality as indicated by the parameter points (c)-(e) demonstrated in Fig. \ref{fig:ConceptFig} of the main text, that a notable decrease in the simulated efficiency is seen which is further reflected experimentally in the spiral bandwidths with the coincidences dropping significantly as the dimensionality increases.}
\begin{figure}[h!]
    \centering
    \includegraphics[width=\linewidth]{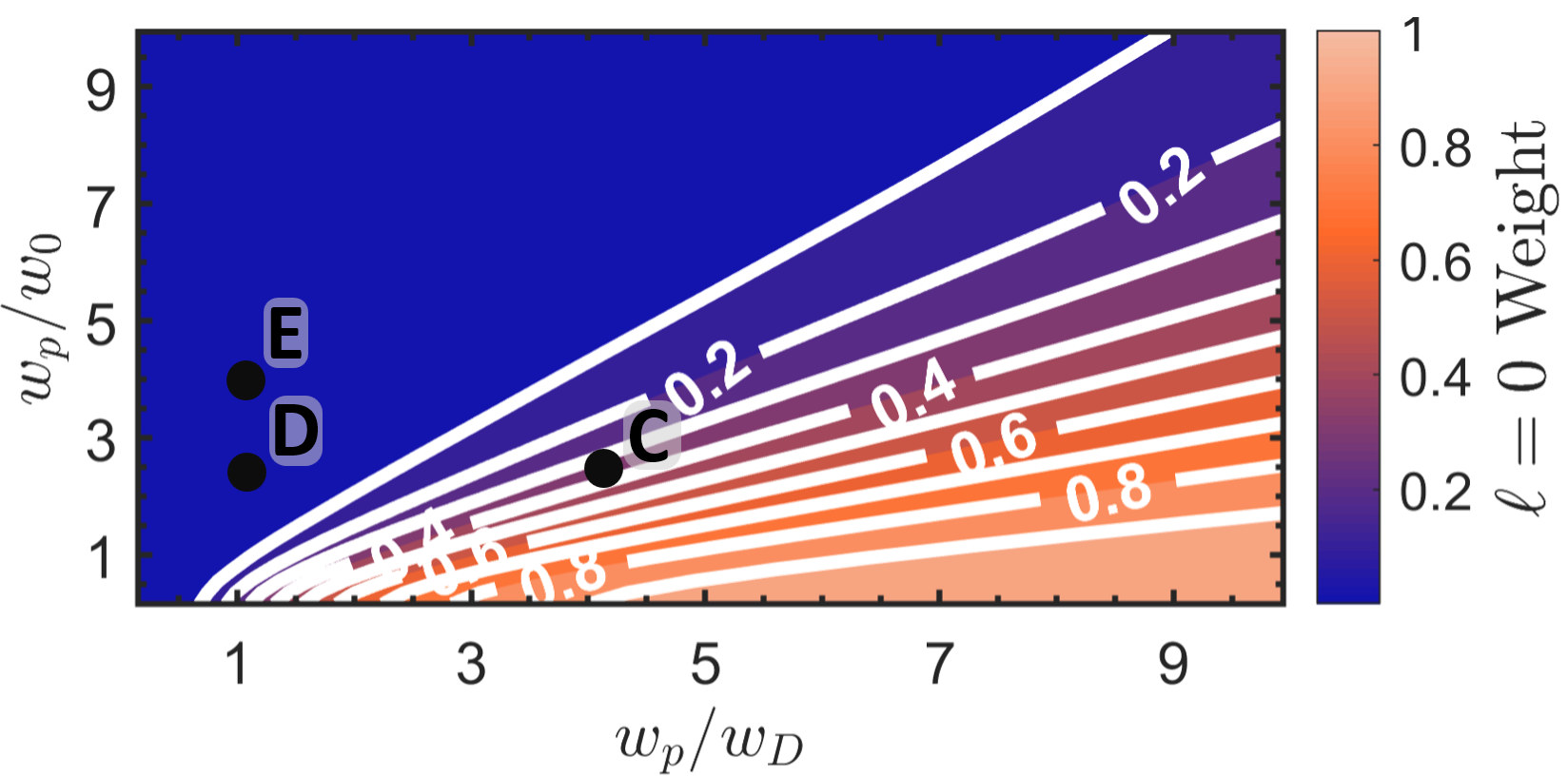}
    \caption{\textbf{Modal detection efficiencies with experimental parameters.} Numerical simulation of the change in detection efficiency for $\ell = 0$ as the experimental parameters are optimised for higher dimensionality. Points C-E indicated correspond to experimental parameters tested in Fig. \ref{fig:ConceptFig} of the main text.}
    \label{modalEfficiency}
\end{figure} 

\BS{This inverse relation between accessing larger dimensions at the expense of lower order mode detection efficiency can be understood as the result of mismatching the lower order spatial mode sizes in favour of the higher-order spatial mode sizes. This occurring both in the up-conversion crystal between the photon C and structured pump photon-state A, as well as the relative detection sizes of the single mode fibres in either arm (detection of photons B and D). Such interplay between accessible dimensionality and detection sizes has also been noted and studied \cite{miatto2012bounds,roux2014projective,nape2020enhancing} when considering similar detection of the direct SPDC modes generated in quantum entanglement sources such as ours and, as such, readily extends to our system.}

\BS{\textit{Sources of noise.} A notable source of noise in the production of entangled photon pairs by pumping a crystal is the generation of additional pairs within the same coincidence window  \cite{takesue2010effects,schneeloch2019introduction,takeoka2015full}. This results in impurity in the detected coincidences as they form a statistical mixture rather than a pure source in which to utilise \cite{graffitti2018independent}. As a result, the event of generating multiple bi-photons serves to reduce the fidelity of the entanglement resource and thus the transferred states. Several works have been presented in an active effort to solve this \cite{pittman2002single, migdall2002tailoring, kaneda2019high}, however this involves generally complex configurations. The straightforward approach to mitigating the additional bi-photon generation events is reduction in the intensity at which the crystal is pumped. While this reduces the efficiency at which the desired single bi-photons are produced, a much larger reduction in the multiple bi-photon probabilities serves to increase the fidelity. It follows that the experiment was carried out at the lowest possible SPDC pump powers ($\sim$1.2 W) in order to mitigate this while maintaining enough signal for detection in arm D, given the maximum SFG pump power allowed by the damage threshold in SLM$_A$.} 
% \av{[Maybe add an explanation of the bad Similarity case, and we needed to increase the SPDC pump to mitigate the noise effect in that case. We'll see about this when we perform the test.]}

\BS{Here, the detected background counts in the up-conversion arm becomes notable so as to maintain higher purities so that higher fidelity quantum transport may be achieved. For instance, the signal-to-noise ratio for the $\ell = 0$ transferred mode in the high-dimensional optimised setup was $\sim$ 500 counts per second (cps) signal: $\sim$ 360 cps noise. The sources of background counts were due to the dark counts from the detector itself as well as unavoidable stray pump light propagating towards the detector. As a result, this signal-to-noise ratio has a notable impact on the detected results due to increased accidentals \cite{ecker2019overcoming,zhu2021high}.}

% \BS{ADAM CHAN YOU CHECK THIS? Additionally, a spread in the spectral correlations of the single bi-photons arise due to a non-unity spectral bandwidth for the pump laser along with properties of the generation crystal. As a result, detection of these different spectral correlations results in a degradation of the purity of the states. In order to mitigate this, a narrow spectral band-pass filter (centred at 532nm with a $\pm3 nm$ FWHM) was placed before the up-conversion detector in order to restrict the spectral width of the photons seen by the detector. The detected coincidences then have higher degrees of correlation and thus purity.}

\BS{A large mismatch in counts between the two detectors is also a direct result of the low up-conversion efficiency currently associated with non-linear processes. As such, a large number of SPDC photons is detected in APD$_B$ whereas a much lower number of SPDC photons are up-converted and consequently detected in APD$_D$. This mismatch means the probability of detecting accidental coincidence counts is higher than if the signals were similar (as in a linear scheme). Here accidentals refer to the event of erroneously detecting a coincidence due to two uncorrelated photons arriving at both detectors at the same time. The number of accidentals ($C_{Acc}$) expected may be calculated using $C_{Acc} = S_D S_BW$ where $S_D$ ($S_B$) are the counts detected in APD$_D$ (APD$_B$) and $W$ is the time window in which coincidences are collected. The number of accidentals thus increases directly with the mismatch in counts as the number of possible coincidences is limited by the lowest signal detected. Subsequently, this varies with the detection efficiency (discussed above) and has an inverse relationship with the dimensionality of the system. For instance, a mismatch of $\sim$ 500 cps in APD$_D$ compared to $\sim$ 650 000 cps in APD$_B$ yields a 1300 times increase in the predicted number of accidentals when our system is optimised for high dimensions as opposed to unitary up-conversion efficiency. This was mitigated by narrowing the coincidence detection window ($W$).}  

\BS{Another factor for consideration is the use of a strong laser pump in the up-conversion process. It has been well documented that additional processes occur with the use of a strong pump due to the high number of input photons \cite{xie2019efficient,pelc2011long,yao2020optimizing}. While choice of a long-wavelength pump relative to the signal wavelength helps suppress the spontaneous Raman scattering contributing to this \cite{kamada2008efficient,pelc2011long,shentu2013ultralow}, factors still remain for consideration. Here the third harmonic of 1565 nm lies around 522 nm and the additional up-conversion of SPDC flourescence as well as the secondary lobes in the SPDC sinc relation with the bandwidth generated would all result in the detection of noise photons that decrease the fidelity of the transferred state. Here we employ the use of a narrow-band band-pass filter centred at 532 nm with an acceptance range of $\pm$3 nm at full width at half maximum (FWHM) before the up-conversion detector, in order to mitigate this effect.} 
% \av{The third harmonic of 1565 nm would be around 522 nm. This could cause a lot of noise, yes. Also, the up-conversion of the SPDC fluorescence or secondary lobes from the sinc squared function. Will write also here something brief explaining the decrease in fidelity when not having a narrow line-width SPDC pump laser (ask for a reference also).}

\begin{figure}[h]
    \centering
    \includegraphics[width=\linewidth]{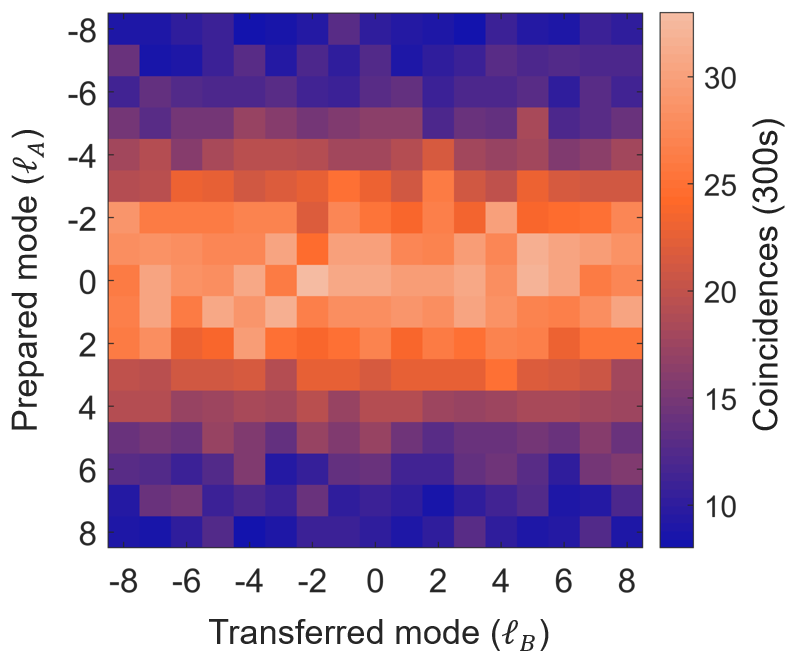}
    \caption{\textbf{Mode dependent detected noise.} Spiral bandwidth plot spanning $\ell = [-8,8]$ showing the measured background noise indicating higher noise for lower order modes.}
    \label{bkrndSBW}
\end{figure} 
\BS{Consequently, it may be noted that due to the presence of mode-dependent detection efficiency, a mode-dependence in the noise detected is present for the transferred states. This is shown in Suppl. Fig. \ref{bkrndSBW} where background noise detected outside of the coincidence window (as per Supplementary Note \ref{background}) is shown for a typical spiral bandwidth. The transferred modes ranged from $\ell = -8$ to $8$ and were scanned for in the same range. A larger noise contribution is shown here for the lower order transferred modes which then falls off as the modal order increases.}
%The optimised parameters of 1 $\mu s$ deadtime, 20 \% efficeincy and 1.2 W SPDC pump power was used here. ADD BACKGROUND SBW

\BS{Furthermore, due to the properties of InGaAs-based SPAD detectors used in the detection of 1565 nm light, a lower efficiency, longer deadtime and more afterpulsing compared to Si-based SPADs for visible photon detection contributes to the noise seen. Here, deadtime refers to the amount of time where no photons can be detected after a previous detection event. The lower limit enabled by the detector is 1 $\mu s$ compared to 22 ns for the visible detector. Afterpulsing refers to additional artificial detections when measuring counts and is an intrinsic property of the device due to trapped electron-hole pairs which causes new avalanches after an actual detection event \cite{lunghi2012free, yan2012ultra}. For our detector, an estimated afterpulsing probability of 5.2 \% at 1 $\mu s$ deadtime and 20\% efficiency results in the detection of additional erroneous signal which relies on the amount of signal being seen. For intance 650 000 counts results in close to 40 000 incorrect counts. Increasing the deadtime of the detector decreases the afterpulsing probability and thus the noise, but then results in a lower rate of detection which can be seen as a decrease in the time-averaged overall detection efficiency with respect to the visible detector. Futhermore, inherent dark counts of the detector occurs due to electrons being set free from vibrational conditions induced by heat and thus generates an undesired avalanche (detection event), despite a temperature of -50 $^{\circ} \text{C}$. This dark count induced noise is independent of the detection rate and sets a lower signal floor of approx. 2000 counts. The tradeoff between deadtime, number of detected coincidences with the InGaAs detector (IDQ220 free-running) and the effect of narrowing the coincidences detection window was analysed using the visibility curves, shown in Suppl. Fig. \ref{VisPara} and Suppl. Table \ref{tab:VisPara}, for the transferred $\ell = \pm 1$ state.}
\begin{figure}[h!]
    \centering
    \includegraphics[width=\linewidth]{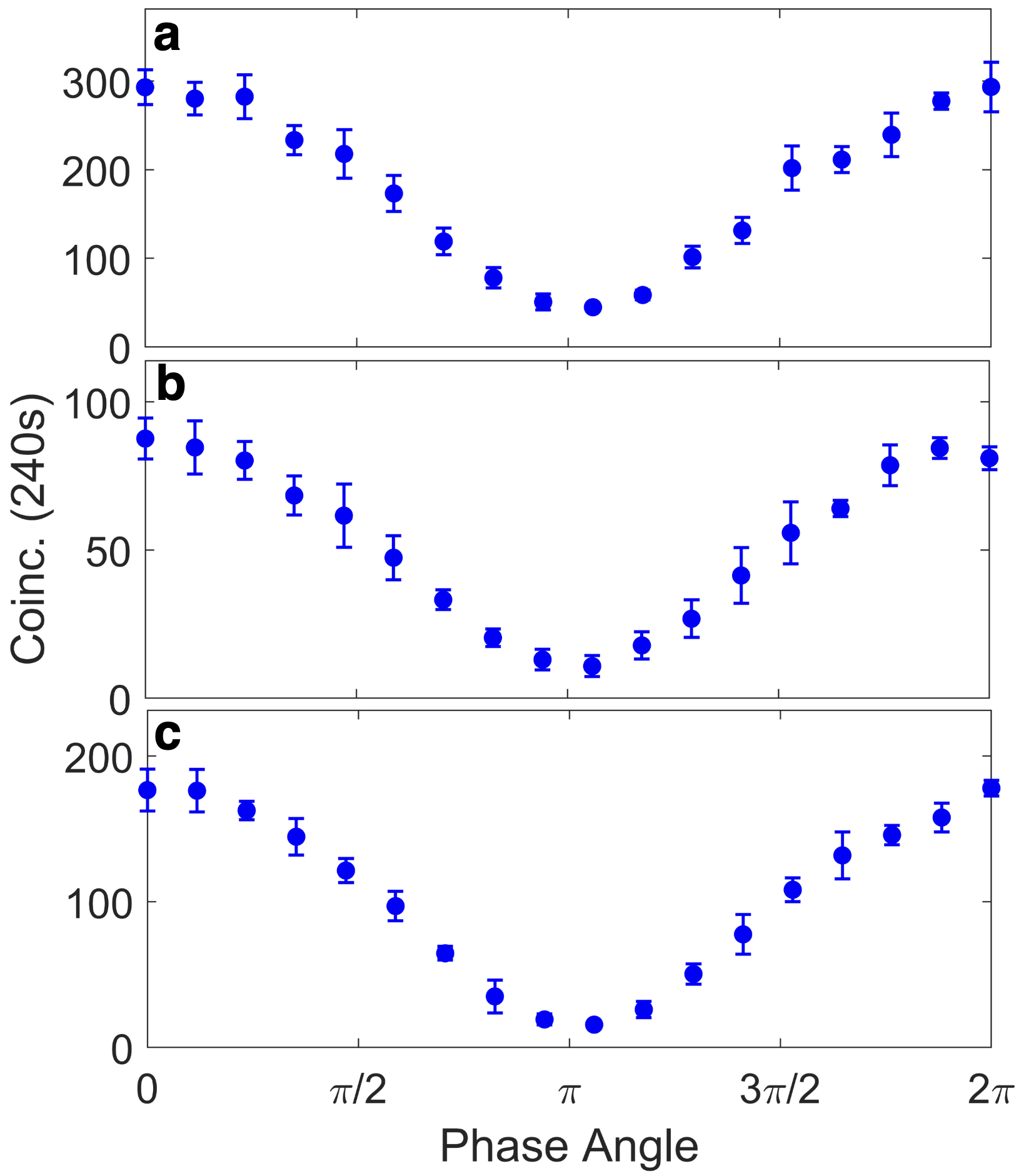}
    \caption{\textbf{Visibilities with detection parameters.} Experimental visibility curves obtained by rotating $\theta = [0,2\pi]$ in the detection mode ($\ket{-1}+ e^{i\theta} \ket{1}$) for transferred state $\ket{\psi} = \ket{-1}+\ket{1}$ for (a) 1 $\mu s$ deadtime and 1.5 ns coincidence window, (b) 5 $\mu s$ deadtime and 1.5 ns coincidence window as well as (c) 1 $\mu s$ deadtime and 0.5 ns coincidence window.}
    \label{VisPara}
\end{figure} 

\begin{table}[h!]
	\begin{tabular}{|c|c|c|c|c|} 
		\hline
		 Fig \ref{VisPara} & Coinc. Window & Deadtime &  Visibility & Max. Coinc. \\
		\hline
		(a) & 1.5 ns & 1 $\mu$ s & 0.74 $\pm$0.10 & 293 $\pm$27.9 \\
		\hline
		(b) & 1.5 ns & 5 $\mu$ s & 0.78 $\pm$0.10 & 87.6 $\pm$6.9 \\
	    \hline
	 	(c) & 0.5 ns & 1 $\mu$ s & 0.84 $\pm$0.10 & 178 $\pm$5.4 \\
	 	\hline
	\end{tabular}
\caption{\textbf{Visibilities for different parameters.} Comparison of the visibility and maximum detected coincidences with different detector deadtimes and coincidence windows for results obtained by rotating $\theta = [0,2\pi]$ in the detection mode $(\ket{-1}+ e^{i\theta} \ket{1})$ for transferred state $\ket{\psi} =\ket{-1}+\ket{1}$.}
	\label{tab:VisPara}
\end{table}

\BS{An increase in the deadtime from $1 \mu s$ to  $5 \mu s$, shown in the comparison between Suppl. Fig. \ref{VisPara} (a) and (b), increased the visibility by $4\%$ as a lower noise contribution was occurring from the InGaAs detector. This, however, also resulted in only a third of the coincidences being retained in the adjustment as a result of a reduction in the efficiency rate. Conversely, when reducing the coincidence detection window from 1.5 ns to 0.5 ns, a much larger increase of 10\% in the visibility was seen with more of the signal being retained (2/3 of the signal in (a)). Such an increase in the visibility can be readily explained as the 'lost' coincidences were the result of reducing the acceptance of additional pairs, spectral spread correlations in time and the probability of measuring accidentals. As such, the photons reducing the fidelity of the transferred state were excluded as opposed to simply reducing the efficiency in order to reduce the number of erroneous detection event due to properties of the detector. Accordingly, the increased signal offset the small loss in visibility for the deadtime, making the 1 $\mu$s the optimal parameter, while the reduction in signal for increased visibility with the lower coincidence window resulted in the 0.5 ns being the optimal measurement setting.} 

\BS{\textit{Sources of experimental error.} Aberrations due to imaging the beam tightly into the crystal, propagating the beam through several imaging systems and crystal inhomogeneity serve to induce errors in the purity of the modes being transferred. Here, higher order modes are also subject to aperture effects in the optical system and due faster expansion upon propagation, 'see' a greater area of the optical components. As such, they encounter more aberrations as propagated throughout the system. Temperature fluctuations due to external temperature variations also cause variations in the alignment, while an air-conditioner is used to try mitigate the effects. The experiment spans a 2 $\times$ 1 m optical table and as such air fluctuations from the conditioner cause beam wander and thus increases fluctuations in the measured coincidences. Isolation of the experiment with a curtain was used to mitigate this along with longer integration times combined with averaging over several measurements.}

\begin{figure}[h!]
    \centering
    \includegraphics[width=\linewidth]{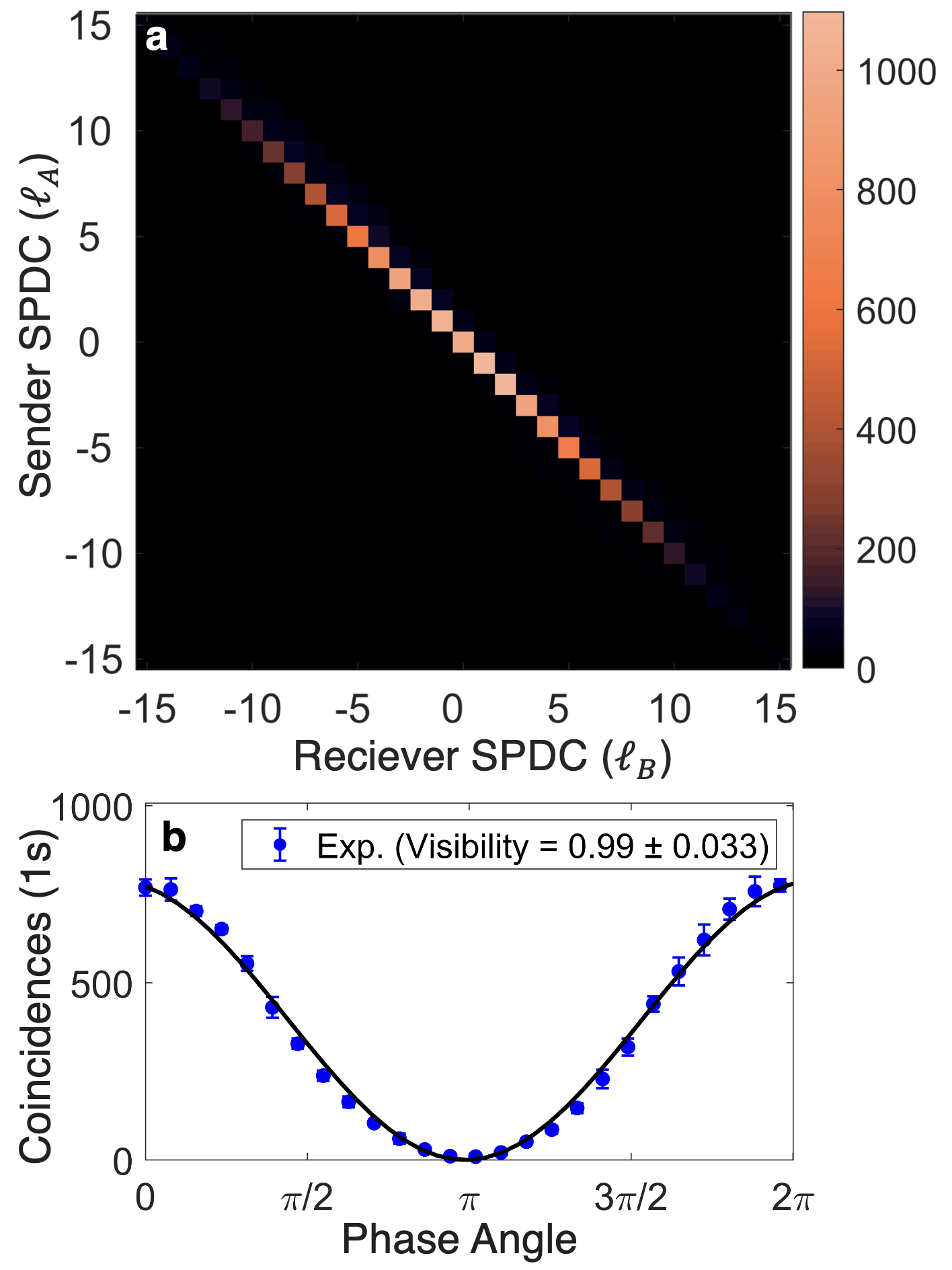}
    \caption{\textbf{Channel SPDC characterisation.} Experimental (a) spiral bandwidth of the channel SPDC and (b) visibility curve obtained by rotating phase angle, $\theta = [0,2\pi]$, in the detection mode ($\ket{-1}+ e^{i\theta} \ket{1}$) in photon B for the state $\ket{\psi} = \ket{-1}+\ket{1}$ in the arm used for quantum transport. No noise subtraction or error correction was performed on the data.}
    \label{SPDCana}
\end{figure} 

\BS{\textit{Quality of the entanglement channel.} The channel SPDC was also briefly measured and analysed for the optimised channel which yielded a quantum transport of capacity of $K$ $\approx$ 15. This was done in terms of an initial spiral bandwidth only considering the SPDC photons (without the SFG process) and then a comparative visibility curve for the Bell state $\ket{\psi} = \ket{-1}+\ket{1}$ projected in the sender arm. Supplementary Figure \ref{SPDCana} shows the experimental results with the bandwidth in (a) giving a Schmidt number of $K_{SPDC} = 17.9$ and visibility curve in (b) giving a visibility of $0.99 \pm 0.033$. It follows that we find the SPDC channel capacity larger than the transferred capacity ($K_{SPDC} \approx 18$ compared to $K \approx 15$), but within a similar range. This may be attributed to the inefficiency of the quantum transport process where lower weightings for the larger order modes caused these to fall below the efficiency required for up-conversion. The Gaussian fall-off of the weightings for the higher-order modes seen here are also reflected in the bandwidth taken for the quantum transport channel. Furthermore, the SPDC visibility for the $\ket{-1}+\ket{1}$ state shows a very high visibility of $0.99 \pm 0.033$, indicating a high fidelity. In comparison to the curves measured  in Suppl. Fig. \ref{background}, we find the visibility comparable to that of the background subtracted value ($0.96 \pm0.044$), showing the measured noise in the system (as described previously) a significant contributor to the loss in fidelity of the transferred states.}

\BS{As a large contribution of the noise factors are due to the use of strong pumps and mismatch in detected counts, it follows that the values shown for the system form a lower bound in the potential performance. Here, improvement in the up-conversion efficiency would serve to decrease the mismatch in counts, increase the signal which results in a lowering of the input coherent state power as well as the SPDC pump power and decrease the barrier for up-conversion of more higher order modes.}

% \av{[Add the spiral bandwidth coming directly from the SPDC and compare it with the teleported Bw. Mostly relevant for the reviewer 1 question about the maximally entangled state we begin with. The qutrit teleportation of Zelinger and JWP also have this discussion in the SI.]}

\begin{center}
    \textbf{Supplementary Note 10 - Stimulating the quantum transport process}
\end{center}

% A bright coherent state produced by a laser was used in order to enhance the up-conversion efficiency of the nonlinear crystal, but with the outcomes conditioned on bi-photon coincidences. \BS{Such an approach was necessary...} 

% \BS{discuss why we cannot use other means to improve the efficiency: *length and *cavity}
\BS{A bright coherent state produced by a laser was used in order to enhance the up-conversion efficiency of the nonlinear crystal, but with the outcomes conditioned on bi-photon coincidences. An intrinsic characteristic of our state transfer process with a bright coherent state is that Alice does not need to know the quantum state to be transferred, a feature that differentiates quantum teleportation from remote state preparation \cite{PhysRevLett.87.077902}. In this sense, the input coherent state and the output single photon state can be considered as {\it carriers} of spatial information, that is the resource being transferred. That is to say, the approach we present is a technical solution to a technological limitation.}
% , and despite encoding the quantum state in many copies, our scheme requires the same ingredients of the quantum teleportation protocols to transfer such spatial information.}

%%% figure
\begin{figure}
\centering
\includegraphics[width=\linewidth]{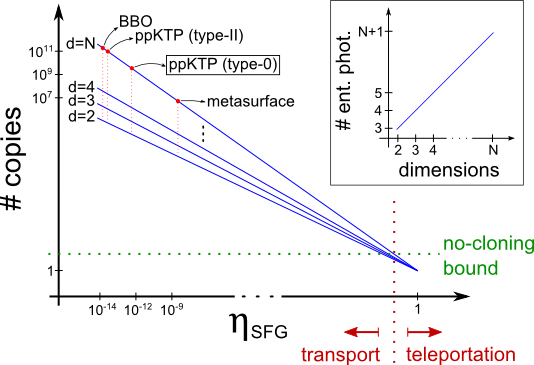}
\caption{\av{\textbf{Nonlinear efficiency road map towards quantum teleportation}. Conceptual plot showing the convergence of the number of copies carrying the $d$-dimensional teleportee state as the SFG is increased. Red dots help to identify the different examples of commercial nonlinear crystals (ppKTP type-0 being our case), and the \emph{all-dielectric} metasurface of Ref. \cite{kivshar2018all} with a notably increased nonlinear coefficient. The plot in the inset shows an example of the amount of extra ancillary photons required to teleport the same $d$-dimensional teleportee state using linear optics.}}
\label{fig:Coherent}
\end{figure}
%%%%%

\av{We designed our experimental setup so we could use the commercial crystal with the highest nonlinear coefficient, considering also a big enough aperture to enlarge the quantum transport channel capacity, i.e., PPKTP crystal for a type-0 three wave mixing processes. Furthermore, lasers working in the CW regime facilitates the concentration of all quantum information in the desired spectral range, while distributing the coincidence events along the whole temporal span being able to reduce the multi-photon accidental events (noise). Despite these advantages, we still need to encode the spatial state to be transferred in $\sim 10^{10}$ photons (3.5 W at 1565 nm) to achieve an up-conversion efficiency of 0.3 \% for the optimum $\ell = 0$ case (as described in Supplementary Notes 8 and 9). However, we expect that this experimental challenge will stimulate further improvements in the field of nonlinear optics rather than placing an upper limit on efficiency in similar future schemes.}

\av{Supplementary Figure \ref{fig:Coherent} presents an intuitive road map towards the perfect quantum teleportation using nonlinear detection systems, taking into account the inevitable improvement of the up-conversion efficiencies in the short future. Recent advances in metasurfaces and metamaterials with nonlinear response \cite{kivshar2018all}, for example, could see physical crystals replaced with these \emph{all-dielectric} meta-optical solutions for even greater efficiency gains (more than 3 orders of magnitude higher than commercial nonlinear crystals). Here we refer to the number of copies as the number of photons per coincidence window, carrying the information of the state to be transferred which is required to obtain a teleportation fidelity above the classical limit. In the case of raw up-conversion efficiencies, without any losses in the transmission and detection sections, the no-cloning bound (green dashed horizontal line) will depend on the system's overall noise. This will dictate what will be the nonlinear efficiency for which we can ensure that the sender cannot keep a better copy than the transferred one and define the conceptual separation between quantum transport and quantum teleportation (red dashed vertical line). It is important to note that the number of copies needed to successfully transfer any arbitrarily high-dimensional quantum state, converges to 1 when the nonlinear efficiency is improved in our scheme. This is not the case when utilising linear optics detection systems, as conceptually plotted in the inset of Suppl. Fig. \ref{fig:Coherent}, where the number of ancillary photons needed grows linearly with the number of dimensions to be teleported.}

\begin{figure}[h!]
    \centering
    \includegraphics[width=0.95\linewidth]{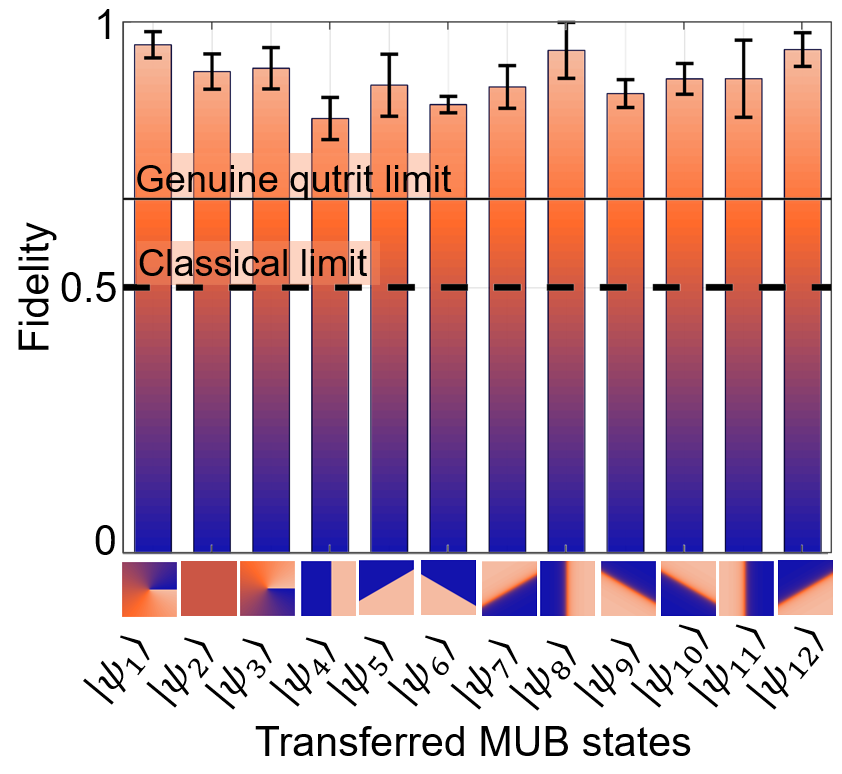}
    \caption{\textbf{Transferred MUB states.} Experimental fidelities measured for state tomography performed on all 12 transferred MUB states for a $d$ = 3 space comprised OAM modes $\ell$ = \{-1,0,1\}. Dashed (solid) lines indicate the classical (genuine qutrit) limit and phase insets along the x-axis show the MUB state phase profiles.}
    \label{MUBFid}
\end{figure} 

\begin{center}
    \textbf{Supplementary Note 11 - Qutrit quantum transport}
\end{center}
\label{QutritMUBS}

\BS{A state tomography on each of the 12 MUB states for a three-dimensional (qutrit) state was performed. The resulting fidelities for each can be seen in Suppl. Fig. \ref{MUBFid}. Each of the MUB states are}
\begin{align}
    \ket{\psi_1} &= \ket{a}, \\
    \ket{\psi_2} &= \ket{b}, \\
    \ket{\psi_3} &= \ket{c}, \\
    \ket{\psi_4} &= \frac{1}{\sqrt{3}}(\ket{a} + \ket{b} + \ket{c}), \\
    \ket{\psi_5} &= \frac{1}{\sqrt{3}}(\ket{a} + \omega\ket{b} + \omega^2\ket{c}), \\
    \ket{\psi_6} &= \frac{1}{\sqrt{3}}(\ket{a} + \omega^2\ket{b} + \omega\ket{c}), \\
    \ket{\psi_7} &= \frac{1}{\sqrt{3}}(\omega\ket{a} + \ket{b} + \ket{c}), \\
    \ket{\psi_8} &= \frac{1}{\sqrt{3}}(\ket{a} + \omega\ket{b} + \ket{c}), \\
    \ket{\psi_9} &= \frac{1}{\sqrt{3}}(\ket{a} + \ket{b} + \omega\ket{c}), \\
    \ket{\psi_{10}} &= \frac{1}{\sqrt{3}}(\omega^2\ket{a} + \ket{b} + \ket{c}), \\
    \ket{\psi_{11}} &= \frac{1}{\sqrt{3}}(\ket{a} + \omega^2\ket{b} + \ket{c}), \\
    \ket{\psi_{12}} &= \frac{1}{\sqrt{3}}(\ket{a} + \ket{b} + \omega^2\ket{c}),
\end{align}

\BS{and were prepared from the $\ell = \{-1,0,1\}$ OAM subspace in our case where $a = -1, b = 0$, $c = 1$ and $\omega = e^{i\frac{2\pi}{3}}$. Here, the phase profiles of each are shown as insets along the x-axis in the figure.}

Based on the MUB tomography projection measurements, the density matrix $\rho_{Ex}$, for each of the transferred MUB states was reconstructed using the maximum likelihood algorithm \cite{JamesMeasurement2001}. The exact fidelities, calculated from $F = \text{Tr}(\sqrt{\sqrt{\rho_{Th}}\rho_{Ex}\sqrt{\rho_{Th}}})^2$ where $\rho_{Th}$ is the theoretical density matrix of the pure MUB state being detected, are then given in Suppl. Table \ref{tab:MUBfidelities}. %Corresponding results for the 12 MUB states reported in Refs. \cite{luo2019quantum, PhysRevLett.125.230501} are shown in columns to the right for comparison and a final fidelity, averaged over all the states, given at the bottom. It follows, that our protocol compares well with those reported where $F_{ave} = 0.895 \pm0.042$ in our scheme compares to $F_{ave} \approx 0.75 \pm 0.055$ \cite{luo2019quantum} and $F_{ave} = 0.70 \pm0.026$ \cite{PhysRevLett.125.230501}.}
\begin{table}[h!]
	\begin{tabular}{|c|c|} 
		\hline
		 State	& Fidelity \\
		\hline
		$\ket{\psi_1}$ & 0.96 $\pm0.025$  \\
		\hline
		$\ket{\psi_2}$ & 0.91 $\pm0.033$ \\
	    \hline
	 	$\ket{\psi_3}$ & 0.91 $\pm0.039$ \\
	 	\hline
	 	$\ket{\psi_4}$ & 0.82 $\pm0.039$ \\
	    \hline
		 $\ket{\psi_5}$	& 0.88 $\pm0.058$ \\
		\hline
		$\ket{\psi_6}$ & 0.84 $\pm0.015$ \\
		\hline
		$\ket{\psi_7}$ & 0.88 $\pm0.041$ \\
		\hline
		$\ket{\psi_8}$ & 0.95 $\pm0.053$ \\
		\hline
		$\ket{\psi_9}$ & 0.87 $\pm0.026$ \\
		\hline
		$\ket{\psi_{10}}$ &  0.89 $\pm0.029$ \\
		\hline
		$\ket{\psi_{11}}$ & 0.89 $\pm0.073$ \\
		\hline
		$\ket{\psi_{12}}$ & 0.85 $\pm0.031$ \\
		\hline
		$F_{ave}$ & 0.90 $\pm0.042$ \\
		\hline
	\end{tabular}
\caption{\textbf{Fidelities for the $d$ = 3 transferred MUB states.} Measured quantum transport fidelities for each one of the 12 MUB states and the qutrit overall fidelity resulting from the average.}
	\label{tab:MUBfidelities}
\end{table}

\begin{center}
    \textbf{Supplementary Note 12 - Unbalanced quantum transport}
\end{center}

\BS{In the following section, four different states of unequal amplitude weightings were constructed and transferred as illustrated in Suppl. Fig \ref{UnevenStates}. Here the states, $\ket{\psi} = 2\ket{-1} + 3\ket{0} + \ket{1}$, $\ket{\psi} = 2\ket{-2} + 3\ket{0} + \ket{2}$, (c) $\ket{\psi} = \ket{-2} + 2\ket{0} + \ket{2}$ and $\ket{\psi} = 2\ket{-3} + \ket{-1} + \ket{1} + 2\ket{4}$
% $\ket{\psi} = \frac{2}{\sqrt{6}}\ket{-1} + \frac{3}{\sqrt{6}}\ket{0} + \frac{1}{\sqrt{6}}\ket{1}$, $\ket{\psi} = \frac{2}{\sqrt{6}}\ket{-2} + \frac{3}{\sqrt{6}}\ket{0} + \frac{1}{\sqrt{6}}\ket{2}$, (c) $\ket{\psi} = \frac{1}{\sqrt{4}}\ket{-2} + \frac{2}{\sqrt{4}}\ket{0} + \frac{1}{\sqrt{4}}\ket{2}$ and $\ket{\psi} = \frac{2}{\sqrt{4}}\ket{-3} + \frac{1}{\sqrt{4}}\ket{-1} + \frac{1}{\sqrt{4}}\ket{1} + \frac{2}{\sqrt{4}}\ket{4}$ 
were prepared and shown in Suppl. Fig \ref{UnevenStates} (a) to (d), respectively. The bar outlines indicate the prepared state weights. Filled-in areas then show the measured values after the quantum transport to photon B. Subsequently, similarities of (a) 0.98 $\pm0.078$, (b) 0.99 $\pm0.072$, (c) 0.99 $\pm0.061$ and (d) 0.95 $\pm0.04$ were calculated, using the equation outlined in the Methods section of the main paper. Good agreement between the prepared and measured values can thus be seen, indicating that general states with different amplitudes may be transferred using our scheme.}
\begin{figure}[h!]
    \centering
    \includegraphics[width=0.9\linewidth]{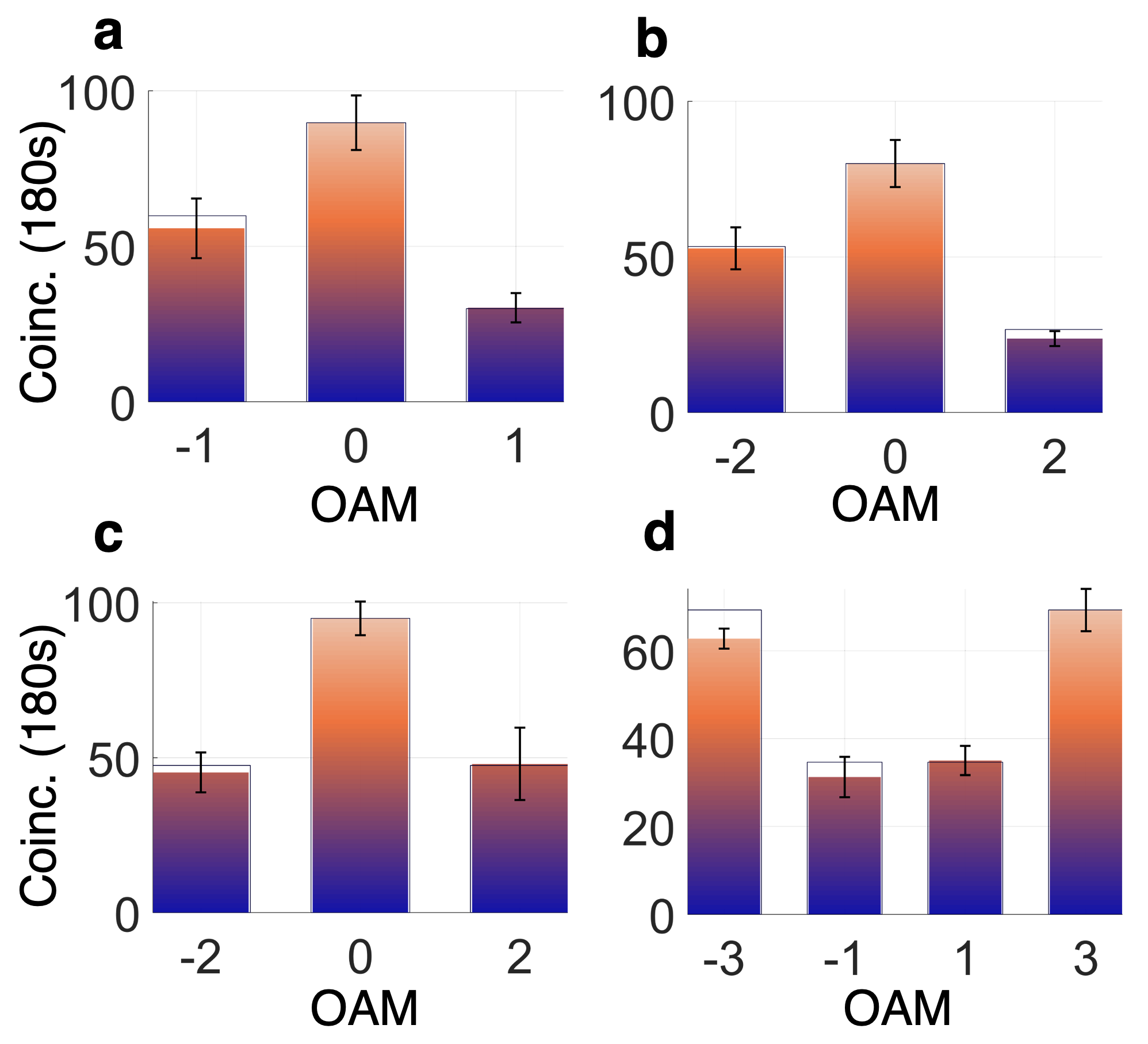}
    \caption{\textbf{Quantum transport of unevenly weighted states.} Experimental measurements (filled bars) of uneven encoded superposition states (bar outlines) for (a) $\ket{\psi} = 2\ket{-1} + 3\ket{0} + \ket{1}$, (b) $\ket{\psi} = 2\ket{-2} + 3\ket{0} + \ket{2}$, (c) $\ket{\psi} = \ket{-2} + 2\ket{0} + \ket{2}$ and (d) $\ket{\psi} = 2\ket{-3} + \ket{-1} + \ket{1} + 2\ket{4}$.}
    \label{UnevenStates}
\end{figure} 

\begin{center}
    \textbf{Supplementary Note 13 - Raw experimental measurements with uncertainties}
\end{center}

\BS{Additional plots are given in this section showing the uncertainties related to measurements in the main text where it was not possible to plot the error bars. \revise{Note that all measurements in this work were repeated between three and five times (limited by time constraints due to long acquisition times) and from this computed the average and standard deviation. A propagation of error analysis was then used to obtain the corresponding uncertainties in all the proceeding values and measures that were computed.} Supplementary Figure \ref{TomoError} shows the tomography data that was taken in order to reconstruct the three-dimensional channel density matrix that was provided in main text Fig. \ref{fig:VisTomoDim}(b). The two-dimensional superposition sub-spaces are indicated by the brackets with the general form of the superposition shown above. The set of values [$0, \frac{\pi}{2}, \pi , \frac{3\pi}{2}$] gives the specific $\theta$ angle used to generate the state prepared and/or measured. The values printed in each of the measurement blocks show the experimental standard deviation.}
\begin{figure}[h!]
    \centering
    \includegraphics[width=\linewidth]{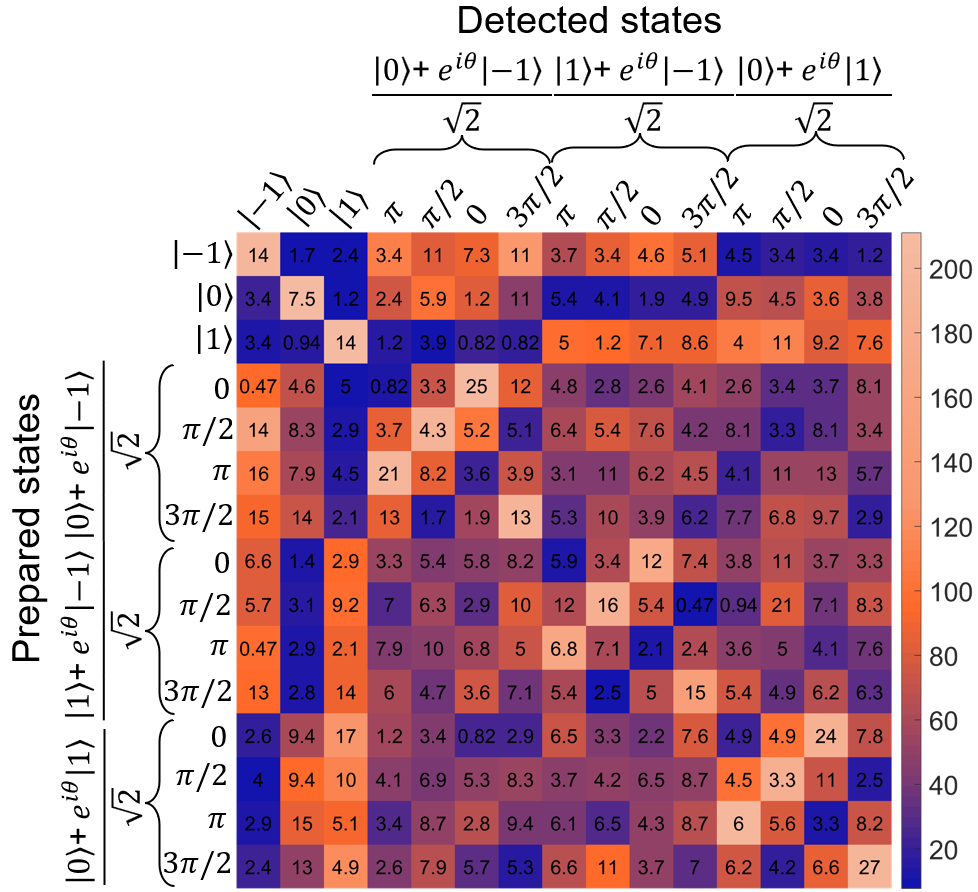}
    \caption{\textbf{Channel tomography measurements with uncertainties.} Experimental measurement plot shown with the detected coincidences given by the false colormap, and the associated uncertainties printed in each measurement block considering an integration time of 120s. $\theta = 0, \frac{\pi}{2}, \pi$ and $\frac{3\pi}{2}$ as indicated for each of the bracketed superposition subspaces.}
    \label{TomoError}
\end{figure} 

\BS{Uncertainties associated with the detection matrix for the four-dimensional MUB constructed from $\ell = [-3, -1, 1, 3]$ in main text Fig. \ref{fig:VisTomoDim}(c) is shown in Suppl. Fig. \ref{MUBError}. Similarly to Suppl. Fig. \ref{TomoError}, the false colormap indicates the coincidences measured with each of the respective errors printed in the measurement blocks.}
\begin{figure}[h!]
    \centering
    \includegraphics[width=0.8\linewidth]{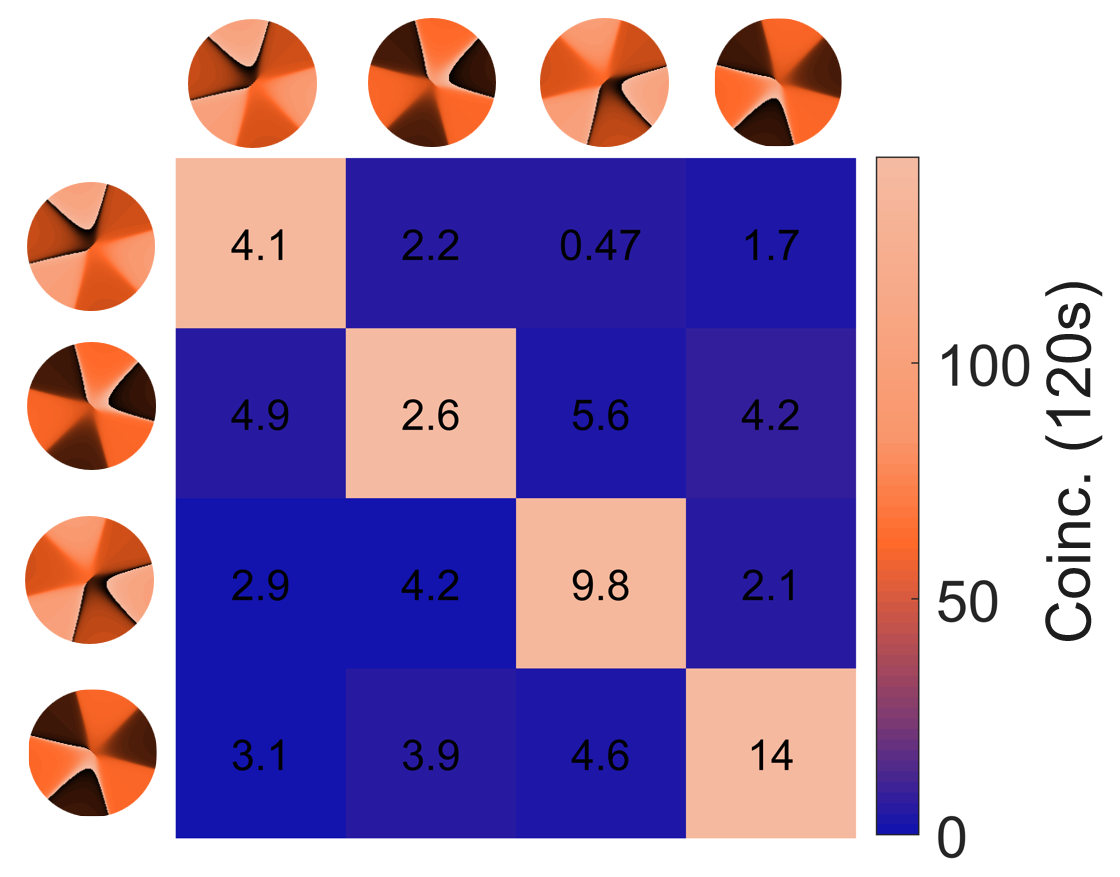}
    \caption{\textbf{Four-dimensional MUB measurements with uncertainties.} Experimental measurement plot shown with the detected coincidences given by the false colormap and the associated uncertainties printed in each measurement block.}
    \label{MUBError}
\end{figure} 

\BS{Supplementary Figure \ref{MUBtomoMatrix} shows the raw averaged measurements taken for the three-dimensional state tomography across all 12 MUB states listed in Supplementary Note 11. Here the colormap indicates the coincidence counts measured over a 120s and the printed values in the measurement blocks indicates the standard deviation associated with each measurement. The phase profiles of each MUB state are shown as insets along the x- and y- axes in the figure. As can be seen, clear detection of the prepared MUB state is obtained which is given by the strong diagonal with close to null values in the off-diagonal terms in each basis. Based on these measurements, the density matrix $\rho_{Ex}$, for each of the transferred MUB states was reconstructed using the maximum likelihood algorithm \cite{JamesMeasurement2001,Agnew2011}. }
\begin{figure}[h!]
    \centering
    \includegraphics[width=\linewidth]{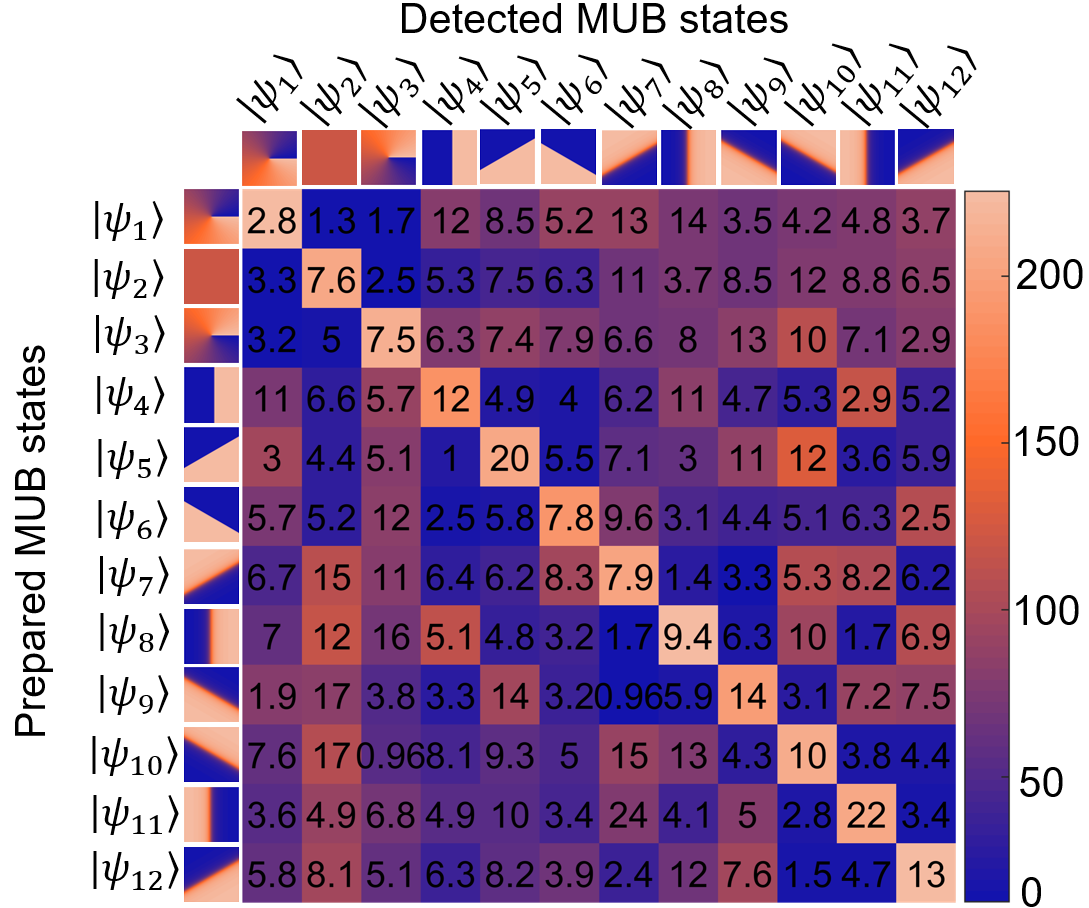}
    \caption{\textbf{Transferred MUB states tomography.} Experimental state tomography measurements considering an integration time of 120s performed on the 12 MUB states for a $d$ = 3 space comprised OAM modes $\ell$ = \{-1,0,1\}. Numbers printed on the measurement blocks indicate the associated standard deviations measured.}
    \label{MUBtomoMatrix}
\end{figure} 

\BS{We give the standard deviations of the largest spiral bandwidth plot taken in Suppl. Fig. \ref{StdDevSBW}. This corresponds to the spiral bandwidth ranging from $\ell = [-8,8]$ shown in the subplot of main text Fig. \ref{fig:DimFid}. The false colour shows the average coincidences detected over a 240s integration time with the numbers again giving the calculated standard deviation obtained from each measurement.}
\begin{figure}[h!]
    \centering
    \includegraphics[width=\linewidth]{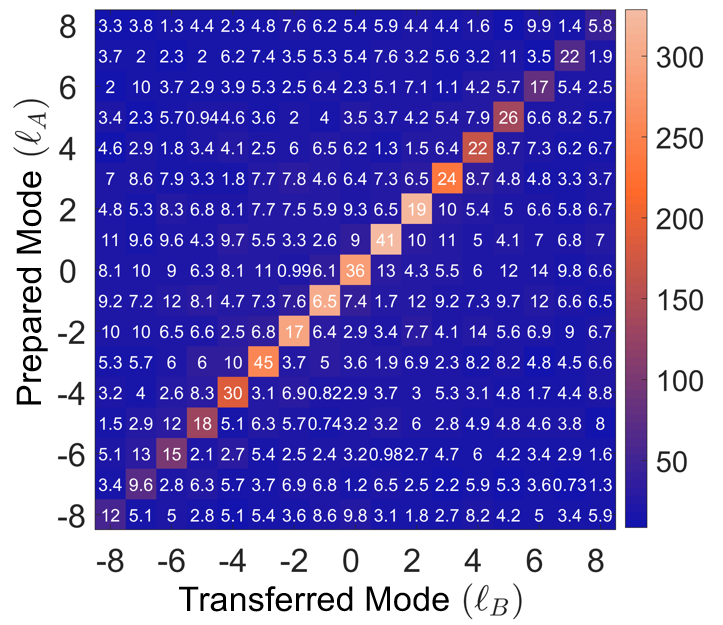}
    \caption{\textbf{Spiral bandwidth measured for high dimensionally tuned setup.} Experimental measurements considering an integration time of 240s with standard deviations printed on the average coincidences shown by the false colormap for the spiral bandwidth of an optimally tuned system where $K$ $\approx$ 15.}
    \label{StdDevSBW}
\end{figure} 

\begin{center}
    \textbf{Supplementary Note 14 - Quantum transport fidelity results}
\end{center}

\begin{figure}[h!]
    \centering
    \includegraphics[width=\linewidth]{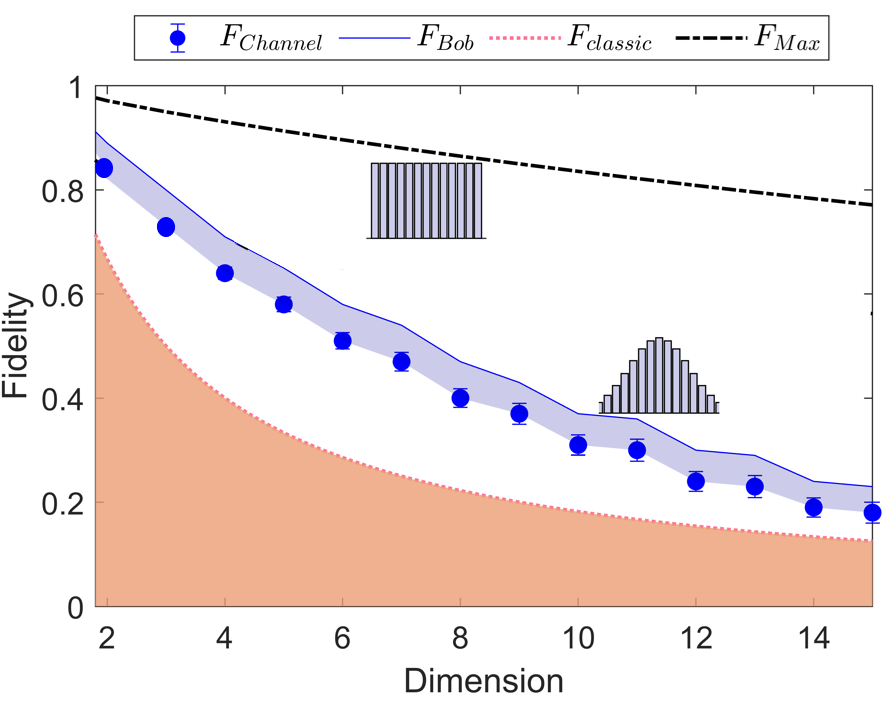}
    \caption{\textbf{\IN{Quantum transport fidelities.}} \IN{The measured quantum transport fidelities for the channel ($F_{channel}$) for states that photon B receives ($F_{B}$) with respect the classical bound ($F_{classic}$) are shown. Our system shows the possibility of transferring up to $d=15$ dimensions. Moreover, since our spectrum for the OAM basis was not flat (maximal correlations), we show how a flat spectrum would improve the quantum transport fidelities ($F_{max}$) under the same experimental conditions.}}
    \label{fig:FidelityCompare}
\end{figure} 

In Suppl. Fig.~\ref{fig:FidelityCompare} we show the fidelities measured from the dimensionality and purity test \cite{nape2021measuring}. Our method extracts the fidelity of the channel $F_{Ch}$ described in the Methods section of the main text. From this we can compute the expected fidelity for each photon B as $\mathcal{F} = \frac{ F_{Ch}d+1}{d+1}$ \cite{horodecki1999general}. Since our spectrum was not uniform, i.e., resembling a system with perfect correlation, we also show the expected fidelity ($F_{max}$) for such a system. Nonetheless, our measurements are all above the classical bound.

\BS{One aspect to note is the Gaussian-like falloff of the experimentally measured correlations as the higher order modes are detected in both the SPDC entangled photons which were measured in Suppl. Fig. \ref{SPDCana} as well as the quantum transport signal shown in figures such as Suppl. Fig. \ref{background}. This feature of such higher-order modes is well-known  and studied due to the detection sizes and effect of the SPDC pump shape \cite{miatto2012bounds,roux2014projective,nape2020enhancing}. In the process of optimising the quantum transport channel bandwidth, however, the modal detection and up-conversion sizes were incidentally adjusted such that the first three modes (i.e. $\ket{-1}, \ket{0}$ and $\ket{1}$) were especially flat with respect to each other. This is shown in the example distribution given in Suppl. Fig. \ref{threeModes} as well as seen in the projective measurements shown in Suppl. Fig. \ref{TomoError}.
\begin{figure}[h!]
    \centering
    \includegraphics[width=0.9\linewidth]{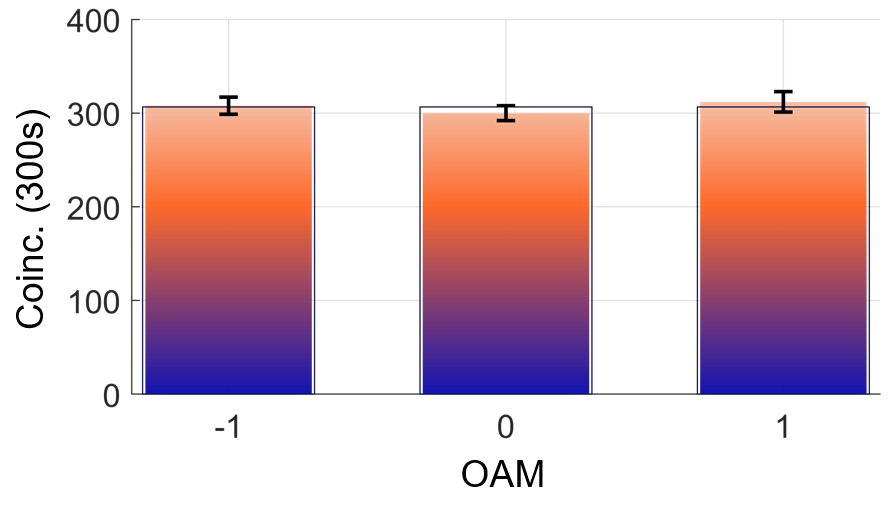}
    \caption{\textbf{\IN{Size-matched flattened modes.}} \IN{Example of the experimentally measured transferred spectrum diagonal for the lowest three modes used to comprise the twelve $d$ = 3 MUB transferred states.}}
    \label{threeModes}
\end{figure}
As such, for these lower order modes, the system correlations resembled the perfect correlations indicated by the $F_{max}$ (dotted line) more closely than the Gaussian fall-off model used to extract the fidelities as given by the blue dots and lines. With this factor, the MUB states comprised of these modes gave higher than predicted fidelities shown in Suppl. Fig. \ref{MUBFid} with values ranging between 0.82 and 0.96 (which is close to the $F_{max}$ value for $d$ = 3 considering a flat spiral bandwidth).} 

\begin{center}
    \textbf{Supplementary Note 15 - From transport to teleport}
\end{center}

There are many configurations in which our scheme could be deployed depending on where we implement the entanglement source, i.e., either in Bob's laboratory, in Alice's or outside of both as a shared resource.  In our considered variation, shown in Suppl. Fig.~\ref{fig:Transport2Tele} (a), Bob and Charlie send their photons to Alice, the former with no information (one photon from an entangled pair) and the latter with the information to be transported (in a bright laser beam). \rvs{We consider in this example a third-party, i.e., Charlie, preparing the state to be transferred, so to emphasize in the similarity between our quantum transport configuration and the quantum teleportation, where such high-dimensional spatial state could be encoded in a single photon (see Suppl. Fig.~\ref{fig:Transport2Tele} (b)). In any of the cases Alice does not need to know the state that is sent to her, never encodes information on any photons, and in our particular case never even sends any photons to Bob.} Instead she makes a measurement on Bob and Charlie's photons with a nonlinear crystal in a manner that is basis, state and dimension independent (she ``simply'' directs the photons to a nonlinear crystal).  As a result of this measurement, high-dimensional information is transferred to Bob. The deployment of our nonlinear quantum transport scheme shown in Suppl. Fig.~\ref{fig:Transport2Tele} (a) has some interesting properties: (1) the information exchange is conditioned on coincidences, so any eavesdropper would intercept a mixed state with no information, (2) Bob is not expecting any photon from Alice. In other words, Alice never sends any photons across the channel, so cannot cheat by making copies of the coherent state and sending them to Bob. The physics of this could be useful in practical setting, e.g., a bank wanting to securely send or receive the fingerprint (high-dimensional spatial information) of a customer without allowing any eavesdroppers to intercept the information. In our scheme this is guaranteed so long as the sender is considered to be a trusted node, similar to the semi-device-independent protocols.
\begin{figure}[h!]
\centering
\includegraphics[width=1\linewidth]{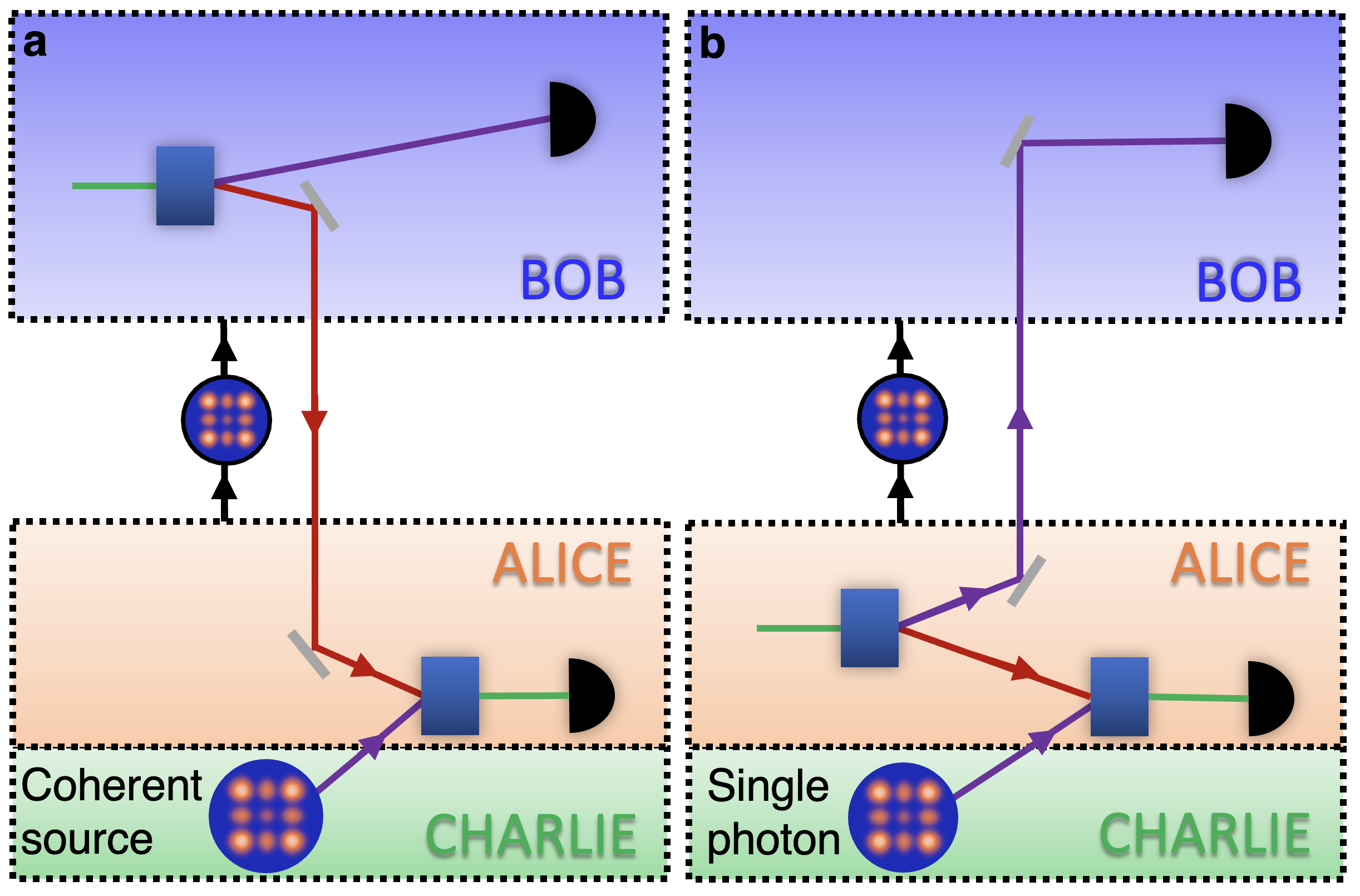}
\caption{\textbf{From quantum transport to quantum teleportation.} (a) In the current experimental configuration, information encoded on a coherent source is necessary to achieve the efficiency required for the nonlinear detector to transfer information.  Information and photons flow in opposite directions.  Alice need not know this information for the process to work, never prepares or sends any photons, and so the state can be arbitrary and unknown.  (b) For sufficiently increased nonlinear efficiency the coherent state could be replaced by a single photon to perform true quantum teleportation without any changes to the physics or conceptualisation of the present scheme.}
\label{fig:Transport2Tele}
\end{figure}

It is interesting to ask what differentiates our scheme from quantum teleportation, as conceptually shown in Suppl. Fig.~\ref{fig:Transport2Tele} (b). What constitutes the core of teleportation is the capacity to transfer an {\it arbitrary} quantum state of a system to another distant system using only entanglement and classical communication as a resource \cite{PhysRevLett.70.1895}. Moreover, the state transmission is accomplished in a secure way by destroying the information during the transfer process. This prevents the possibility of creating copies of the transmitted state elsewhere, safeguarding the privacy of the transmitted quantum state \cite{barnett2022single}. The quantum teleportation protocol also forms the basis of quantum repeaters and distributed quantum networks. From this it is clear that our scheme is neither quantum teleportation, because from a strict theoretical point of view the state is not completely destroyed after being transferred (although she cannot use them, Alice has more copies), nor remote state preparation \cite{PhysRevLett.87.077902}, in which the input state {\it must be} known by Alice for it to work.

Here, it is essential to acknowledge that our current experiment falls short of meeting the full requirements for applications in quantum repeaters and quantum computation due to the use of a bright and in principle knowable coherent state as the teleportee.
% However, despite this technological limitation, our experiment is built on the same physics, leveraging entanglement and classical communication to securely transfer a quantum state from one system to another. 
This nonlinear quantum transport can be seen as inspired by the principles of teleportation, and although it may not fulfill the complete requirements due to current technological limitations in nonlinear optics, e.g. for application in quantum repeaters, it is intriguing to contemplate its potential applications in quantum technology. It is important to note also that from a practical point of view the fact that Bob is not expecting any photon, makes our scheme robust against any potential cheating sender.

%Both linear teleportation and our nonlinear quantum transport scheme require an entangled photon pair as a quantum resource, but linear optical approaches for qudit teleportation require ancillary photons in a manner that the experimental configuration is hard-coded for a particular channel's dimensionality. 
We use a bright coherent state produced by a laser as the input source in order to enhance the up-conversion efficiency of the nonlinear crystal (full details in the Supplementary Note 8 and 10), but with the outcomes still conditioned on bi-photon coincidences: single photon processes at both crystals. Notably, our efficiency hurdle is dimension independent, $\gamma^2$, where $\gamma$ is the entanglement generation and nonlinear detection efficiency (assuming SPDC and anti-SPDC as the processes). %whereas the linear approach scales exponentially with dimension as $\gamma^{(d-1)/2}$ so that even small inefficiencies are heavily penalised. %As an example, if materials forming the non-linear platforms for creation and detection reach $\gamma = 0.1$ then for $d$ = 15 our approach would have an efficiency of $0.1^2 = 10^{-2}$ while with the linear case it is $10^{-7}$; 100 dimensions again returns $10^{-2}$ for our approach (crucially, it does not depend on $d$) but now $10^{-49}$ for the linear case. Thus while the need for ancillary photons is fundamental to linear optical teleportation schemes, 
Thus the need for additional photons in the nonlinear case is a technological need because of present low efficiency, and may improve in the future, e.g. already by orders of magnitude by using artificial nonlinear media such as metasurfaces and metamaterials \cite{kivshar2018all}, being used also recently to generate complex quantum states \cite{santiago2022resonant}, although still far from the efficiency needed for single photons.  Such an advance would allow our approach to seamlessly transition from quantum transport to quantum teleport, as shown graphically in Suppl. Fig.~\ref{fig:Transport2Tele} (b), for true high-dimensional teleportation of quantum states. 

%\bibliographystyle{ieeetr}
%\bibliography{mybibfile.bib}

}
\end{document}